\documentclass[a4paper, 12pt, oneside]{article}
\usepackage [english]{babel}
\usepackage{natbib} %Serve per la bibliografia
\usepackage{authblk}%author w/affiliation
\usepackage[T1]{fontenc}
\usepackage[utf8]{inputenc}

\usepackage{amsmath,amssymb}
\usepackage{mathabx}
\usepackage{tabularx}
\usepackage{latexsym}
\usepackage{amsfonts}
\usepackage{mathtools} %to be used to define =^H
\usepackage{bm}%bold greek letters
\usepackage{dsfont}%indicator operator
\usepackage{bbm}%indicator operator

\usepackage{amsmath,amsthm,amssymb}
\usepackage{amscd}
\usepackage[inline]{enumitem}%easy to customize than \enumerate

\usepackage{multirow}
\usepackage{graphicx,graphics,rotating}
\usepackage[graphicx]{realboxes} %rotate tables
\usepackage{epstopdf} % for eps files
\usepackage[font=scriptsize,labelfont=bf,labelsep=period,textfont=it]{caption}

\usepackage{mathrsfs}
\usepackage[a4paper,top=2.5cm,bottom=2.5cm,left=2cm,right=2cm,height rounded,bindingoffset=0.5cm]{geometry}

\usepackage{etoolbox}
\usepackage{xpatch}  
\usepackage[gen]{eurosym}
\usepackage{lscape}
\usepackage[title,titletoc]{appendix}

\usepackage{sgame}
\usepackage{setspace}
\usepackage[font={small,sl}]{caption}
\usepackage{booktabs,caption} %to be used w/ \captionsetup
\usepackage[para,online,flushleft]{threeparttable}	%to align notes under table

\usepackage{blindtext}
\usepackage{fancyhdr}
\usepackage{setspace}
\usepackage[explicit]{titlesec}
\usepackage[normalem]{ulem}
\usepackage{lipsum}
\usepackage{mwe}  
\usepackage{verbatim}
\usepackage{booktabs,array}  
\usepackage{listings}
\usepackage{subfig}
\usepackage[colorlinks=true,citecolor=blue,linkcolor = blue,urlcolor  = blue]{hyperref}
\usepackage{xcolor}
\usepackage{amssymb}

\usepackage{booktabs}
\usepackage[hang,flushmargin]{footmisc}

\usepackage[linesnumbered,lined,ruled,commentsnumbered]{algorithm2e}

\usepackage{setspace}

\makeatletter

\newcommand{\distas}[1]{\mathbin{\overset{#1}{\kern\z@\sim}}}%
\newsavebox{\mybox}\newsavebox{\mysim}
\newcommand{\distras}[1]{%
  \savebox{\mybox}{\hbox{\kern3pt$\scriptstyle#1$\kern3pt}}%
  \savebox{\mysim}{\hbox{$\sim$}}%
  \mathbin{\overset{#1}{\kern\z@\resizebox{\wd\mybox}{\ht\mysim}{$\sim$}}}%
}

\newtheorem{assumption}{Assumption}
\newtheorem{proposition}{Proposition}

\newtheorem{remark}{Remark}
\newtheorem{definition}{Definition}

\renewcommand{\theequation}{\arabic{section}.\arabic{equation}}

\ifx\csname urlstyle\endcsname\relax
  \newcommand\@doi[1]{doi:\discretionary{}{}{}#1}\else
 \newcommand\@doi{doi:\discretionary{}{}{}\begingroup
\urlstyle{tt}\Url}\fi

\def\vector#1{\mbox{\boldmath{$#1$}}} % Vectors.
\def\matrix#1{\mathbf #1} % Matrices.
\newcommand{\pmat}[1]{\begin{pmatrix} #1 \end{pmatrix}}

\newcommand{\Bmat}[1]{\begin{Bmatrix} #1 \end{Bmatrix}}
\newcommand{\thetano}{\widehat{\theta}}

\newcommand{\Xit}{\mathbf{X}_{it}}

\newcommand{\Dit}{D_{it}}
\newcommand{\dit}{d_{it}}

\newcommand{\Yit}{Y_{it}}
\newcommand{\yit}{y_{it}}

\newcommand{\zit}{Z_{it}}

\newcommand{\uit}{U_{it}}

\newcommand{\vit}{V_{it}}

\newcommand{\rit}{R_{it}}

\newcommand{\bmeta}{\bm{\eta}}

\newcommand{\bmpsi}{\bm{\psi}}

\newcommand{\Var}{\mathrm{Var}}

\newcommand{\inv}{^{-1}}

\newcommand{\E}{\mathbb{E}}
\newcommand{\R}{\mathbb{R}}  %uncomment
\newcommand{\W}{\mathrm{W}}

\newcommand{\indep}[3]{#1 \perp\kern-5pt \perp #2 \mid #3}
\newcommand{\one}{\mathds{1}}

\newcommand{\zero}{\bm{0}}

\newcommand{\Wk}{\mathcal{W}_k}

\title{\Large Double Machine Learning for Static Panel Data with Instrumental Variables: New Method and Applications}

\author{
Anna Baiardi$^{\dagger}$\,
Paul Clarke$^{\ddagger}$\,
Andrea A. Naghi$^{\star}$\,
Annalivia Polselli$^{\ddagger}$%
\thanks{%This paper is Annalivia Polselli's job market paper. 
The authors thank Frank Windmeijer, Federica Liberini, Luca Favero, Angelina Nazarova, and Daniela Sonedda for their helpful comments, and the participants at the QMUL Causal Machine Learning Workshop, 11th ICEEE, 36th EC$^2$ Conference.  
Annalivia Polselli acknowledges support of the British Academy through the Postdoctoral Fellowship (grant number PFSS24/240003). 
The authors acknowledge the use of the High Performance Computing Facility (Ceres) and its associated support services at the University of Essex in the completion of this work.}
}

\date{Latest update: \today}

\affil[$\dagger$]{\footnotesize Erasmus School of Economics, Erasmus University and Tinbergen Institute, Rotterdam, Netherlands.}
\affil[$\star$]{\footnotesize Department of Business Analytics and Applied Economics, Queen Mary University of London, UK}
\affil[$\ddagger$]{\footnotesize Institute for Social and Economic Research, University of Essex, Colchester, UK.}

\begin{document}
\maketitle

%%%%%%%%%%%%%%%%%%%%%%%%%%%
\vspace{-1cm}

\begin{abstract}
%179 words
\setstretch{1.1}
Panel data methods are widely used in empirical analysis to address unobserved heterogeneity, but causal inference remains challenging when treatments are endogenous and confounding variables high-dimensional and potentially nonlinear. Standard instrumental variables (IV) estimators, such as two-stage least squares (2SLS), become unreliable when instrument validity requires flexibly conditioning on many covariates with potentially non-linear effects. %Leveraging machine learning to approximate complex high-dimensional functions of the confounders, 
This paper develops a Double Machine Learning estimator  for static panel models with endogenous treatments (panel IV DML), %The method combines first-difference transformation with Neyman-orthogonal score functions to obtain consistent estimates and valid inference of the structural parameter. 
and introduces weak-identification diagnostics for it. % (first-stage F-statistic and Anderson-Rubin test and confidence sets). 
%Applications to three prominent studies on the effects of immigration on voters' attitudes towards politics and economics reveal stronger effects in one case and weak-instrument concerns in the others due to undetected nonlinearities.
We revisit three influential migration studies that use shift-share instruments. In these settings, instrument validity depends on a rich covariate adjustment.
In one application, panel IV DML strengthens the predictive power of the instrument and broadly confirms 2SLS results. In the other cases, flexible adjustment makes the instruments weak, leading to substantially more cautious causal inference than conventional 2SLS. Monte Carlo evidence supports these findings, showing that panel IV DML improves estimation accuracy under strong instruments and delivers more reliable inference under weak identification. %The proposed framework broadens the set of tools for applied research with endogenous treatments in high-dimensional panel settings.

\medskip
\noindent \textbf{Keywords:} Anderson-Rubin test, causal effects, double machine learning, Neyman orthogonality, panel data, shift-share instrument, weak identification.\\
\textbf{JEL codes:} C14, C18, C33, C36, C45, C52.
\end{abstract}
%%%%%%%%%%%%%%%%%%%%%%%%%%%
% 1. INTRO
\section{Introduction}\label{sec:intro}
Panel data are the backbone of empirical research in modern economics. 
The structure of panel data enables researchers to account for time-invariant unobserved heterogeneity and exploit within-unit variation for causal inference. However, causal estimation remains challenging when treatment variables are endogenous and identification hinges on the validity of instrumental variables (IVs). 
In many applications, instruments are only valid conditional on rich sets of covariates \citep{angrist1995,abadie2003}.
Traditional econometric methods, such as two-stage least squares (2SLS), cannot handle the complexities introduced by high-dimensional confounders or nonlinear relationships. For instance, least squares estimation may be infeasible in high-dimensional settings due to rank deficiency induced by many irrelevant regressors, and it cannot capture nonlinear relationships unless these are explicitly parameterized. This creates a tension between the need to control for a rich set of confounders to justify instrument validity and the limitations of conventional methods under sparsity and model-selection constraints.

This paper presents a new method to address these challenges. We propose a Double Machine Learning (DML) procedure for static panels with endogenous treatments and instrumental variables, which we call `Panel IV DML'. 
Our approach extends the framework of \citet{chernozhukov2018} to accommodate serial dependence and unobserved individual heterogeneity, features typical of panel datasets, and that of \citet{clarke2023} to allow for treatment endogeneity. 
First, we derive a novel DML estimator for static panel data models with endogenous treatments based on Neyman orthogonal score functions that account for unobserved individual heterogeneity. The proposed method allows for flexible predictions of the nuisance functions with various machine learning algorithms. It further delivers first-order normal statistical inference for treatment effects, adjusted for regularization and over-fitting bias through the use of Neyman orthogonal score functions and \emph{block-k-fold} cross-fitting, where each unit's entire time series is assigned to the same fold. 
Second, to ensure credible inference in the possible presence of weak instruments, we propose first-stage {\em F}-statistics and Anderson-Rubin (AR) test statistics and confidence sets allowing us to detect weak identification issues and, thus, ensure valid inference even when the instrument strength is limited. To our knowledge, this is among the first work to develop weak IV tests for DML procedures, providing empirical researchers with new tools for robust instrumental variables analysis.
%Then, $68\% (=108/477)$ of these empirical studies use panel data against only $29\% (=108/477)$ which employ cross-sectional data.
% XS: (140/477)*100 = 29%; Panel: (326/477)*100 = 68%
% IV Panel: 45 + 63 = 108 (w/TS); 8 + 15 = 23 (w/XS) => 131/326*100 = 40%

This framework is particularly relevant in applied economics, as panel data models estimated with an instrumental variable method (e.g., 2SLS, generalized method of moments) account for 40\% of the {\it American Economic Review} (AER) articles using panel data, published in the  between 2011-2018.\footnote{This figure is based on a review of 477 empirical articles published in the \emph{American Economic Review} between 2011 and 2018. In particular, $68\%(=326/477)$ of these inspected articles employ panel data methods. Among the set of these panel data articles, $40\%(=131/326)$ implement an instrumental variables estimation strategy. Detailed information on data collection is provided in Appendix~\ref{sec:meta}.}
We hence illustrate the empirical relevance of our method by revisiting three highly influential studies in migration and political economy that rely on shift-share instruments. Such instruments typically require conditioning on rich sets of covariates to be plausibly exogenous, or \emph{as good as randomly assigned} \citep{borusyak2025practical}, hence these applications provide a natural setting for panel IV DML, which flexibly controls for high-dimensional and nonlinear confounding.

The first empirical study we revisit, \citet{tabellini2020},  studies the political and economic effects of immigration in early 20th-century U.S. cities using a shift-share IV. Our panel IV DML increases the predictive power of the instrument (relevance) since we allow for flexible adjustment in the instrument equation. The panel IV DML estimates broadly align with conventional 2SLS in the baseline specifications for the political outcome, but find no effect for the economic outcome, unlike 2SLS. When additional controls are included in the specification as a robustness check, the main effects found in the original study for both economic and political outcomes disappear under both 2SLS and panel IV DML, suggesting that the results are sensitive to the inclusion of confounders rather than to the estimation method \emph{per se}.
%\red{our Panel IV DML estimator uncovers systematically larger effects on conservative voting patterns than conventional 2SLS, when no other covariates are included, which is consistent with the presence of additional nonlinear confounding that standard methods may not fully capture.}
%\red{We hence provide an additional robustness by including more control variables and we find that the main effect disappears. We strengthened IV, we confirm results for welfare state, disagree with original study on public education; but adding more (possibly nonlinear) controls both 2SLS and panel IV DML estimators yield the same conclusion which disagrees with original results.} 

In contrast, the re-analyses of \citet{moriconi2019,moriconi2022}, which examine the impact of immigration on political and economic attitudes of individual voters and parties, reveal weak-instrument concerns and, therefore, second-stage results should be interpreted with caution.  Weak-identification diagnostics indicate that, while some reduced-form relationships may be present, the available variation is insufficient to support identification of the treatment effect. In these cases, panel IV DML helps assess instrument validity through flexible adjustment, while potentially reducing effective instrument relevance. In these exercises, panel IV DML estimator leads to more cautious inference, emphasising the value of robust weak-identification diagnostics, especially when instruments are only moderately strong with conventional 2SLS. 

%\red{DML really questions strength of IV so then the sign of 2nd stage coefs can be questioned. Without AR in 2SLS regressions, we would have concluded that IV strong and there is no effect, but it is weak and no effect whereas panel IV DML never.}

Our Monte Carlo simulation results support the findings of the empirical applications. Through simulation exercises calibrated to studied empirical designs, we show that panel IV DML outperforms conventional 2SLS in both bias and rate of convergence when confounding is nonlinear and the instrument is strong. When instruments are weak, our Panel IV DML method provides more reliable inference under weak identification than conventional 2SLS, and AR-based confidence sets more reliably reflect true uncertainty than standard inference.
%%%% COMMENTED FOR JMP %%%%
% \newpage
% \begin{center} {\em Contribution to the literature} \end{center}

Our work intersects with several strands of the econometric literature.
First, the paper is situated within the broad field of causal inference methods that employ doubly/debiased (orthogonal) estimation to adjust for high-dimensional confounders \citep{belloni2014, belloni2014inference, belloni2016,chernozhukov2018,chernozhukov2022,knaus2022double,huber2023,scaillet2025,chernozhukov2025fisher,scaillet2025}.
The canonical article is \citet{chernozhukov2018}, who introduced the DML method for cross-sectional models by combining the predictive power of machine learning (ML) algorithms to flexibly predict the functional form of the covariates with statistical methods to obtain valid statistical inference of the structural parameter. %They provide the conditions which must be satisfied by the data generating process and the ML algorithms chosen by the analyst for the resulting DML estimator to be $\sqrt{N}$-consistent and converge in distribution to normality, thus permitting regular normal-based statistical inference.
From this seminal paper, we extend the DML procedure for partially linear regression (PLR) models with instrumental variables to panel data. With repeated observations for each subject in the sample, the presence of the unobserved individual heterogeneity, usually assumed to be correlated with the confounding variables and treatment (i.e.\ fixed effects), poses serious challenges for machine learning and DML, which were not developed for serially correlated data.\footnote{For instance, naively applying to panel data the DML for partially linear regression models with instrumental variables set out by \citet[Sec.\ 4.2]{chernozhukov2018} (i.e.\ using `pooled' estimation) would ignore this heterogeneity and, thus, be inconsistent.}
We thus use conventional panel data transformations (such as first-differences) for static panels to remove the linear effect of the unobserved heterogeneity. Then, provided that we can learn new nuisance functions (that is, transformations of the nuisance functions of the original partially linear panel model), we show how DML can be used to obtain a consistent estimator of the structural parameter with desirable asymptotic properties.  

Second, within the literature on causal machine learning, our contribution relates to a growing strand that extends the DML framework to both static and dynamic panel data models targeting different causal estimands. In static panel settings, \citet{klosin2022} propose an estimator for the average partial effect in models with continuous exogenous treatments, using a first-difference transformation to eliminate unobserved time-invariant heterogeneity. 
Extensions of DML to dynamic panel data models address additional sources of endogeneity arising from predetermined regressors and lagged outcomes. \citet{semenova2023} develop a procedure for dynamic panel models with predetermined variables, binary treatments and fully heterogeneous treatment effects, relying on weak sparsity and linearity assumptions suitable for Lasso estimation. \citet{chernozhukov2024} further extend the framework by combining the Arellano-Bond estimation strategy with Lasso to estimate structural parameters in dynamic linear panel models with lagged dependent variables, predetermined covariates, and unobserved individual and time fixed effects.
In contrast to the rest of the DML literature, \citet{arganaraz2025} allow for multivariate and general functionals of unobserved heterogeneity in high-dimensional panel settings, and by providing a full characterization of Neyman-orthogonal moments in models with nonparametric unobserved heterogeneity, even when nuisance components or the distribution of the heterogeneity are not fully identified.

%Other studies have incorporated difference-in-differences (DID) identification strategies into the DML context. For instance, \citet{chang2020} and \citet{zhang2025} focus on the canonical two-period DID design (one pre-treatment and one post-treatment period) within the DML framework. In contrast, \citet{haddad2024} generalize the approach to settings with multiple time periods and potentially time-varying treatment intensities (i.e., a staggered adoption design).  Finally, other works have examined the consequences of various sampling splitting strategies with serially dependent data \citep{fuhr2024,cerqua2025}. 

The work most closely related to ours in this strand of the literature is \citet{clarke2023}. The authors develop DML methods for partially linear panel regression models while accounting for low-dimensional unobserved individual heterogeneity, without imposing sparsity on the nuisance functions, thereby allowing the use of general machine learning algorithms within the DML framework. %\blue{Similarly, \citet{clarke2023} adapt standard panel estimators (namely, correlated random effects, within-group, and first-difference) to static partially linear regression models with fixed effects, without imposing sparsity on the nuisance functions, thereby allowing the use of general machine learning algorithms within the DML framework.} 
We extend their framework by relaxing the exogeneity assumption imposed on the treatment variable in their setting. Allowing for potentially non-random instruments requires explicitly modelling the structural equation of the instrument (uncommon in conventional IV procedures) and, consequently, learning an additional nuisance function for that equation. This IV setting also necessitates the construction of new Neyman-orthogonal score functions tailored to our panel data structure to remove regularization bias and deliver valid asymptotic inference under flexible machine learning estimation of the nuisance functions.
%The closest article to ours in this group is \citet{clarke2023} in the way nuisance functions are learnt from the observed data while accounting for the presence of the low-dimensional unobserved individual heterogeneity. We build on \citet{clarke2023} by relaxing the exogeneity assumption of the treatment variable in their partially linear panel regression (PLPR) model. The structural equations are modelled to allow for potentially non-random instrumental variables, requiring a large set of confounding variables to approximate exogeneity. This new setting requires that we construct Neyman orthogonal score functions (one for the structural model and one for the reduced form model) suitable for our panel data setting to remove the regularisation bias and obtain valid asymptotic.

A complementary article to ours is \citet{marquez2025}, who proposes a DML procedure for panel data to capture nonlinearity in the effects of otherwise valid instrumental variables which would be speciously weak using 2SLS.  %Rather than imposing an additive linear form for the treatment equation, as we do, the author follows the alternative specification by \citet{chernozhukov2018} for PLR models with IV, where 
She uses a control function approach in which ML is used to learn the functional form of the instrument(s) and exogenous regressors in the first-stage.
In contrast, we adopt a first-stage specification for the treatment which is linear in the instrument and nonlinear in the exogenous regressors. This setting gives us a direct connection to conventional 2SLS for scenarios where established weak IV procedures can be used. %for settings where the instruments are weak and/or not plausibly randomly assigned. 
Our approach hence prioritizes improving instrument validity by flexibly controlling for a rich set of confounding variables using ML algorithms, making the instrument \emph{as good as randomly assigned}, and by developing inference procedures that remain valid under weak identification. % In addition,  we do not impose any linearity assumption in the functional form of the covariates in the outcome equation, which allows us to accommodate more complex data structures with machine learning algorithms. 
%By allowing the instrument's effect to enter nonlinearly, this approach addresses a key weakness of standard 2SLS: when the true first-stage relationship is nonlinear in the instrumental variables, a linear specification can severely understate the instrument's predictive power, making even a strong instrument appear weak. 
%Overall, the two approaches are complementary in targeting different potential concerns arising from the use of instrumental variables.  \citet{marquez2025}'s framework is more appropriate when instrument strength is not in question but potential nonlinearities in the treatment equation may obscure identification, whereas our framework is more suitable when instrument strength may be limited, but the structural form is correctly specified, and the functional form of the covariates is potentially nonlinear.
%If the researcher suspects significant nonlinearities in how the instrument affects the treatment (but believes instrument strength is valid), the control-function approach is preferable; if instead the form is correctly specified but instrument strength is in doubt, our panel IV DML framework is more appropriate.

We also interact with the literature on valid inference in IV settings under weak identification. Some studies focus primarily on developing inference tools (including $F$ and $t$-tests, adjusted critical values) for detecting weak instruments in low-dimensional structural models \citep{cragg1993,stock2005,kleibergen2006,olea2013,lee2022,windmeijer2025}.
Others built on Anderson-Rubin \citep[henceforth AR]{anderson1949} test statistics to provide robust inference regardless of instrumental relevance in settings with few instruments \citep{moreira2019} or in settings where the number of instruments grows with the sample size \citep[e.g.,][]{anatolyev2011,carrasco2016,crudu2021,mikusheva2022,dovi2024}. %
%In practice, the AR test remains underused in empirical research, despite evidence that 2SLS inference can be severely distorted under weak identification, leading recent work to call for its broader adoption even when instruments appear strong \citep{davidson2014,andrews2019,keane2023,keane2024}.
%
%A central concern highlighted by the rest of this literature is that standard 2SLS inference can exhibit substantial size distortions when instruments are weak or when OLS bias is large, even under heteroskedasticity-robust standard errors \citep[see][]{keane2023,keane2024}. In such cases,  first-stage $F$-statistics may overstate identification strength, critical values become invalid, and robust alternatives, such as the Anderson-Rubin \citep[henceforth AR]{anderson1949} type tests,  provide more reliable inference \citep{davidson2014,andrews2019,moreira2019}. Despite its robustness properties, the AR test remains underused in applied empirical research, where reporting tends to focus on first-stage {\em F}-statistics rather AR test. However, recent work has called for broader adoption of the AR framework, even when instruments appear strong \citep{keane2024}. 
%
Our paper contributes to this conversation by proposing AR-based inference for the panel IV DML estimator with few instruments and under heteroskedasticity, aligning with best practices emerged in the econometrics literature \citep{davidson2014,andrews2019,keane2023,keane2024}.\footnote{The AR test statistic and confidence set are not as commonly used in applied work as the first-stage F-statistic, and is rarely reported in standard 2SLS regression tables. Some examples of empirical studies that report AR test statistics and/or AR confidence sets alongside the conventional weak IV tests are: \citet{clark2024} in labour economics; \citet{ronconi2012} and \citet{abayasekara2025} in health economics; \citet{cruz2005} in the economics of education, and \citet{khammo2024} in public economics; and \citet{fumagalli2021} in household behaviour and family economics. }
Building on this established discussion,  we leverage the desirable asymptotic properties of the DML estimator and the linearity of instrumental variable in the first-stage equation to derive statistical tests for the panel IV DML estimator under weak identification (first-stage {\em F}-statistic, Anderson-Rubin test and confidence sets), not originally proposed for the cross-sectional case.
The proposed statistics incorporate flexible information on confounding variables, cross-fitting strategy, and orthogonal residuals that mitigate the bias arising from misspecified confounding structures. The linearity of the instrumental variable in the first-stage relationship between the instrument and the treatment is fundamental for invoking conventional weak IV asymptotic theory and associated inference procedures.\footnote{A methodological challenge for future research is extending weak IV asymptotics to more general scenarios where the first-stage relationship is nonlinear, as in \citet{chernozhukov2018} and \citet{marquez2025}.}

Finally, we contribute to the broader discussion on the value added of machine learning methods to causal analysis and policy evaluation in economics. A growing body of empirical work \citep[e.g.,][]{deryugina2019,langen2023,strittmatter2023,baiardi2024b,baiardi2024a} documents promising results from the use of causal machine learning techniques (e.g., causal forest, generic machine learning, and double machine learning) to address empirical research questions by uncovering nonlinearities and heterogeneous effects that would otherwise remain undetected with conventional estimation methods. These applications use cross-sectional data or panel data reshaped into a cross-sectional form. Our work complements this literature by extending these insights to panel data settings, where unobserved heterogeneity, nonlinearities, and high-dimensional confounding pose additional challenges for causal inference. Although panel data with instrumental variables is a common estimation strategy in empirical economics, the application of causal machine learning methods to this setting remains largely underdeveloped — a gap this paper directly addresses.
%In this article, we document the value of DML compared to conventional 2SLS in capturing nonlinearities in the data and potentially weak instruments in a panel data framework. 
%Although the setting considered in this paper is arguably more relevant for current empirical applications in economics than standard cross-sectional data, the application of causal machine learning methods to panel data with instrumental variables remains largely unexplored in practice. %due to limitations of ML algorithms, which are not designed to handle serially correlated data and the unobserved individual heterogeneity. %We use \emph{block-k-fold} sample splitting to allow for correlation within the series and adapt the learning algorithm as in \citet{clarke2023} to address the unobserved individual heterogeneity.

The remainder of the paper is structured as follows. Section~\ref{sec:frame} introduces the notation and presents the econometric framework for panel IV DML. Section~\ref{sec:dml_est} describes the estimation and inference with panel IV DML by defining the Neyman orthogonal score functions, the estimators, and the (robust) tests under weak identification.   Section~\ref{sec:empirical} applies the method to three empirical applications with shift-share instruments in the context of the economics of migration. Section~\ref{sec:mcsimul} illustrates the finite sample properties of the panel IV DML estimator with the Monte Carlo simulation exercises. Section~\ref{sec:conclusion} concludes.

% 2. Econometric Framework for Panel IV: 
\setcounter{equation}{0}
\setcounter{theorem}{0}
\setcounter{lemma}{0}
\setcounter{proposition}{0}
% Machine learning; Model and Assumptions; transformation; the bias of XS-DML
\section{Econometric Framework for Panel IV}\label{sec:frame}
\setcounter{equation}{0}
%%%%%%%%%%%%%%%%%%%%%%%%%%%%%%%%%%%%%%%%%%%%%%%%%%%%%%%%%%%%%%%%%%%%%%%%%%%%%%%%%
\subsection{Notation}
Suppose that individual $i$ is randomly drawn from a population and that measures are collected for everyone in the sample over multiple time periods $t$ (or waves in survey studies). %\footnote{The estimation problem and results hold with unbalanced panel with appropriate modifications in the notation.}  
Let $\{(\Yit,\Dit,\vector Z_{it},\vector X_{it}) : \ t=1,\ldots,T\}_{i=1}^{N}$ be  independent and identically distributed (\emph{iid}) random vectors for each of the $N$ individuals across all $T$ time points, where $\Yit \in \mathcal{Y}$ is the outcome variable, $\Dit\in{\cal D}$ a continuous or binary endogenous treatment  variable (or intervention), $\vector Z_{it}= (Z_{it,1},\dots,Z_{it,r})'\in{\cal Z}$ is a $r\times1$ vector of valid instruments (continuous or binary), and $\vector X_{it} = (X_{it,1},\dots,X_{it,p})'\in \mathcal{X}$ a $p\times 1$ vector of control (pre-determined) variables, usually including a constant term, able to capture time-varying confounding induced by non-random treatment selection.\footnote{Throughout, we use ${\bf v}^\prime$ to indicate the matrix transpose of arbitrary vector {\bf v} and take vectors to be column vectors unless the opposite is explicitly stated.} We denote the realizations of these random variables by $\{(\yit, \dit, \vector z_{it},\vector x_{it})\}$, respectively. For continuous $\Dit \in \mathcal{D}\subset\R$, if $\dit\geq 0$ then it is assumed that a dose-response relationship is maintained with $\Dit=0$ indicating null treatment; otherwise, $\Dit$ is taken to be centered around its mean $\mu_D$ such that $\Dit \equiv \Dit-\mu_D$. For binary $\Dit\in\{0,1\}$, $\Dit=0$ is taken to indicate the absence and $\Dit=1$ the presence of treatment.
In the interval between times $t-1$ and $t$, it is assumed that the realizations of time-varying predictor $\vector X_{it}$ and instrument $\vector Z_{it}$ precede that of $(Y_{it},\Dit)$.

%%%%%%%%%%%%%%%%%%%%%%%%%%%%%%%%%%%%%%%%%%%%%%%%%%%%%%%%%%%%%%%%%%%%%%%%%%%%%%%%%%%%

\subsection{Model and Assumptions}
Consider the following partially linear panel regression (PLPR) model with instrumental variables, based on extending the cross-sectional model by \citet{robinson1988} (see also \citet[sec.\ 4.2]{chernozhukov2018}):
\begin{align}
% \Yit &= \rit\theta_0 + l_0(\vector X_{it}) + \alpha_i + \uit, \label{eqn:yit} \\
 \Yit &= \{\Dit-r_0(\Xit)\}\theta_0 + l_0(\vector X_{it}) + \alpha_i + \uit, \label{eqn:yit} \\
 \Dit & = \vector \vit'\bm{\pi}_0 + r_0(\vector X_{it}) + \zeta_i + \rit, \label{eqn:dit}\\
 %\vector Z_{it} & =  \vector m_0(\vector X_{it}) + \bm{\gamma_i} +  \vector \vit, 
 \vector \vit & = \vector Z_{it} - \vector m_0(\vector X_{it}) - \bm{\gamma_i}, \label{eqn:zit}
\end{align}
\noindent  where $\Yit$ is the outcome of interest, $\Dit$ an \emph{endogenous} treatment variable (or policy intervention), $ \vector Z_{it}$ is a $r\times1$ vector of \emph{exogenous excluded} instruments which is \emph{not fixed}, $\vector X_{it}$ is a $p\times1$ vector of \emph{exogenous included} regressors; $\bmeta_0 = (l_0, r_0, \vector m_0)$ is a vector of possibly nonlinear nuisance functions of the covariates to be estimated using ML algorithms; $\theta_0$ is the structural parameter of interest to estimate consistently to conduct causal inference, and $\bm{\pi}_0$ a $r\times1$ vector of parameters for the \emph{excluded exogenous} variables (or instruments). The $\alpha_i$, $\zeta_i$ and $\bm{\gamma_i}$ are unobserved individual-level, or fixed, effects; all three are functions of the omitted (time-invariant) individual-level omitted variables $\xi_i$.

Endogenous treatment selection induces a correlation between errors $\uit$ and $\rit$ such that  $\E(U_{it}\mid D_{it},\vector X_i,\xi_i)\neq 0$ but \mbox{$\E(U_{it}\mid \vector Z_{it},\vector X_{it},\xi_i)=0$.}  Unlike conventional 2SLS settings, the instruments $\vector Z_{it}$ are not treated as fixed %[and potentially non-randomly assigned; Equation~\eqref{eqn:zit} hence explicitly models the stochastic process generating the instruments.] 
but as wave $t$ realizations conditional on $\Xit$ but preceding $(\Yit,\Dit)$.
Based on (\ref{eqn:zit}), %[the residual ${\vector V}_{it}$ in \eqref{eqn:zit} serves as a vector of orthogonalized instruments, where the effect of ${\vector X}_{it}$ is partialled-out from ${\vector Z}_{it}$, which in turn allows both \eqref{eqn:yit} and \eqref{eqn:dit} to be parametrized in terms of the same nuisance parameter $r_0(\Xit)$.  %\vector\vit\}$  %For simplicity, we assume that $r=1$ but the results can be generalized to multiple instrumental variables.
the role of instrumental variable residuals ${\vector V}_{it}$ is crucial to the construction of our DML procedure. These are required to ensure a) the score function is Neyman orthogonal and, hence, that DML controls the size of the errors introduced by using ML-based estimation of the nuisance functions, and b) the nuisance functions for the first-difference estimator can be consistently learnt (see Remark \ref{rmk:2.4} below). Moreover, using $\vector \vit$ as the instrumental variable rather than ${\vector Z}_{it}$ enables \eqref{eqn:yit} and \eqref{eqn:dit} to be parametrized in terms of the same nuisance parameter $r_0(\Xit)$, which simplifies the resulting DML algorithm.

The assumptions which must be satisfied by the data generating process in order that \eqref{eqn:yit}-\eqref{eqn:dit} holds, and the special role played by \eqref{eqn:zit}, are now set out:
\begin{assumption}\label{item:asm_feedback} 
   (a) \textsc{(No feedback from outcome and treatment to predictors and instruments given omitted variables)} For all $t=1,\dots,T$ with $T\ge2$ 
    \begin{align*}
    \Pr\!\big(\vector{X_{it}}, \vector{Z_{it}}\mid Y_{it-1}, D_{it-1}, \vector{X_{it-1}}, \vector{Z_{it-1}},  &\ldots,Y_{i1}, D_{i1}, \vector{X_{i1}}, \vector{Z_{i1}}, \xi_i \big) \\
    &=\Pr\!\left( \vector{X_{it}}, \vector{Z_{it}}\mid \vector{X_{it-1}}, \vector{Z_{it-1}}, \ldots, \vector{X_{i1}}, \vector{Z_{i1}}, \xi_i\right).
    \end{align*} 
        (b) \textsc{(Local no-feedback assumption)}
        \begin{equation*}
            \Pr\!\left(\vector{X_{it}} \mid \vector{Z_{it-1}}, \vector X_{it-1},\xi_i \right) = \Pr\!\left(\vector X_{it} \mid \vector X_{it-1},\xi_i \right).
        \end{equation*}
    \end{assumption}

\begin{assumption}
\label{item:asm_nolags}
    \textsc{(Static structural model for outcome)} 
    \begin{equation*}
    Y_{it}=g^*_0(\vector X_{it},\xi) + D_{it}\theta_{it} + \epsilon_{it},
    \end{equation*}
    where $\E\left(\epsilon_{it}\mid D_{it},\vector X_i,\xi_i\right)\neq 0$ but $\E\left(\epsilon_{it}\mid \vector Z_{it},\vector X_{it},\xi_i\right)= 0$. $\theta_{it}$ is the causal effect of $D_{it}$ on $Y_{it}$ specific to individual $i$ at wave $t$.
\end{assumption}

\begin{assumption} \label{item:asm_effect_fe}   
    \textsc{(Homogeneity of the treatment effect)} The individual-time treatment effects are mean-independent of the predictors, instruments and omitted time-invariant variables such that
    \begin{equation*}
    \E\left(\theta_{it}\mid D_{it}=d,\vector X_{it},\vector Z_{it},\xi_i\right)=d\theta_0.
    \end{equation*}
\end{assumption}

\begin{remark}
    \rm{$\E\left(\theta_{it}|D_{it}=d\right)=d\theta_0$ without further assumptions, where $\theta_0$ is the average treatment effect on the treated (ATET) if $\Dit$ is binary; and if it can be assumed that  $\E\left(\theta_{it}|D_{it}=0\right)=\E\left(\theta_{it}|D_{it}=1\right)$ then $\theta_0$ is the average treatment effect (ATE).}
\end{remark}

\begin{remark}
\rm{Under Assumptions \ref{item:asm_feedback}(a) and Assumptions \ref{item:asm_nolags}-\ref{item:asm_effect_fe}, it follows that
\begin{equation*}
g^*_0\left(\vector X_{it},\xi_i\right) = \E\left( Y_{it}\mid \vector X_{it},\xi_i\right) - \E\left( D_{it}\mid \vector X_{it},\xi_i\right)\theta_0.
\end{equation*}
Hence, if $l^*_0\left(\vector X_{it},\xi_i \right)\equiv\E\left( Y_{it}\mid \vector X_{it},\xi_i\right)$, we have that
\begin{equation*}
  Y_{it}=l^*_0\left(\vector X_{it},\xi_i\right)+\left\{ D_{it} - \E\left( D_{it}\mid \vector X_{it},\xi_i\right)\right\}\theta_0 + U_{it}
\end{equation*}
\noindent where $U_{it}=\epsilon_{it}+(\theta_{it}-\theta_0)D_{it}$ satisfies $\E\left(U_{it}\mid \vector Z_{it},\vector X_{it},\xi_i\right)= 0$. 

We now require three further assumptions about the relationship between the instruments and predictors, the treatment and instruments:}
\end{remark}

\begin{assumption} \label{item:iv_effect}   
    \textsc{(Nonlinear instrumental variable model)}
    \begin{equation*}
    \vector Z_{it} = {\vector m}^*_0\left(\vector X_{it},\xi_i\right)+\vector V_{it},
    \end{equation*}
where $\E\left(\vector V_{it}\mid \vector X_{it},\xi_i\right)={\bf 0}$.
\end{assumption}

\begin{assumption} \label{item:linear_ivs}   
    \textsc{(Linear effects of instruments on treatment)}
    \begin{equation*}
    D_{it} =\vector V_{it}^{\prime}\bm{\pi}_0 + r^*_0\left(\vector X_{it},\xi_i\right)+R_{it},
    \end{equation*}
where $r^*_0\left(\vector X_{it},\xi_i\right)=\E(\Dit\mid \vector X_{it},\xi_i)$ and $\E\left(R_{it}\mid \vector X_{it},\xi_i\right)=0$.\footnote{Generally, $\E\left(R_{it}\mid \vector{Z}_{it},\vector X_{it},\xi_i\right)=0$ but because Assumptions \ref{item:iv_effect} and \ref{item:linear_ivs} are taken to hold together it requires only $\E\left(R_{it}\mid \vector X_{it},\xi_i\right)=0$.}
\end{assumption}

\begin{remark}
\rm{ The treatment model assumes there are no interactions between the instruments and exogenous regressors, and that the effects of the instruments are linear. \citet{marquez2025} allows a more general specification in which $\E[\Dit\mid\mathbf{Z}_{it},\Xit,\xi]=r_1(\mathbf{Z}_{it},\Xit)+\gamma_i$, where $r_1$ is a second treatment-related nuisance parameter in addition to $r_0(.)$ but Assumption \ref{item:iv_effect} is not needed; this specification allows the analyst to address the problem of weak instruments induced by 2SLS applied to instrumental variables with nonlinear effects on the treatment. In contrast, the setup here is based on a common specification for multiple instruments and permits the use of established techniques for weak instruments, namely, the \cite{stock2005} {\em F}-statistic and Anderson-Rubin confidence intervals, as set out in Section~\ref{sec:dml_est}. }
\end{remark}

\begin{assumption}\label{item:asm_fe}
\begin{enumerate*}[label=(\alph*)]
    \item  \textsc{(Additive separability)} Each of the nonlinear nuisance functions is additively separable such that $l^*_0(\vector X_{it},\xi_i)=l_0(X_{it})+\alpha_i$, ${\vector m}^*_0(\vector X_{it},\xi_i)={\vector m}_0(\vector X_{it})+{\bm\gamma}_i$ and $r^*_0(\vector X_{it},\xi_i)=r_0(\vector X_{it})+\zeta_i$, where $\alpha_i=\alpha(\xi_i)$, ${\bm \gamma}_i={\bm \gamma}(\xi_i)$ and $\zeta_i=\zeta(\xi_i)$ are measurable functions of $\xi_i$. \label{item:additive}
    \item \textsc{(Fixed effects)} The unobserved individual heterogeneity is correlated with the included variables such that $\E(\alpha_i|\vector D_{it}, \vector X_{it})$,  $\E({\bm \gamma}_i|\vector X_{it})$ and $\E(\zeta_i| \vector Z_{it}, \vector X_{it})$ are nonzero.\label{item:fe}
\end{enumerate*}
\end{assumption}

\begin{remark}
\label{rmk:2.4}
\rm{ Fixed effects panel model \eqref{eqn:yit}-\eqref{eqn:dit} follows straightforwardly from Assumption \ref{item:asm_feedback}(a), Assumptions \ref{item:asm_nolags}-\ref{item:asm_effect_fe} and Assumptions \ref{item:linear_ivs}-\ref{item:asm_fe} alone. Assumptions \ref{item:asm_feedback}(b) and \ref{item:iv_effect} are needed to permit consistent learning of the nuisance parameters. Assumption  \ref{item:iv_effect} is not simply a model of the association between instrument and regressor, but an assumption about the data-generating process for the instrument. Together with Assumption~\ref{item:asm_feedback}.b, it ensures that the first-difference estimator (introduced in Section~\ref{sec:transformed_models}) is consistent. Specifically, the two assumptions together block pathways between $Y_{it-1}$ and $\Xit$ induced by omitting $\mathbf{Z}_{it-1}$ from the conditioning set so that $\E[Y_{it-1} \mid \vector{X_{it}}, \vector X_{it-1},\xi_i] = l_0(\vector{X_{it-1}}) + \alpha_i$ and, hence, $\E[Y_{it} - Y_{it-1} \mid \vector{X_{it}}, \vector X_{it-1}] = l_0(\vector{X_{it}}) - l_0(\vector{X_{it-1}})$; it also follows that $\E[D_{it-1} \mid \vector{X_{it}}, \vector X_{it-1},\xi_i] = r_0(\vector{X_{it-1}}) + \zeta_i$ and $\E[\mathbf{Z}_{it-1} \mid \vector{X_{it}}, \vector X_{it-1},\xi_i] = {\vector m}_0(\vector{X_{it-1}}) + \mathbf{\gamma}_i$ so that the nuisance function contrasts involving $r_0$ and ${\vector m}_0$ can be learnt consistently. }
\end{remark}

%In the PLPR model~\eqref{eqn:yit}-\eqref{eqn:yit}, the instrument Z is `as good as random' after conditioning on a set of covariates. Some famous examples in economics include:

%%%%%%%%%%%%%%%%%%%%%%%%%%%%%%%%%%%%%%%%%%%%%%%%%%%%%%%%%%%%%%%%%%%%%%%%%%%%%%%%%%%%%%
%% TRANSFORMED MODEL
%%%%%%%%%%%%%%%%%%%%%%%%%%%%%%%%%%%%%%%%%%%%%%%%%%%%%%%%%%%%%%%%%%%%%%%%%%%%%%%%%%%%%%
\subsection{First-differencing for Panel Data with IV}\label{sec:transformed_models}
We employ the first-difference (FD) transformation to remove the unobserved individual heterogeneity (or fixed effects) in the PLPR Model~\eqref{eqn:yit}-\eqref{eqn:zit}. The FD transformation is preferable over other traditional panel data techniques, such as the within-group (or fixed effects) transformation and correlated random effects \citep{mundlak1978,chamberlain1984}, because it imposes the fewest constraints on the data generating process and permits feasible estimation of the nuisance functions.%\footnote{This is also supported by our Monte Carlo simulation results for panel IV DML with the fixed effects estimation displayed in Tables~\ref{tab:mc_strong_wg}-\ref{tab:mc_weak_wg} in printed Appendix~\ref{sec:app_tabs}. Thus, our recommendation is to employ the FD (exact) approach as described in Section~\ref{sec:algorithm}.}

Let $\widetilde{Y}_{it} = \Yit-Y_{it-1}$ be the first-difference of the random variable $Y_{it}$, and $\widetilde{l}_0(\vector X_{it}) = l_0(\vector X_{it})-l_0(\vector X_{it-1})$ be the first-difference of the nuisance function $l_0$ for all $i=1,\dots,N$ and $t=2,\dots,T$.  The first differences of the other random variables and nuisance functions are similarly defined using \ $\widetilde{}$ \ notation. Then, the first-differenced PLPR model with IV model based on model equations~\eqref{eqn:yit}-\eqref{eqn:zit} is
\begin{align}
 \widetilde{Y}_{it} &= \big\{\widetilde{D}_{it}- \widetilde{r}_0(\vector X_{it})\big\}\theta_0 + \widetilde{l}_0(\vector X_{it}) +\widetilde{U}_{it}, \label{eqn:yit2} \\
 \widetilde{D}_{it} & = \widetilde{\vector V}_{it}'\bm{\pi}_0 + \widetilde{r}_0(\vector X_{it}) + \widetilde{R}_{it}, \label{eqn:dit2}\\
% \widetilde{\vector Z}_{it} & =  \widetilde{\vector m}_0(\vector X_{it}) + \widetilde{\vector V}_{it}, \label{eqn:zit2}
 \widetilde{\vector V}_{it} & = \widetilde{\vector Z}_{it} - \widetilde{\vector m}_0(\vector X_{it}), \label{eqn:zit2}
\end{align}
\noindent where $\alpha_i,\zeta_i$ and $\bm{\gamma_i}$ are now absent from the system, and
\begin{equation}
 \widetilde{Y}_{it} = \widetilde{{\bm V}}_{it}^\prime{\bm\delta}_0 + \widetilde{l}_0(\vector X_{it}) +\widetilde{U}^*_{it}, \label{eqn:redform2} \\
\end{equation}
where ${\bm\delta}_0={\bm\pi}_0\theta_0$ and $\widetilde{U}^*_{it}=\widetilde{U}_{it}+\widetilde{R}_{it}\theta_0$. Following \citet{andrews2019}, we refer to \eqref{eqn:yit2} as the structural model, \eqref{eqn:dit2} as the first-stage model, and \eqref{eqn:redform2} as the reduced-form model.  The estimator of reduced-form model parameter ${\bm\delta}_0$ plays a crucial role in deriving asymptotic results for weak instruments. 

Under the model above, naively using two-stage least squares (2SLS) to estimate the effect of endogenous treatment $\widetilde{D}_{it}$ on $\widetilde{Y}_{it}$ using $\widetilde{{\bf Z}}_{it}$ as an instrumental variable leads to bias. For example, in the simple $T=2$ case and letting $\widetilde{{\bf d}}=(\widetilde{D}_{12},\ldots,\widetilde{D}_{n2})^\prime$, $\widetilde{Z}=(\widetilde{{\bf Z}}_{12},\ldots,\widetilde{{\bf Z}}_{n2})^\prime$ and $\widetilde{X}=(\widetilde{{\bf X}}_{12},\ldots,\widetilde{{\bf X}}_{n2})^\prime$, it can be shown that
\begin{equation*}
\widehat{\theta}^{2SLS}-\theta_0=\frac{(M_{X}\widetilde{\bf d})^\prime M_{\hat{Z}}}{(M_{X}\widetilde{\bf d})^\prime M_{\hat{Z}}(M_{X}\widetilde{\bf d})}M_X\left(\widetilde{{\bf e}}+\widetilde{{\bf g}}_2\right)\equiv {\bf b}\left(\widetilde{{\bf e}}+\widetilde{{\bf g}}_2\right), 
\end{equation*}
where $\widetilde{{\bf e}}$ is the residual of the population-level linear projection of $\widetilde{{\bf y}}=(\widetilde{Y}_{12},\ldots,\widetilde{Y}_{n2})^\prime$ 
on $\widetilde{X}$ and $\widetilde{Z}$, and $\widetilde{{\bf g}_2}=(\widetilde{g}_0({\bf X}_{12})\ldots,\widetilde{g}_0({\bf X}_{n2}))^\prime$; $M_X\widetilde{\bf d}$ is the residual of the linear projection of $\widetilde{\bf d}$ onto the space spanned by the columns of $\widetilde{X}$, or ${\rm span}(\widetilde{X})$, $M_X(\widetilde{{\bf e}}+\widetilde{{\bf g}_2})$ is that for $\widetilde{{\bf e}}+\widetilde{{\bf g}_2}$ onto ${\rm span}(\widetilde{X})$, and $M_{\hat{Z}}{\bf a}$ is the residual of the linear projection of any ${\bf a}$ onto ${\rm span}(M_X\widetilde{Z})$. The usual bias term for 2SLS is ${\bf b}\widetilde{{\bf e}}$, induced by the sample correlation between $\widetilde{{\bf Z}}_{it}$ and the structural error being non-zero, albeit negligible for strong $\widetilde{{\bf Z}}_{i2}$; but ${\bf b}\widetilde{{\bf g}}_2$, the (scaled) linear projection of $\widetilde{{\bf g}_2}$ onto ${\rm span}(\widetilde{X})$, is potentially non-zero whether $\widetilde{{\bf Z}}_{i2}$ is strong or weak.

The main challenge for implementing DML is to learn the transformed \emph{ex-ante} unknown nuisance functions $\widetilde{l}_0(\vector X_{it})$, $\widetilde{r}_0(\vector X_{it})$, and $\widetilde{\vector m}_0(\vector X_{it})$.  For this, we adopt the \emph{FD (exact) approach} proposed by \citet{clarke2023} to learn these functions from the observed transformed data on $\widetilde{Y}_{it}$, $\widetilde{D}_{it}$ and $\widetilde{{\bf Z}}_{it}$, respectively, given ${\bf X}_{it-1}$ and $\Xit$. In the next sections, we handle the first source of bias and derive a consistent estimator of the target (or causal) parameter of interest when the treatment variable is endogenous.

%%\input{002_ml_for_causality}
% 3. Estimator; Inference; Weak IV tests
\section{Estimation and Inference %with Panel IV DML Method
}\label{sec:dml_est}
\setcounter{equation}{0}
\setcounter{theorem}{0}
\setcounter{lemma}{0}
\setcounter{proposition}{0}

This section presents the estimation and inference of the panel IV DML estimator and the weak identification diagnostics. A detailed discussion of the algorithm is in the Online Appendix~\ref{sec:algorithm}.

%%%%%%%%%%%%%%%%%%%%%%%%%%%%%%%%%%%%%%%%%%%%%%%%%%%%%%%%%%
\subsection{Panel IV DML Estimator}
In this section, we set out DML estimation of the panel IV model parameters first by deriving a Neyman-orthogonal score function for estimating structural model parameters %$\widehat{\vector b} = %\widehat{(\theta,\bm{\pi})'}$, 
${\vector b}_0=(\theta_0,{\vector{\pi}}_0^\prime)^\prime$ needed for DML and second by establishing its asymptotic properties.\footnote{Equivalent expressions for reduced-form model parameters ${\vector b}_0^{\rm rf}=(\mathbf{\delta}_0^\prime,{\vector{\pi}}_0^\prime)^\prime$ to facilitate the use of weak instrument techniques are derived, but omitted from the main text.}

%To simplify notation, we stack the random variables by time and individual dimensions. 
Denote the  $N(T-1)\times 1$ random vectors of first differences %for individual $i$ by %
$\widetilde{\vector Y}  = (\widetilde{\vector Y}_1,\dots,\widetilde{\vector Y}_{N})'$, with 
$\widetilde{\vector Y}_i  = (\widetilde{ Y}_{i2},\dots,\widetilde{Y}_{iT})'$ for individual $i$, and the first differences of the nuisance function %${\widetilde l}_0({\bf X}_{it})$ by %is a $(T-1)\times 1$ vector, 
$\widetilde{\vector l}_0 = (\widetilde{\vector l}_{01},\dots,\widetilde{\vector l}_{0N})'$, with $
\widetilde{\vector l}_{0i} = \big(\widetilde{l}_0({\bf X}_{i2}),\ldots, \widetilde{l}_0({\bf X}_{iT})\big)'$ for $i=1,\dots,N$. %and $t=2,\dots,T$.  
The other $N(T-1)$-vectors of first differences %$\{\widetilde{\vector Y}, \widetilde{\vector D},\widetilde{\vector U},\widetilde{\vector R},\widetilde{\vector U}^*,\widetilde{\vector r}_0\}, $ 
$\widetilde{\vector D},\widetilde{\vector U},\widetilde{\vector R},\widetilde{\vector U}^*$ and $\widetilde{\vector r}_{0} $ 
are similarly defined. The $N(T-1)\times r$ matrices relating to instrumental variable model \eqref{eqn:zit2}, where $r$ is the number of valid instrumental variables, are  $\widetilde{\matrix Z}= (\widetilde{\matrix Z}_1,\dots,\widetilde{\matrix Z}_{N})'$ with  $\widetilde{\matrix Z}_i  = (\widetilde{\vector Z}_{i2},\dots,\widetilde{\vector Z}_{iT})'$, and $\widetilde{\matrix V}= \widetilde{\matrix Z}-\widetilde{\matrix M}_{0}$, where $\widetilde{\matrix M}_{0} = (\widetilde{\matrix M}_{01}, \dots,\widetilde{\matrix M}_{0N})'$ with $\widetilde{\matrix M}_{0i} =  \big(\widetilde{{\vector m}_0}(\vector X_{i2}),\ldots,\widetilde{{\vector m}_0}(\vector X_{iT})\big)'$.\footnote{$\widetilde{\matrix M}_{0}$ should not be confused with projection matrix residual $M_X$ defined in the previous section.} %are $(T-1)\times (r+1)$ matrices  for all $i=1,\dots,N$ and $t=2,\dots,T$, 
Finally, let $\matrix X$ be a $NT\times p$ matrix of untransformed confounding variables. The complete set of random vectors and matrices with observed realizations $\matrix  W=\{\widetilde{\vector Y},\widetilde{\vector D},\widetilde{\matrix Z},{\matrix X} \}$. %for individual $i$ is $W_i=\{\widetilde{\vector Y}_i,\widetilde{\vector D}_i,\widetilde{\matrix Z}_i,{\matrix X}_i \}$. %For notational convenience, let $\zero_r$ be a conformable vector or matrix of zeros.

%{\color{red} [PC: Not sure why the "For notational convenience" sentence is needed so have commented it out...?][AP: The dimension of the matrices below may not clear otherwise]}
%For notational convenience, we suppress the index $i$ by stacking the  $(T-1)\times 1$ vectors into $N(T-1)\times 1$ vectors,  the $(T-1)\times r$ matrices into $N(T-1)\times r$ matrices, and the $(T-1)\times r$ matrix of covariates into $N(T-1)\times p$ matrix.

\begin{proposition}{(Neyman-orthogonal Score for Panel IV DML)}
\label{prop:score}
Suppressing subscript $i$, the contribution of a single individual to a Neyman orthogonal score $\bmpsi$ for (finite-dimensional) parameters $\theta_0$ and $\vector\pi_0$ of structural model~\eqref{eqn:yit2}-\eqref{eqn:zit2} is 
\begin{equation}\label{eqn:prop_score1} 
%\bmpsi(W; \theta_0, \vector \pi_0;\bmeta_0)%
\bmpsi_0=\bmpsi(W; \vector b_0;\bmeta_0)%
=- \pmat{ %
    \vector V_0^\perp \Omega_{\theta\theta}\inv \widetilde{\vector U} \\  %
    \widetilde{\matrix V}'\Omega_{\pi\pi}\inv \widetilde{\vector R}},
\end{equation} 
%and the Neyman orthogonal score $\psi^{\rm rf}$ for parameters $\vector\delta_0$ and $\vector\pi_0$ of reduced-form model~\eqref{eqn:dit2}-\eqref{eqn:redform2} is
%\begin{equation}\label{eqn:prop_score2} 
%\bmpsi^{\rm rf}(W; \vector \delta_0, \vector \pi_0;\bmeta_0)%
%\bmpsi^{\rm rf}_0=\bmpsi^{\rm rf}(W; \vector b_0^{\rm rf}, \vector \pi_0;\bmeta_0)%
%=- \pmat{ %
%    \widetilde{\matrix V}' \Omega_{\delta\delta}\inv \widetilde{\vector U}^*  \\  %
%    \widetilde{\matrix V}'\Omega_{\pi\pi}\inv \widetilde{\vector R}}.
%\end{equation} 
where $\bmeta_0=\big(\widetilde{\vector l}_0,\widetilde{\vector r}_0,\widetilde{\matrix M}_0\big)$ is the true (infinite-dimensional) nuisance parameter, and row vector $\vector V_0^\perp = \vector \pi_0' \widetilde{\matrix V}'$ is the combined effect of the $r$ instrumental variables on $\widetilde{D}_{it}$ in model~\eqref{eqn:dit2}.
The residual variance-covariance matrices in the above expressions are $\Omega_{\theta\theta}= \E[\widetilde{\vector U}\widetilde{\vector U}'|\matrix X]$ and $\Omega_{\pi\pi} = \E[\widetilde{\vector R}\widetilde{\vector R}'|\matrix X]$. These scores lead to locally efficient estimators %$\vector b_0 = (\theta_0,\vector \pi_0')'$ and $\vector b_0^r = (\vector \delta_0',\vector \pi_0')'$, respectively, 
in the sense of being semi-parametrically efficient if %$\bmeta = \bmeta_0$, 
$\E[\widetilde{\vector U}{\widetilde{\vector U}}'| \widetilde{\matrix Z},\matrix X]= \Omega_{\theta\theta}(\matrix X)$, $\E[\widetilde{\vector R}\widetilde{\vector R}'| \widetilde{\matrix Z},\matrix X]= \Omega_{\pi\pi}(\matrix X)$, and $\E[\widetilde{\vector U}\widetilde{\vector R}'| \widetilde{\matrix Z},\matrix X]={\bf 0}$. 
The Regularity Conditions are that nuisance parameter estimates $\bmeta =\big(\widetilde{\vector l},\widetilde{\vector r},\widetilde{\matrix M}\big)\in T$, where $T$ is a convex subset of some normed vector space of square-integrable functions, $\|{\widetilde{\matrix Z}}\|^4$ and the singular values of $\|\Omega_{jj}(\matrix{X}_i)\|$, for $j=\theta,\delta,\pi$, have finite expectation under data generating process $P$, there exists finite $C_{\bmeta}$ such that for any ${\bmeta}\in T$, 
%$\Pr(\| \E[\bmpsi^\perp | \widetilde{\matrix Z}, \matrix X] \| < C_{\bmeta}) = 1$, where $\Pr$ and norm $\|.\|$ are taken with respect to $P$, for $\vector \psi^{\perp}=\vector \psi$ and $\vector \psi^{\perp}=\vector \psi^{\rm rf}$.  Then (\ref{eqn:prop_score1})-(\ref{eqn:prop_score2}) are Neyman orthogonal at true  %$(\theta_0,{\vector\pi}_0)'$ 
%$\vector b_0$ and $\vector b_0^{\rm rf}$, respectively, %$({\vector\delta}_0,{\vector\pi}_0)'$ 
%for all ${\bmeta}\in T_N$, where $T_N\subset T$ is a proper shrinking neighbourhood of ${\bmeta_0}$.
$\Pr(\| \E[\bmpsi\mid \widetilde{\matrix Z}, \matrix X] \| < C_{\bmeta}) = 1$, where $\Pr$ and norm $\|.\|$ are taken with respect to $P$. Then (\ref{eqn:prop_score1}) is Neyman orthogonal at true $\vector b_0$ for all ${\bmeta}\in T_N$, where $T_N\subset T$ is a proper shrinking neighbourhood of ${\bmeta_0}$.%
\footnote{For the reduced-form model parameter $\vector b_0^{\rm bf}$, the score function is
\[
\bmpsi^{\rm rf}_0=\bmpsi^{\rm rf}(W; \vector b_0^{\rm rf}, \vector \pi_0;\bmeta_0)%
=- \pmat{ %
    \widetilde{\matrix V}' \Omega_{\delta\delta}\inv \widetilde{\vector U}^*  \\  %
    \widetilde{\matrix V}'\Omega_{\pi\pi}\inv \widetilde{\vector R}},
\] 
where $\Omega_{\pi\pi} = \E[\widetilde{\vector R}\widetilde{\vector R}'|\matrix X]$ and  $\Omega_{\delta\delta} = \E[\widetilde{\vector U}^*\widetilde{\vector U}^{*\prime}|\matrix X]$. The score leads to locally efficient estimators in the sense of being semi-parametrically efficient if $\E[\widetilde{\vector R}\widetilde{\vector R}'| \widetilde{\matrix Z},\matrix X]= \Omega_{\pi\pi}(\matrix X)$, $\E[\widetilde{\vector U}^*\widetilde{\vector U}^{*\prime}| \widetilde{\matrix Z},\matrix X]= \Omega_{\delta\delta}(\matrix X)$ and $\E[\widetilde{\vector U}^*\widetilde{\vector R}'| \widetilde{\matrix Z},\matrix X]={\bf 0}$.}
%%%%%%%%%%%%%%%%%%%%%%%%%%%%%%%%%%%%%%%%%%%%%%%%%%%%%%%%%
%\noindent where $\vit^\perp$ is an {\em orthogonalized regressor} of the instrument chosen to ensure Neyman orthogonality, and $ {\vector r}_{it} = \big(\uit \,,\, \rit \,,\, \uit^* \big)'$ is a row vector of the residual terms, where $\uit^*= \Yit - l(\vector X_{it}) - \vit\delta - \alpha_i^*$ is the reduced form equation.
\end{proposition}
\noindent A proof is given in printed Appendix~\ref{sec:proof_score}. A formal definition of $T_N$ is later provided in the proof of Proposition~\ref{prop:dml}.

\begin{remark}{(Linear score function).}
\rm{
 The Neyman orthogonal score is linear in $\vector b$: 
\begin{equation*}\label{eqn:prop_score1b} 
\bmpsi = %
 \pmat{ \vector V_0^\perp  \widetilde{\vector U} \\  %
    \widetilde{\matrix V}' \widetilde{\vector R}} %
= \pmat{ %
    \vector V_0^\perp (\widetilde{\vector Y}-\widetilde{\vector l}_0) \\  %
    \widetilde{\matrix V}'(\widetilde{\vector D}-\widetilde{\vector r}_0)} %
    -\pmat{ %
    \vector V_0^\perp (\widetilde{\vector D}-\widetilde{\vector r}_0)& \zero_{r} \\  %
    \zero_{r} & \widetilde{\matrix V}'\widetilde{\matrix V}}%
    \pmat{\theta_0\\\vector \pi_0}%
    \equiv \bmpsi^b - \bm{\Psi}^a\vector b_0,
\end{equation*}
\noindent where the true $\bmpsi$, $\bmpsi^b$ and $ \bm{\Psi}^a$ at $\bmeta=\bmeta_0$ are $\bmpsi_0$, $\bmpsi^b_0$ and $\bm{\Psi}^a_0$, respectively,\footnote{For reduced-form model, the Neyman orthogonal function is linear in parameters $\vector b_0^{\rm bf}$:
\[
\bmpsi^{\rm rf} = %
 \pmat{ \widetilde{\matrix V}'  \widetilde{\vector U}^* \\  %
    \widetilde{\matrix V}' \widetilde{\vector R}} %
= \pmat{ %
    \widetilde{\matrix V}' (\widetilde{\vector Y}-\widetilde{\vector l}_0) \\  %
    \widetilde{\matrix V}'(\widetilde{\vector D}-\widetilde{\vector r}_0)} %
    -\pmat{ %
    \widetilde{\matrix V}' \widetilde{\matrix V}& \zero_{r} \\  %
    \zero_{r} & \widetilde{\matrix V}'\widetilde{\matrix V}}%
    \pmat{\vector \delta_0\\\vector \pi_0}%
    \equiv \bmpsi^{b{\rm rf}} - \bm{\Psi}^{a{\rm rf}}\vector b_0^{\rm rf},
\]
where the true $\bmpsi^{b{\rm rf}}$ and $ \bm{\Psi}^{a{\rm rf}}$ at $\bmeta=\bmeta_0$ are $\bmpsi_0^{b{\rm rf}}$ and $\bm{\Psi}^{a{\rm rf}}_0$, respectively.} and $\bf 0_r$ is a conformable vector of zeros.
}
\end{remark}

%%%%%%%%%%%%%%%%%%%%%%%%%%%%%%%%%%%%%%%%%%%%%%%%%
%%% UNCOMMENTED AP: to check %%%%
%%%%%%%%%%%%%%%%%%%%%%%%%%%%%%%%%%%%%%%%%%%%%%%%%
\begin{remark}{(Properties of the score function).}
\rm{The score functions satisfy: \\ 
(a) $\partial_{\eta} \E\big[\bmpsi^\perp(W; \vector b_0;\bmeta_0)\big][\bmeta-\bmeta_0]=\zero$, where expectations are with respect to the data generating process ${\cal P}$; and (b) the moment condition $\E[\bmpsi^\perp(W; \vector b_0;\bmeta_0)]=\zero$ for unique $\vector b_0$ at $\bmeta_0$. Neyman orthogonality (a) reduces the first-order bias introduced by regularising machine learning algorithms (\emph{regularization bias}), and guarantees that the estimated effect is insensitive to small errors in predicting the nuisance parameters.}
\end{remark}

\begin{remark}{ (The Panel IV DML Estimator).}
\rm{}
Moment condition (b) guarantees a closed-form solution of the structural parameters such that
%%%%%%%%%%%%%%%
\begin{equation}\label{eqn:score_moment}
   {\vector b_0} = \pmat{\theta_0\\\vector \pi_0} = %
   \E\Bigg[ \pmat{\vector V_0^{\perp}(\widetilde{\vector D}-\widetilde{\vector r}_0)& \zero_{r} \\  %
   \zero_{r} & \widetilde{\matrix V}'\widetilde{\matrix V}}
   \Bigg]\inv%
   \E\Bigg[\pmat{ \vector V_0^\perp (\widetilde{\vector Y}-\widetilde{\vector l}_0) \\  %
  \widetilde{\matrix V}' (\widetilde{\vector D}-\widetilde{\vector r}_0)}\Bigg].
\end{equation}
\noindent Solving the moment condition~\eqref{eqn:score_moment}, based on the Neyman-orthogonal score from Proposition~\ref{prop:score}, with respect to $\vector b_0$ yields the panel IV DML estimator 
\begin{equation}\label{eqn:dml_estimator}
   \widehat{\vector b} = \pmat{\thetano\\\widehat{\vector \pi}} = %
   \pmat{\vector V^\perp (\widetilde{\vector D}-\widehat{\vector r})& \zero_{r} \\  %
   \zero_{r} & \widetilde{\matrix V}'\widetilde{\matrix V}}
   \inv%
   \pmat{ \vector V^\perp (\widetilde{\vector Y}-\widehat{\vector l}) \\  %
  \widetilde{\matrix V}' (\widetilde{\vector D}-\widehat{\vector r})},
\end{equation}
\noindent with asymptotic variance-covariance matrix
\begin{equation}\label{eqn:Sigma}
{\matrix \Sigma}  = %
\pmat{\Sigma_{\theta\theta} &0\\
\zero_{r} & \matrix\Sigma_{\pi\pi} }
=%
{\matrix Q}\inv\lim _{N\to\infty } %
   \Var\Bmat{ %
   \pmat{\vector V_0^\perp \big(\widetilde{\vector Y}-\widetilde{\vector l}_0)\\
   \widetilde{\matrix V}'\big(\widetilde{\vector D}-\widetilde{\vector r}_0\big)
   }}{\matrix Q}\inv
\end{equation}
\noindent where ${\matrix Q}\inv  =  
\begin{pmatrix}
\vector Q_{v^{\perp}r}\inv&\zero_{r}\\
\zero_{r}&  \matrix Q_{vv}\inv\\
\end{pmatrix}$, $\matrix Q_{v^{\perp}r}=\E[\widetilde{\vector V}^{\perp}\big(\widetilde{\vector D}-\widetilde{\vector r}_0\big)]$ and  $\matrix Q_{vv}=\E\big(\widetilde{\matrix V}'\widetilde{\matrix V}\big)$, ${\matrix Q}\inv$ is the inverse of ${\matrix Q}$, and $N$ is the cross-sectional sample size of the fold used to estimate the interest parameters. A consistent estimator of the scalar $\vector Q_{v^{\perp}r}$ is $\widehat{\vector Q}_{v^{\perp}r} = \widehat{\vector V}^{\perp}\big(\widetilde{\vector D}-\widehat{\vector r}\big)$, and of the $r\times r$ matrix $\matrix Q_{vv}$ is $\widehat{\matrix  Q}_{vv} = \widehat{\matrix V}' \widehat{\matrix V}$.\footnote{ The variance-covariance matrix of the Panel IV DML estimators is robust to heteroskedasticity, and has the conventional Eiker-Huber-White \emph{sandwich} formula.}
\end{remark}

%%%%%%%%%%%%%%%%%%%%%%%%%%%%%%%%%%%%%%%%%%%%%%%%%%%%%
To present our main result for the estimation and inference of $\mathbf{b}_0$ using the panel IV DML estimator, let $\{\delta_N\}_{N=1}^{\infty}$ and $\{\Delta_N\}_{N=1}^{\infty}$ be two sequences of positive constants tending to zero as $N$ increases such that $\delta_N\ge N^{-1/2}$. Let $c,C>0$ be arbitrary fixed constants denoting lower and upper bounds, respectively. Further fixed constants:\ recall that $r\ge1$ is the number of instrumental variables, and let $K\ge2$ be the  number of cross-validation folds chosen by the analyst.

\begin{proposition}{\textbf{(Asymptotic Distribution of the Panel IV DML Estimator)}} \label{prop:dml}
Suppose that the Regularity \mbox{Conditions~\ref{item:a}-\ref{item:e}} below are all satisfied for all probability laws $P \in \mathcal{P}$ for random vectors and matrices \mbox{$\matrix W = \{\widetilde{\vector Y},\widetilde{\vector D},\widetilde{\matrix Z},\matrix X\}$}:
\begin{enumerate}[label=(\alph*)]
    \item The model defined in Equations~\eqref{eqn:yit2}-\eqref{eqn:redform2} holds for all $P\in \mathcal{P}$;\label{item:a}
    \item $\|\widetilde{\vector Y}\|_{P,q} + \|\widetilde{\vector D}\|_{P,q} + \|\widetilde{\matrix Z}\|_{P,q}\le C$ for all $q\ge1$;\label{item:b}
    \item The singular values of $\E[\widetilde{\matrix V}'_0\widetilde{\matrix V}_0]$ all exceed some arbitrary $c>0$ as does $\big| \E\big[\vector \pi_0' \widetilde{\matrix V}'_0 (\widetilde{\vector D}-\widetilde{\vector r_0})\big]\big|$;\label{item:c}
    \item All of $\big\|\E\big[\widetilde{\vector U}_0^2|\matrix X\big]\big\|_{P,\infty}$, $\big\|\E\big[\widetilde{\vector R}_0^2|\matrix X\big]\big\|_{P,\infty}$ and $\big\|\E\big[\widetilde{\matrix V}_0^2|\matrix X\big]\big\|_{P,\infty}<C$;\label{item:d}
    \item Given first-stage base learner prediction $\widehat{\bmeta}$ based on block-k-fold cross fitting, the following hold with probability no smaller tha $1-\Delta_N: \|\widehat{\bmeta}-\bmeta_0\|_{P,q}\le C$ for $q>2$ and $\|\widehat{\bmeta}-\bmeta_0\|_{P,q}\le \delta_N$ with $\|\widetilde{\matrix m}-\widetilde{\matrix m}_0 \|_{P,2}\times \Big\{\|\widetilde{\vector l}-\widetilde{\vector l}_0 \|_{P,2} + \|\widetilde{\vector r}-\widetilde{\vector r}_0 \|_{P,2} + \|\widetilde{\matrix m}-\widetilde{\matrix m}_0 \|_{P,2} \Big\} \le \delta_N N^{-1/2}$;\label{item:e}
\end{enumerate}

%hold and there exist suitable ML algorithms to obtain ${\widehat\bmeta}$ from learning $\bmeta_0$ with a combined convergence rate of $o_p(N^{1/4+\epsilon})$ for some $\epsilon>0$. %In addition, let $\widehat{\Omega(V,X)}$ be the ${\sqrt N}$-consistent estimator for $\Omega(V,X)$. 
\noindent Then, the panel IV DML estimator $\widehat{\vector b}$ based on the Neyman-orthogonal score from Proposition~\ref{prop:score} obeys $\Sigma\inv\sqrt{N} \big(\widehat{\vector b}-\vector b_0\big)\distas{}{}N(\zero, \mathbf{I})$ uniformly over $P\in\mathcal{P}$, where the variance-covariance matrix is $\Sigma = \big(\E[\matrix \Psi_0^a]\big)\inv\E[\bmpsi^b_0\bmpsi_0^{b'}]\big(\E[\matrix \Psi_0^a]\big)\inv$. Moreover, this result holds if $\Sigma$ in Equation~\eqref{eqn:Sigma} is replaced by a $\sqrt{N}$-consistent estimator $\widehat{\Sigma}$.
\end{proposition} 
%%%%%%%%%%%%%
\begin{remark}
   \rm{Norms $\|.\|_{P,q}$ in $L_q(P)$ are defined over $\mathcal{T}$ space of $P$-square-integrable nuisance functions $\bmeta=\bmeta(W)$ for $P\in \mathcal{P}$, where $\mathcal{T}_N\subset \mathcal{T}$be a properly vanishing neighbourhood of true $\bmeta_0$ as defined by bounds on $\widehat{\bmeta}$, $\mathcal{P}_N \subset \mathcal{P}$ is the corresponding vanishing neighbourhood of data generating process $P$. Norms over the counting measure are denoted by $\|.\|_{q}$ in the usual way, and those with matrix arguments are Schatten norms.} 
\end{remark}

\begin{remark}
   \rm{A proof extending that of \citet[Theorem 4.2]{chernozhukov2018}, in which Regularity Conditions~\ref{item:a}-\ref{item:e} are shown to satisfy the assumptions listed in their Assumptions~3.1 and~3.2 required for Theorem 3.1 to hold, can be found in the Online Appendix~\ref{sec:dml}.}
\end{remark}

%The proof of Proposition~\ref{prop:dml} is in the Appendix~\ref{sec:dml}.

%%%%%%%%%%%%%%%%%%%%%%%%%%%%%%%%%%%%%%%%%%
\subsection{Inference under Weak Identification}
% scalar `rkf' =	r(chi2)/(`N'-1) *				///
% 				(`N'-`iv1_ct'-`sdofminus') *	///
% 				(`N_clust'-1)/`N_clust' /		///
% 				`exex1_ct'
As discussed above, the partially linear first-difference model \eqref{eqn:yit2}-\eqref{eqn:zit2}  (and \eqref{eqn:zit2}-\eqref{eqn:redform2}) holds under Assumptions~\ref{item:asm_feedback}-\ref{item:asm_fe}, where Assumptions~\ref{item:asm_feedback}(b) and~\ref{item:iv_effect} permit $\widehat{\vector{l}}_{0i}$, $\widehat{\vector{r}}_{0i}$ and $\widehat{\mathbf{M}}_{0i}$ to be consistently learnt.  Furthermore, {Propositions~\ref{prop:score} and~\ref{prop:dml} present the Neyman orthogonal score and Regularity Conditions under which DML yields consistent and asymptotically normal estimators of $\theta_0$ and $\bm{\pi}_0$ (or $\bm{\delta}_0$ and $\bm{\pi}_0$).
If we have $\sqrt{N}$-consistent estimators of $\Omega_{\theta\theta}$ and $\Omega_{\pi\pi}$ then, without loss of generality, we can set both equal to identity matrices in the following explanation.  Then, following \citet[Sect.~1]{chernozhukov2018}, a second-order Taylor series expansion of score~\eqref{eqn:prop_score1} around $\bm{\psi}=\bm{\psi}_0$, where 
$\bm{\psi}_0 = (\theta_0, \bm{\pi}_0, \widetilde{l}_0, \widetilde{r}_0, \widetilde{m}_0)$, gives
\begin{equation*}
\sqrt{N}\,(\widehat{\theta}_{\mathrm{DML}} - \theta_0) =
\left[ \frac{1}{N}\bm{\pi}_0' \widetilde{\matrix V}_{0}'\bigl(\widetilde{\vector D} - \widetilde{\vector r}_{0}\bigr) \right]^{-1} \frac{1}{\sqrt{N}} \bm{\pi}_0'\widetilde{\matrix V}_{0}' \widetilde{\vector U}_{0} + o_p(1),
\end{equation*}
and
\begin{equation*}
\sqrt{N}\,(\widehat{\bm{\pi}}_{\mathrm{DML}} - \bm{\pi}_0) =
\left[ \frac{1}{N} \widetilde{\matrix V}_{0}\widehat{\matrix V}_{0}'\right]^{-1}
\frac{1}{\sqrt{N}}\widetilde{\matrix V}_{0}\widetilde{\matrix R}_{0}'+ o_p(1),
\end{equation*}
where DML allows us to lump into \(o_p(1)\) all cross-products involving at least one residual from
\(\widetilde{\vector D} - \widetilde{\vector r}_{0}\), \(\widetilde{\vector R}_{0}\), \(\widetilde{\matrix V}_{0}\), and \(\widetilde{\vector U}_{0}\),
and at least one of \(\widehat{\bm{\pi}}_{\mathrm{DML}} - \bm{\pi}_0\),
\(\widehat{\vector l} - \widetilde{\vector l}_{0}\),
\(\widehat{\vector r} - \widetilde{\vector r}_{0}\),
and \(\widehat{\matrix M} - \widetilde{\matrix M}_{0}\)
in the Taylor series
(\(\widehat{\vector l}\), \(\widehat{\vector r}\), and \(\widehat{\matrix M}\) are, respectively, predictions of \(\widetilde{\vector l}\), \(\widetilde{\vector r}\), and \(\widetilde{\matrix M}\) from DML stage one);
and the requirement that the base learners satisfy
\(o_p(N^{-1/4})\)
ensures the same for cross-products involving pairs from
\(\widehat{\bm{\pi}}_{\mathrm{DML}} - \bm{\pi}_0\),
\(\widehat{\vector l} - \widetilde{\vector l}_{0}\),
\(\widehat{\vector r} - \widetilde{\vector r}_{0}\),
and \(\widehat{\matrix M} - \widetilde{\matrix M}_{0}\).
This allows weak-instrument asymptotics to be applied in the usual manner for both the first-stage \(F\) and Anderson-Rubin statistic. 

Assessing instrument relevance is routine in applied work, as weak instruments bias 2SLS estimator and invalidate its normal asymptotic properties \citep[see, e.g.,][]{bound1995}. In practice, applied researchers typically rely on the first-stage {\em F}-statistic, which is routinely reported in empirical tables. We therefore begin by showing how conventional first-stage {\em F}-statistic can be used within our DML framework, providing a diagnostic that is familiar and immediately interpretable for applied researchers. We then turn to the Anderson-Rubin test statistic and confidence set which, although more robust to weak identification, remains underutilized in empirical practice.

%%%%%%%%%%%%%%%%%%%%%%%%%%%%%%%%%%%%%%%%%%%%%%%%%%%%%%%%%%%%%%%%%
%%% FIRST-STAGE F STAT
%%%%%%%%%%%%%%%%%%%%%%%%%%%%%%%%%%%%%%%%%%%%%%%%%%%%%%%%%%%%%%%%%
\subsubsection{First-stage {\em F}-statistic for Panel IV DML}
The F-statistic from the linear regression of the endogenous variable on the instrument(s) (first-stage regression) is routinely reported in empirical IV applications and often used as a diagnostic for instrument relevance. In practice, instruments are typically considered sufficiently strong when this statistic exceeds conventional critical values.\footnote{Conventional \citet{stock2005}'s critical values can be used for inferences in just-identified cases. In practice, a widely used rule-of-thumb is to consider an instrument \emph{strong} if the first-stage {\em F}-statistic exceeds the value of 10 \citep[][pp.470-471]{stock2019}. A more rigorous approach is based on thresholds for first-stage F derived by \citet{stock2005}, who consider the maximal tolerable t-test size distortion, or alternatively \emph{how often a 5\% t-test rejects a true hypothesis}. Recently, \citet{lee2022} claim these thresholds should be reconsidered. They show that only when the {\em F}-statistic exceeds 104.7 the t-tests are ensured not to reject the null hypothesis at a rate higher than the desired one (e.g., 5\% rate).}
Given the importance of assessing weak identification for valid inference, we derive expressions for a first-stage F-statistic adapted to our panel IV DML framework.\footnote{Our tests for weak IV can be easily adapted to the partially linear regression model with linear instrumental variables within cross-sectional DML framework  by \citet{chernozhukov2018}. These tests have not been yet developed in that setting at the time of writing.} 

%\red{The first-stage {\em F}-statistic from the linear regression of the endogenous variable on the instrumental variable(s) is usually reported in regression tables when an IV procedure is employed, and conclusions are drawn on an instrument being \emph{strong} if it exceeds some critical values. Given the practical importance of testing for potentially weak instruments to conduct valid inference,} we provide formulae of the first-stage {\em F}-statistic suitable for our panel IV DML framework.\footnote{Our tests for weak IV can be easily adapted to the partially linear regression model with linear instrumental variables within cross-sectional DML framework  by \citet{chernozhukov2018}. These tests have not been yet developed in that setting at the time of writing.} 

\begin{definition}{(First-stage {\em F}-statistic for Panel IV DML)}\label{def:fstat}
    Let $r\ge1$ be the number of excluded exogenous instruments, and $N$ the number of subjects in the estimation sample. Assume the conditions in Proposition~\ref{prop:dml} hold such that the DML first-stage estimator, $\widehat{\bm{\pi}}$, is well-behaved. Under the null hypothesis $H_0: \bm{\pi} = \zero$, the {\em F}-statistic is 
    \begin{equation}\label{eqn:Fstat}
    F^{DML} = \frac{1}{r} \widehat{\bm{\pi}}' \widehat{\matrix \Sigma}^{-1}_{\pi\pi}\widehat{\bm{\pi}}
    \end{equation}
    \noindent where $\widehat{\matrix\Sigma}_{\pi\pi}= \widehat{\matrix Q}_{vv}\inv \,\Big\{ N\inv\widehat{\matrix V}'\big(\widetilde{\vector D}-\widehat{\vector r}\big) \big(\widetilde{\vector D}-\widehat{\vector r}\big)'\,\widehat{\matrix V} \Big\}\,\widehat{\matrix Q}_{vv}\inv$ with $\widehat{\matrix Q}_{vv} = N\inv\widehat{\matrix V}'\widehat{\matrix V}$ is  a consistent estimator of the asymptotic variance-covariance matrix ${\matrix \Sigma}_{\pi\pi}$. %, and is distributed as an F distribution with r degrees of freedom at the numerator and N-1 at the denominator (due to clustering).
\end{definition}
%\red{[PROOF in the appendix. Using the result from Prop 3.2, $\widehat{\bm{\pi}}$ is asymptotically normally distributed. The estimated variance converges to a finite variance; by Slutsky thm and continuous mapping thm the inverse is defined.]}

\begin{remark}
\rm{The first-stage {\em F}-statistic~\eqref{eqn:Fstat} relies on the asymptotic normality guaranteed under the DML conditions stated in Proposition~\ref{prop:dml}.}
\end{remark}

\begin{remark}
\rm{The first-stage {\em F}-statistic~\eqref{eqn:Fstat} is robust to non-homoskedastic errors (i.e., serially correlated errors, heteroskedasticity, and clusters), and corresponds to the robust version of the \citet{kleibergen2006} {\em F}-statistic in a conventional 2SLS setting.%
\footnote{In conventional IV settings, the Kleibergen-Paap {\em F}-statistic (robust F) is usually reported in tables to argue the strength of the instrument. Alternative statistics suggested by the literature are effective {\em F}-statistic \citep{olea2013} which is centred around the correct population parameter, unlike the conventional robust {\em F}-statistic, and the efficient GMM {\em F}-statistic \citep{windmeijer2025}. In the just-identified case with one instrument, like in this case, the effective $F$ and the robust $F$ coincide and conventional \citet{stock2005} critical values can be used \citep{andrews2019}. We do not implement the effective {\em F}-statistic for the panel IV DML estimator because the calculation of the critical values is computationally demanding. In addition, Anderson-Rubin robust test statistic should be preferred.}
}
\end{remark}

% The effective F-statistic for the  Panel IV DML setting is
% \begin{equation*}
%    F^{eff} = %\frac{\widehat{\bm{\pi}}' \widehat{\Sigma}^{-1}_{N,\pi\pi}\widehat{\bm{\pi}}}{tr( \widehat{\Sigma}_{\pi\pi}\widehat{Q}_{vv})} =
%    \frac{N \widehat{\bm{\pi}}' \widehat{\matrix Q}_{vv}\widehat{\bm{\pi}}}{tr( \widehat{\matrix \Sigma}_{\pi\pi}\widehat{\matrix Q}_{vv})}.
% \end{equation*}
%\red{However, we have not yet implement the effective AR statistic.}

%%%%%%%%%%%%%%%%%%%%%%%%%%%%%%%%%%%%%%%%%%%%%%%%%%%%%%%%%%%%%%%%%%%%%%%%%%%%%%%%
%%% ANDERSON-RUBIN
%%%%%%%%%%%%%%%%%%%%%%%%%%%%%%%%%%%%%%%%%%%%%%%%%%%%%%%%%%%%%%%%%%%%%%%%%%%%%%%%
\subsubsection{AR Test Statistic and Confidence Sets for Panel IV DML}
It is well established that statistical inference based on conventional 2SLS estimator is invalid when instruments are weak, especially when the OLS bias is large \citep[see][]{keane2024}. %In particular, standard errors, heteroskedasticity-robust ones, tend to be downward biased, leading to overrejection of null hypotheses in t-tests, especially when the OLS bias is large \citep[see][]{keane2024}. 
Therefore, conventional F- and t-based inference becomes unreliable, even when instruments appear strong, whereas tests that are robust to weak identification and do not rely on assumptions about instrument relevance, such as the Anderson-Rubin (AR) test \citep{anderson1949}, remain valid \citep{andrews2019,moreira2019,keane2023}.

The AR test evaluates the null hypothesis on the structural parameter by testing the corresponding reduced-form restriction. Specifically, it assesses whether the instrument(s) have a statistically significant reduced-form effect on the outcome under the assumption of instrument exogeneity (i.e., that instruments are as good as randomly assigned). This approach provides valid inference on the structural parameter without requiring strong identification (i.e., relevance). %, since the reduced-form coefficient equals the product of the first-stage and structural coefficients.

\begin{definition}{(Anderson-Rubin Test for Panel IV DML)}\label{def:archi2}
 Let $r\ge1$ be the number of excluded exogenous instruments, $N$ the number of subjects in the estimation sample, and $\widehat{\bm{\delta}}$ the panel IV DML estimator of the reduced-form equation satisfying the moment condition $\E({\matrix V}'\vector U^*)=\zero$ such that the equality $\bm{\delta} = \bm{\pi}\theta$ holds.  When $\theta=\theta_0$, the null hypothesis is $\bm{\delta}(\theta_0)=\zero$ and the Anderson-Rubin Wald test statistic for panel IV DML at the $1-\alpha$ level is
\begin{equation}\label{eqn:archi2}
    \mathrm{AR}(\theta_0)^{DML} =  \widehat{\bm{\delta}}'(\theta_0)  \widehat{\matrix \Sigma}_{\delta\delta}\inv \widehat{\bm{\delta}}(\theta_0)\distas{a}{}\chi^2(r)_{1-\alpha}
\end{equation}
\noindent where  $\widehat{\matrix \Sigma}_{\delta\delta}= \widehat{\matrix Q}_{vv}\inv \,\Big\{ N\inv \widehat{\matrix V}'\big(\widetilde{\vector Y}-\widehat{\vector l}\big)\big(\widetilde{\vector Y}-\widehat{\vector l}\big)' \, \widehat{\matrix V} \Big\}\,\widehat{\matrix Q}_{vv}\inv$ with $\widehat{\matrix Q}_{vv} = N\inv\widehat{\matrix V}'\widehat{\matrix V}$ is a consistent estimator of the asymptotic variance-covariance matrix ${\matrix \Sigma}_{\delta\delta}$. %, and is distributed as $\chi^2$ with r degrees of freedom.  
\end{definition}

\begin{remark}
\rm{The AR test statistic constructed from DML estimator relies on the asymptotic normality guaranteed under the DML conditions stated in Proposition~\ref{prop:dml}.}
\end{remark}

%\red{[PROOF in the appendix. Similarly to Prop 3.3., we use use the results from Prop 3.2, $\widehat{\bm{\delta}}$ is asymptotically normally distributed. The estimated variance converges to a finite variance; by Slutsky thm and continuous mapping thm the inverse is defined.]}

%Under the null hypothesis, the AR Wald test statistic converges in distribution to a $\chi^2(r)$ with $r$ degrees of freedom. AR Wald test statistic is distributed as F distribution in finite samples with $r$ degrees of freedom at the numerator and $(N-1)$  degrees of freedom at the denumerator, when multiplied by a finite sample correction for clustered data $c = (N-1)/N$.

\begin{definition}{(AR Confidence Sets).}
\rm{Using the Anderson-Rubin test statistic in~\eqref{eqn:archi2}, it is possible to construct the set of parameter values $\theta_0$ for which the null hypothesis $H_0: \, \theta=\theta_0$ is not rejected. The AR confidence set (CS) for panel DML IV estimator is obtained by test inversion \citep[following][]{davidson2014} and defined as
\begin{equation}
CS_{AR}(1-\alpha)= \big\{\theta_0\in \R : \mathrm{AR}(\theta_0)^{DML}\le\chi^2(r)_{1-\alpha}\big\}.
\end{equation}
}
\end{definition}
\noindent As in the conventional 2SLS estimation method, the solutions of the quadratic form associated with the AR test statistic are: (a) a bounded interval $[x,y]$, where $\{x,y\}$ with two real roots $x\le y$; (b) a disjoint set $(-\infty,x] \cup [y,+\infty)$ with two real roots $x< y$; (c) the real line $(-\infty,+\infty)$; and (d) the empty set, which is only possible in overidentified settings. AR CS~(c) and~(d) have no real roots.

The type of AR CS informs about the relevance of the instrument. When instruments are strong, the AR CS is bounded, as in case~(a), and the model is well-identified. %A bounded AR CS is asymmetric being constructed from the two real roots of the quadratic form of the AR test statistic.
When instruments are sufficiently weak, the AR CS are usually unbounded (cases b-c) and uninformative, since a bounded set generally does not exist locally when the parameter is not identified \citep{davidson2014}. Empty sets (case d) occur when the equality $\bm{\delta} = \bm{\pi}\theta_0$ fails due to either (a) treatment effect heterogeneity when the equality holds for $\theta\ne\theta_0$, or (b) invalidity of the instruments -- i.e., the failure of the overidentifying restrictions -- when there is no value of $\theta$ that satisfies the inequality  \citep[as in][]{andrews2019}.

%\red{[ADD EXPLANATION FROM LITRATURE] See DM2014. (a) IV is strong and the effect is the range; (b) and (c) means that the IV is weak; (d) means that either  there is no solution to the quadratic form or no effect.}

% 4. APPLICATIONS
\section{Empirical Applications: Shift-Share IVs}\label{sec:empirical}
\setcounter{equation}{0}

%\red{[Find common pattern in empirical applications. The impact of the flow of immigrants in Europeans countries and US states on political and socio-economic preferences of voters and parties.]}

In this section, we showcase the applicability of our panel IV DML method.\footnote{The panel IV DML estimation is conducted in \textsc{R} using the latest version of the \texttt{xtivdml} package accessible in its latest version at \url{https://github.com/POLSEAN/xtivdml} at the time of writing, which is built on  \textsc{R} packages \texttt{DoubleML} \citep{DoubleML} and \texttt{xtdml} \citep{xtdml}. The conventional 2SLS estimation is implemented in \textsc{Stata 19.5} using the community-contributed commands \texttt{xtivreg2} \citep{schaffer2005} and \texttt{twostepweakiv} \citep{sun2018}}  
For a comprehensive illustration, we revisit three empirical studies that examine whether and how immigration affects the political views of native voters towards immigration in the United States (\citealp{tabellini2020}) and Europe (\citealp{moriconi2019, moriconi2022}). 

As typical in the migration literature, an instrumental variable strategy (2SLS) is employed to account for the potentially endogenous decision of immigrants to settle across areas: on the one hand, they may be more likely to settle in areas where the economy or attitudes towards immigration are more favourable; on the other hand, they may decide to settle in places where housing is more affordable and which are economically declining. 
%%%%
These three studies adopt modified versions of the widely used shift-share instrument \citep{card2001}, which exploits the tendency of immigrants to settle in areas with larger pre-existing communities from the same country of origin or ethnic group.\footnote{The focus on articles that exclusively use a shift-share IV is motivated by the fact that the instrument is not randomly assigned, by construction, and a rich set of covariates is typically required to make the instrument approximately exogenous, namely \emph{as good as randomly assigned}. This setting is well suited to our panel IV DML approach, which flexibly adjusts for high-dimensional confounding.}  
The version of the shift-share instrument used in these articles is constructed as the weighted average of new national inflows from each immigrant’s country of origin (the `shifts'), using fixed pre-existing local settlement patterns (the `shares') as weights. In formulae, 
\begin{equation}\label{eqn:ssh}
    Z_{rt} = \frac{1}{Pop_{rT_0}} \sum_{c\in\mathcal{C}} Sh_{crT_0}\sum_{r\in\mathcal{R}} M_{crt}
\end{equation}
where $Pop$ is the total population in city/region $r$ in the pre-settlement period $T_0$,  $Sh_{crT_0}= M_{crT_0} / \sum_r M_{crT_0}$ is a fixed share of pre-existing immigrants,  $M_{crt}$ is the national stock (inflows) of immigrants, $\mathcal{C}$ is the set of countries of origin, and $\mathcal{R}$ is the set of receiving cities/regions.

Regarding the validity of the assumptions underlying the shift-share instrument, historical settlement patterns (the `shifts') are generally strong predictors of current immigrant inflows and therefore are often found to satisfy the relevance condition in conventional 2SLS settings.
The exogeneity of these instruments and the validity of the exclusion restriction have recently been at the center of a number of papers, which highlight that one of the two components (either the `shifts' or the `shares') must be exogenous \citep[e.g.,][]{adao2019shift,borusyak2022quasi,goldsmith2020bartik}. In our empirical applications, we abstract from this debate and take as given the arguments provided in the original papers in support of these assumptions. Instead, we focus on how our panel IV DML method can address the potential presence of a large number of possibly nonlinear confounders.
Depending on the setting, the exogeneity of the instrument may only be valid once conditioning on certain controls \citep[see, for instance,][]{borusyak2025practical}. These controls may enter the true nuisance functions nonlinearly, and ignoring such nonlinearities can affect the estimation of treatment effects due to model misspecification.

In our re-analysis, we compare the results from our panel IV DML estimator to those obtained from conventional 2SLS with FD, which is the closest panel data estimation method to our estimation approach.\footnote{As previously discussed in Section~\ref{sec:transformed_models}, the FD (exact) approach allows us to approximate the transformed unknown nuisance functions in nonlinear settings without imposing many constraints on the fixed effects, unlike the within-group transformation (or fixed effects) and the correlated random effects device. The other two approaches are possible but may lead to inconsistent estimates when the true functional form is highly nonlinear \citep[for further details, see][]{clarke2023}.} 
As the original estimation approaches are either 2SLS with FE \citep{tabellini2020} or pooled ordinary least squares with region and year fixed effects \citep{moriconi2019,moriconi2022} due to the structure of the data,  we additionally contrast the estimates obtained from a 2SLS regression with FE with 2SLS with FD in each empirical application. This preliminary step allows us to verify that any differences with panel IV DML method arise only from the use of a different estimation methodology and not from different specification choices (see Online Appendices~\ref{sec:app_tabellini} and~\ref{sec:app_moriconi} for an exhaustive discussion of conventional panel data estimation results).

In our panel IV DML regressions, the functional form of the confounding variables is learned flexibly using machine learning algorithms from each major class (i.e., Lasso for L-norm regularized linear models, gradient boosting with 1000 trees for tree-based methods, and a single-hidden-layer neural network for deep learning) allowing the model to capture a broad range of nonlinearities in the data. The hyperparameters of the base learners are tuned with grid search \citep{bergstra2012} (see Table~\ref{tab:hyperpara} in the Online Appendix~\ref{sec:tuning}). 
The set of covariates employed as inputs in panel IV DML estimation with neural network and gradient boosting includes raw covariates only, no interaction terms or polynomials are included because these base learners are designed to automatically capture nonlinearities in the data. Conversely, panel IV DML with Lasso uses an extended dictionary of nonlinear terms of the raw covariates (i.e., polynomials up to order three and interaction terms of all covariates) to satisfy weak sparsity assumption. The inputs used in panel IV DML estimation also include one-period lags of all control variables, following the FD (exact) approach discussed in Section~\ref{sec:transformed_models}. 
In addition, the sample in panel IV DML regressions is divided in two folds, where the number of folds is chosen to account for the small cross-sectional dimension ($N$) in all empirical applications.\footnote{Dividing the sample into a higher number of folds reduces the size of the estimation sample considerably, which may differ in terms of observable characteristics from the prediction sample. In such cases, the learner would need to extrapolate the information, but is not desirable for flexible tree-based learners and neural networks. We allow for cross-fitting as desirable to restore efficiency.}

%%%%%%%%%%%%%%%%%%%%%%%%%%%%%%%%%%%%%%%%%%%%%%%%%%%%%%%%
\subsection{Empirical Example: \citet{tabellini2020}}\label{sec:tabellini}

The first empirical application revisits the study by \citet{tabellini2020}, which examines the political and economic effects of restrictive immigration policies induced by World War~I and the Immigration Acts of the 1920s in the United States (U.S.) that reduced the quota of European immigrants. The analysis exploits the exogenous variation in immigration from European countries to 180 U.S. cities over three census years 1910, 1920 and 1930. %The final sample consists of a balanced panel of 180 U.S. cities observed over three census years. 
Data are collected from various sources: U.S. Census of Population, Voteview, Census of Manufactures.

The author employs an instrumental variable strategy to address the potentially endogenous settlement patterns of European immigrants across U.S. cities. The instrument used is a leave-out version of the shift-share measure defined in equation~\eqref{eqn:ssh}, which interacts historical settlement patterns of different ethnic groups in each city in 1900 with contemporary immigration flows of the same group, excluding immigrants who ultimately settle in the same city. The identifying assumption for the validity of the instrument is verified by the author with a pre-trend test, which confirms that, prior to 1900 (before any networks formation), European immigrants did not disproportionately settle in cities experiencing economic growth or political change. The original analysis is conducted using conventional 2SLS estimation with fixed effects (FE). The main findings of the article suggest that immigration induced hostile political reactions, such as the election of more conservative legislators, stronger support for anti-immigration legislation, and lower redistribution.

 We re-analyze the baseline specifications from Table~3 (Column 4, Panel B) and Table~5 (Column 1, Panel B)  in \citet{tabellini2020}  using conventional 2SLS with FD and our panel IV DML method with the FD approach. We focus on two outcomes: for the political effects, we look at the \emph{Poole-Rosenthal DW Nominate Score}, which ranks congressmen on an ideological scale from liberal to conservative using voting behaviour on previous roll-calls; for the economic effects, we consider the \emph{(log) occupational score}, which is a proxy for natives' income and does not capture within occupation changes in earnings.\footnote{As explained by the author, wage data are not available since until 1940. Occupational scores are commonly used in the literature to proxy lifetime earnings, which are calculated by assigning the median income of an individual job category in 1950 to them.} 
The baseline specifications include only interaction terms of region and year fixed effects. 

We then depart from the original analysis by augmenting the baseline specifications with additional control variables: predicted industrialization, immigrant and city population, value added manufacturing, skill ratios, fraction of blacks, value of products, employment share in manufacturing. In the original analysis, each of these control variables is added one at a time to the baseline specification to evaluate the robustness of the results.\footnote{%Our Tables~\ref{tab:tabellini2020_D2}-\ref{tab:tabellini2020_D3} in the Appendix~\ref{sec:app_tabs} display our corresponding estimates obtained from 2SLS regressions with FD and panel IV DML regressions with FD (exact) approach after including one control variable at a time in the baseline specifications. 
For instance see Tables D2-D3 in the Online Appendix of \citet[][]{tabellini2020}. We thank Marco Tabellini and Francesco Maria Toti Ognibene for providing us with the material to replicate Tables~D2-D3 in the Online Appendix of \citet{tabellini2020}.} 
In practice, the identification of the effect with inclusion of many irrelevant controls in the estimating equation may not be possible with conventional estimation methods, such as least squares, due to matrix singularity and multicollinearity issues. In contrast, DML methods (for both cross-sectional and panel data) can handle high-dimensional covariate spaces.%\footnote{To clarify, Panels~B of our Table~\ref{tab:tabellini2020_D} in the main text and Table~\ref{tab:tabellini2020_fefd} in the Online Appendix report 2SLS estimates even with the augmented set of covariates because the estimation is conducted via the statistical software \proglang{Stata} using the \texttt{xtivreg2} command, which automatically detects and drops multicollinear covariates estimating the model on a smaller set of variables. In contrast, the \texttt{plm} function in \proglang{R} does not handle multicollinearity in the internal estimation routine, and the estimation fails when the design matrix is not full rank.} 

%%% 2SLS FD vs FE %%%
Before discussing panel IV DML results, we briefly compare conventional 2SLS estimates obtained using fixed effects (FE) and first-differences (FD) %, which is the panel data approach most closely aligned with our panel IV DML method presented in Section~\ref{sec:transformed_models} 
(see the Online Appendix~\ref{sec:app_tabellini} for a detailed discussion). This preliminary step ensures that any differences observed against our panel IV DML estimator reflect methodological rather than specification choices.  In the baseline specifications, the shift-share instruments appear  strong with both FE and FD estimators for the political outcome (\emph{DW Nominate Score}) and the economic outcome (\emph{Log Occupational Score}). 
The second-stage coefficients are positive and significant at least at 5\% level. When additional controls are included, an extension not implemented in the original analysis, the shift-share instrument becomes weak in the specification of the political outcome while remaining strong in the specification of the economic outcome with both panel data estimators. Regardless of the strength of the instrument, both estimators produce statistically insignificant effects, contrasting the results found in the baseline regression.
In general, the two estimators produce very similar results in terms of sign, magnitude and statistical significance, ensuring that subsequent differences between 2SLS and panel IV DML can be explained by differences in the methodology. %The baseline (significant and positive) findings of the original study from FE regressions are confirmed with FD (Panel~A of Table~\ref{tab:tabellini2020_fefd}).

\begin{table}[t!]
\centering
 \caption{The political and economic effects of immigration}\label{tab:tabellini2020_D}
 \vspace{-2mm}
\scalebox{.59}{
\begin{threeparttable}	%use it for notes
  \begin{tabular}{lcccccccc}   
\vspace{-3mm}\\   
\hline\hline
\vspace{-3mm}\\	
\emph{Dependent variable: }&\multicolumn{4}{c}{DW Nominate Score}&\multicolumn{4}{c}{Log Occupational Score}\\
\cmidrule(l{.35cm}r{.25cm}){2-5}\cmidrule(l{.35cm}r{.25cm}){6-9}
&2SLS&  DML-Lasso&DML-NNet& DML-Boosting&2SLS&  DML-Lasso&DML-NNet& DML-Boosting\\
&(1)&(2)&(3)&(4)&(5)&(6)&(7)&(8)\\
\vspace{-3mm}\\
\hline
\vspace{-3mm}\\
\multicolumn{9}{c}{\textbf{Panel A:} \emph{Baseline specification with fixed-effects interactions }}\\	
\multicolumn{9}{l}{\emph{Second-stage results}}\\
Fr. Immigrants	&	1.772**	&	2.738***	&	2.359***	&	2.84***	&	0.095**	&	0.042	&	0.01	&	0.065	\\
	&	(0.829)	&	(0.777)	&	(0.79)	&	(0.874)	&	(0.042)	&	(0.041)	&	(0.056)	&	(0.053)	\\
%p-value theta	&		&	0	&	0.003	&	0.001	&		&	0.304	&	0.857	&	0.221	\\
AR 95\% CS	&	[0.049, 3.495]	&	[0.382, 4.462] 	&	[-0.05, 4.559] 	&	[0.008, 4.921] 	&	[0.017, 0.173]	&	[-0.103, 0.166] 	&	[-0.121, 0.144] 	&	[-0.133, 0.175] 	\\

\vspace{-3mm}\\
\multicolumn{9}{l}{\emph{First-stage results}}\\
Shift Share IV	&	0.965***	&	1.091***	&	1.051***	&	0.967***	&	0.933***	&	0.982***	&	0.98***	&	0.957***	\\
	&	(0.226)	&	(0.154)	&	(0.158)	&	(0.156)	&	(0.094)	&	(0.11)	&	(0.112)	&	(0.123)	\\
%p-value pi	&		&	0.000	&	0.000	&	0.000	&		&	0.000	&	0.000	&	0.000	\\
F stat	&	18.23	&	49.284	&	43.838	&	38.101	&	99.45	&	78.489	&	74.733	&	59.899	\\
AR $\chi^2$	&	3.84*	&	11.221***	&	7.869***	&	9.439***	&	5.40**	&	0.858	&	0.029	&	1.031	\\
%p-value ARchi2	&	0.05	&	0.001	&	0.005	&	0.002	&	0.02	&	0.354	&	0.865	&	0.31	\\

\vspace{-3mm}\\
\multicolumn{9}{l}{\emph{Quality of learners}}\\
Model RMSE	&	0.232	&	0.406	&	0.414	&	0.419	&	0.019	&	0.039	&	0.039	&	0.044	\\
MSE of l	&		&	0.291	&	0.297	&	0.3	&		&	0.024	&	0.023	&	0.027	\\
MSE of r	&		&	0.037	&	0.035	&	0.034	&		&	0.035	&	0.035	&	0.035	\\
MSE of m	&		&	0.021	&	0.02	&	0.021	&		&	0.024	&	0.023	&	0.023	\\
Observations	&	303	&	303	&	303	&	303	&	342	&	342	&	342	&	342	\\
No. clusters	&	157	&	157	&	157	&	157	&	125	&	125	&	125	&	125	\\
%No. variables	&	77	&	1974	&	84	&	84	&	77	&	1974	&	84	&	84	\\

\vspace{-3mm}\\
\multicolumn{9}{c}{\textbf{Panel B:} \emph{Specification with all controls (not originally implemented)}}\\																
\multicolumn{9}{l}{\emph{Second-stage results}}\\	
Fr. Immigrants	&	0.928	&	2.184	&	1.807	&	0.465	&	0.094	&	0.05	&	0.001	&	-0.039	\\
	&	(2.103) 	&	(1.507)	&	(1.85)	&	(1.52)	&	(0.060) 	&	(0.08)	&	(0.081)	&	(0.125)	\\
%p-value theta	&		&	0.147	&	0.329	&	0.76	&		&	0.529	&	0.995	&	0.755	\\
AR 95\% CS	&	[-4.276, 4.883]	&	[-2.252, 9.237] 	&	[-4.938, 10.666] 	&	[-4.131, 6.268] 	&	[-0.032, 0.207]	&	[-0.15, 0.241] 	&	[-0.158, 0.183] 	&	[-0.271, 0.255] 	\\

\vspace{-3mm}\\
\multicolumn{9}{l}{\emph{First-stage results}}\\
Shift Share IV	&	0.541***	&	0.528***	&	0.47***	&	0.657***	&	0.745***	&	0.792***	&	0.85***	&	0.589***	\\
	&	(0.178)  	&	(0.112)	&	(0.115)	&	(0.156)	&	(0.117) 	&	(0.06)	&	(0.116)	&	(0.139)	\\
%p-value pi	&		&	0.000	&	0.000	&	0.000	&		&	0.000	&	0.000	&	0.000	\\
F stat	&	9.26	&	21.987	&	16.531	&	17.587	&	40.89	&	173.047	&	52.899	&	17.575	\\
AR $\chi^2$	&	0.18	&	2.003	&	1.011	&	0.148	&	2.2	&	0.44	&	0.001	&	0.066	\\
%p-value ARchi2	&	0.668	&	0.157	&	0.315	&	0.701	&	0.138	&	0.507	&	0.981	&	0.797	\\

\vspace{-3mm}\\
\multicolumn{9}{l}{\emph{Quality of learners}}\\
Model RMSE	&	0.218	&	0.406	&	0.447	&	0.425	&	0.019	&	0.039	&	0.039	&	0.044	\\
MSE of l	&		&	0.292	&	0.319	&	0.306	&		&	0.024	&	0.024	&	0.027	\\
MSE of r	&		&	0.028	&	0.027	&	0.031	&		&	0.031	&	0.031	&	0.031	\\
MSE of m	&		&	0.018	&	0.017	&	0.016	&		&	0.022	&	0.021	&	0.023	\\
Observations	&	297	&	297	&	297	&	297	&	338	&	338	&	338	&	338	\\
No. clusters	&	154	&	154	&	154	&	154	&	125	&	125	&	125	&	125	\\
%No. variables	&	96	&	4026	&	110	&	122	&	92	&	4416	&	114	&	128	\\

\hline
\end{tabular} 	
\footnotesize
\textbf{Note:}
Panel A reports our estimates based on baseline specifications in Table 3 (Column 4, Panel B) and Table 5 (Column 2, Panel B) of \cite{tabellini2020}. Columns 1 and 5 display our estimates from conventional 2SLS regression with FD transformation, and Columns 2-4 and 6-8 our estimates from  panel IV DML estimation with different base learners. Panel B is not implemented in the original article, and adds several controls to the baseline specifications in Panel A including: predicted industrialization, log 1900 city and immigrant population, log of value added per establishment in 1904, natives’ 1900 skill ratios, 1900 fraction of blacks, 1904 log value of products per establishment, and the 1904 employment share in manufacturing. 
%%%%%
The number of raw control variables from the original analysis is 77 in Panel A (Columns 1-4), 79 Panel A (Columns 5-8), 96 in Panel B (Columns 1-4), 92 in Panel B (Columns 5-8). Slight differences in the number of covariates in the panel IV DML estimates are due to the type of learner and the FD (exact) approach. Specifically, the set of confounding variables in panel IV DML estimations with NNet and Boosting (Columns 3-4, 7-8) does not include any interaction terms, unlike the original analysis, because these base learners are designed to capture possible nonlinearities in the data. Panel IV DML estimation with Lasso (Columns 2 and 6) uses an extended dictionary of the raw variables, which includes polynomials up to order three and interaction terms between all the covariates, to satisfy Lasso's weak sparsity assumption. The number of covariates used in panel IV DML estimation (Columns 2-4, 6-8) also includes the lags of all covariates, following the FD (exact) approach.
%%%%%%%%%%%%%%%%%%%
Observations for which data is unavailable are dropped from the final estimation sample; therefore, when there are only two time periods the unit with at least a missing case is dropped after the first-difference. The difference in the number of observations between estimation methods is explained by missing values generated after transforming the covariates to use as inputs in the machine learning algorithms.
% In Panel A, the number of observations is 303 with 157 groups (Columns 1-4), and  the number of observations 356 is with 126 groups (Columns 5-8). In 
% Panel B, the number of observations is 297 with 154 groups (Columns 1-4), and
% the number of observations 348 with 125 groups (Columns 5-8).
%%%%%
Panel IV DML technical note: 2 folds, cross-fitting, hyperparameters are tuned as per Table~\ref{tab:hyperpara}. 
%%%%%
Standard errors in parenthesis are clustered at the metropolitan area in Columns 1-4 and at the city code level in Columns 5-8. Significance levels: * p $<$ 0.10, ** p $<$ 0.05, *** p $<$ 0.01.
\end{threeparttable}
}
\end{table}
  %%new

%\textbf{2SLS vs panel IV DML.} 
We now proceed to evaluate how our panel IV DML estimates compare to the corresponding 2SLS with FD. Table~\ref{tab:tabellini2020_D} reports the results of the baseline specifications in Panel A, and the augmented specifications with all control variables in Panel~B for the two outcome variables of interest. Columns~1 and 5 in Table~\ref{tab:tabellini2020_D} report the 2SLS estimates with FD, and Columns 2-4 and 6-8 the panel~IV~DML estimates with FD using Lasso, Neural Network (NNet), and Gradient Boosting (Boosting). 
%While the instrument does not appear weak in the conventional 2SLS specifications, it is important to 

For each specification, we need to verify whether the shift-share instruments in panel IV DML regressions still predict immigrant shares across cities (i.e., instrument relevance) within the more flexible DML framework because our estimator controls for covariates in a more flexible way and, therefore, the effective strength of the instrument may differ.
%%%%
We start by discussing the baseline regression estimates (Panel A of Table~\ref{tab:tabellini2020_D}) focusing on one outcome at a time.  For the political outcome (Columns 2-4), the first-stage F-statistics from panel IV DML regressions are much larger than those from 2SLS, which slightly exceeds \citet{stock2005}'s cut-off of 16.30, but never exceed \citet{lee2022}'s threshold of 104.70. 
In this case, Panel IV DML strengthens the relevance of the shift-share instrument.
The AR test statistics for the relevance of the reduced-form stage, when there is no second-stage effect regardless of the strength of the instrument, strongly reject the null hypothesis with panel IV DML, suggesting some effect of immigration on the political outcome.
This is confirmed by the AR confidence sets (CS) at 95\% level which are bounded, include the estimated second-stage coefficient and do not include zero, with the exception of neural network due to small sample size (further reduced by sample splitting). Given the strength of the instrument (based on the \citet{stock2005}'s threshold), we can comment on the statistical significance of the second-stage coefficient and make inference about the effect of immigration. All second-stage coefficients are positive and significant at 1\% level; specifically, panel IV DML coefficients are larger in magnitudes than 2SLS, suggesting a stronger effect of the fraction of immigrants on the political outcome than originally found. 
Overall, panel IV DML findings indicate that a greater inflow of immigrants leads to the election of more conservative congressmen; the direction of the effect is aligned with the original 2SLS findings, which seem to be underestimated.

%%%%
Moving to the economic outcome (Columns~5-8 of Panel~A), the first-stage F-statistics from panel IV DML are smaller than those from 2SLS, but always well above the conventional critical value of 16.30 by \citet{stock2005} and below 104.70 by \citet{lee2022}'s. Therefore, the instrument can still be considered strong with panel IV DML. Unlike 2SLS results, the AR test statistic and AR CS from panel IV DML regressions highlight the absence of any effect for the economic outcome. That is,  in Columns 6-8, the AR test never rejects the null hypothesis of no reduced-form effect when the treatment effect is assumed to be absent, and the AR CS always include zero as a possible value of the treatment effect. Given the strength of the IV and the results of the AR diagnostics, we can interpret the second-stage coefficients as not statistically different from zero with panel IV DML. By contrast, 2SLS finds a positive and statistically significant (at 5\% level) second-stage coefficient.
Overall, panel IV DML findings contradict conventional 2SLS results, suggesting that there is no evidence that employment gains for natives, induced by immigration, were accompanied by occupational or skill upgrading, as found in the original article. 
%Overall, these findings indicate that a greater inflow of immigrants leads to the election of more conservative congressmen whereas there is no evidence that immigration-induced employment gains for natives were accompanied by occupational or skill upgrading, as found in the original article. 

We now focus on Panel~B of Table~\ref{tab:tabellini2020_D}, when all controls are included simultaneously (not originally implemented). In both specifications (for political and economic outcomes), the first-stage F-statistics generally become smaller than those from Panel A. In Columns~2-4 of Panel~B, the shift-share instrument in the panel IV DML regressions produces stronger first-stages ($F>16.30$) relative to the corresponding 2SLS regression, where the instrument is clearly weak ($F=9.26$). In Columns~5-8 of Panel~B, the first-stage F-statistics obtained from both 2SLS and panel IV DML estimators are largely above \citet{stock2005}'s threshold, and in the case of the panel IV DML regression with Lasso even above \citet{lee2022}'s threshold of 104.70. Overall, results for Columns~2-8 provide supporting evidence of a strong instrument. This pattern suggests that allowing for flexible and potentially nonlinear control adjustment in the instrument equation strengthens the predictive power of the instrument with panel IV DML. The AR test statistics are insignificant with both estimators in both political and economic specifications, suggesting we cannot reject no second-stage effects. The absence of a causal effect for either outcome variable is supported by the AR CS, which include zero as a plausible value of the treatment effect, as well as by the statistical insignificance of the second-stage coefficients with 2SLS and panel IV DML estimators. 

This first empirical application highlights that our method can enhance the plausibility of instrument's validity assumption, while controlling for more and potentially nonlinear confounding variables. We show that our panel IV DML estimator generally confirms the conventional 2SLS results, with the only exception of the economic outcome in the baseline specifications (Panel A of Table~\ref{tab:tabellini2020_D}). 
Our robustness analysis with additional control variables (Panel B  of Table~\ref{tab:tabellini2020_D}) illustrates how the panel IV DML method in shift-share designs, where conventional estimators may fail. We find that the main effect of the original article (Columns 1-5 in Panel A of Table~\ref{tab:tabellini2020_D}) for both outcomes disappears in both 2SLS and panel IV DML regressions.

\subsection{Empirical Examples: \citet{moriconi2019} and \citet{moriconi2022}}\label{sec:moriconi2019_2022} 

In this section, we revisit two related studies by \cite{moriconi2019, moriconi2022} on the impact of high-skilled (HS) and low-skilled (LS) immigration on support for redistribution policies and perceived attitudes in European countries. 
Similarly to \citet{tabellini2020}, endogeneity concerns arise as immigrants may decide to settle in areas with more favourable policies and attitudes towards immigration.
%a simple OLS regression of the outcomes on local immigration would raise endogeneity concerns, as immigrants may settle in areas with more favourable policies and attitudes towards immigration.
To address these issues, an IV approach is employed with a modified version of the shift-share instrumental variable~\eqref{eqn:ssh} by skill-specific group (i.e., HS and LS). %to capture the skill-origin specific effects of immigration. 
The instruments are constructed interacting the aggregate immigrant flows by skill group and country of origin (the `shift') with the initial distribution of immigrants by nationality across regions (the `share'), as in \citet{mayda2022}. The authors show that the skill-specific shift-share instruments are uncorrelated with economic and demographic regional trends in the period before the analysis, so that the exclusion restriction is satisfied. %This supports the validity of the exclusion restriction. 
%The original studies find that the shift-share instruments are generally strong, with large first-stage F-statistics when used individually ($16.30<F\ll104.7$), but weaker when included jointly in the same regression ($10<F\le16.3$). %When the instruments for high-skilled (HS) and low-skilled (LS) immigrants are used together, concerns of weak IV may arise because the first-stage F statistic is much smaller, although the authors exclude this is the case. \red{However, the authors claim that concerns of weak IV should be reasonably minor.} 

The data for the analyses in both articles are obtained from the European Social Survey (ESS), the European Labor Force Survey (EULFS), the Manifesto Project Database and Eurostat. The main analyses are implemented on a dataset of individual voters from twelve European countries observed over the election years between 2007 and 2016. The final dataset is not a conventional panel dataset, where the same subject is observed over multiple time periods, as contains information of randomly sampled native voters across regions of the same twelve 12 EU countries in each election year.
In contrast, our panel IV DML method requires (balanced or unbalanced) panel data where the same subject is followed over at least two consecutive periods. To satisfy this requirement, we construct an aggregated balanced panel dataset from the individual data by averaging variables at the NUTS2 region-year level.\footnote{In the original analyses both the endogenous variables and the instruments are at the NUTS2 region-year level. Therefore, the level of aggregation of these variables does not change in our aggregated data, but it affects the outcome variables and the individual-level controls only.}  %After the aggregation, we observe each European country in at least two election years. , allowing the use of the panel IV DML method. 
%The dataset of party platforms used in \citet{moriconi2019} is an unbalanced panel in which political parties from different European countries are observed over up to three election years. However, the relative specifications in the original article is estimated through pooled least squares regressions with country and election-year fixed effects rather than applying the within-group panel transformation at the party level to address unobserved heterogeneity.

Both studies estimate the main specifications through pooled ordinary least-squares (POLS) regressions with regional and election-year fixed effects, using individual voters' data.  Therefore,  we first re-estimate the key specifications using both 2SLS with fixed effects (FE) and 2SLS with first-differences (FD) estimation approaches using our aggregated (unbalanced) panel dataset to establish a consistent comparison with our panel IV DML method. %While 2SLS with FE is the standard estimation method for panel data with this type of analysis, 2SLS with FD is the panel data technique most comparable to our proposed panel IV DML approach. This preliminary step allows us to compare results across panel estimators that remove idiosyncratic regional variation in different ways, and to ensure that any differences with the panel IV DML method arise from the use of a different estimation methodology. Finally, we apply our panel IV DML method using three base learners (Lasso, Gradient Boosting, and Neural Network) to model potentially nonlinearities in the nuisance functions.

\subsubsection{Empirical Application:  \citet{moriconi2019}}\label{sec:moriconi2019}
The article investigates the effect of skilled and unskilled immigration on individual preferences for the expansion of the welfare state and public education in twelve EU countries. While the overall effect of migration varies, their main findings highlight a pro-redistribution effect with HS immigration from natives’ voters. % \red{and an anti-redistribution effect with LS immigration from party platforms}.

In this second empirical application,  we re-examine the main results of \citet{moriconi2019} relative to  Columns 2, 4 and~6 (Panels A and B) of their Table~4 (on individual voters' data).\footnote{In the Online Appendix~\ref{sec:app_moriconi2019}, we revisit the specifications in Table~5 of \citet{moriconi2019}, estimated using parties' data. We thank the authors for providing us with the entire replication package to reproduce the main analysis.}  
The dependent variables are: \emph{Net Welfare State} and \emph{Net Public Education}.
The control variables included in the regressions are the same as in the original article and include: the share of women, average age, share of tertiary/post-tertiary education, average GDP per capita (in log), share of tertiary sector (in log), average unemployment rate, and election year dummies.  The regression analysis is separately estimated by skill-specific group of immigrants.

%%%%%%%%%%%%%%%%%%%%
%\textbf{2SLS (FD vs FE) vs original.} 
Before discussing the panel IV DML results, we briefly compare the results from conventional 2SLS estimators with FE and FD (see Online Appendix~\ref{sec:app_moriconi2019} for more details). In general, both FE and FD estimators confirm that the instrument is strong in the HS immigration sample ($F\ll16.30$), but borderline weak in the LS immigration sample  (as $10<F<16.30$) for both specifications. AR diagnostics, not computed in the original study, find the presence of a second-stage effect of HS immigrants and, unlike the original article, of LS immigrants on both outcomes. The AR CS from both panel estimators agree on the direction of the effect of HS immigrants on welfare expansion, and of LS immigrants on education expansion. Therefore, any difference observed between 2SLS with FD and panel IV DML later can be explained by differences in the adopted methodology, i.e. 2SLS versus panel IV DML.

%%%%%%%%%%%%%%%%%%%%%%
%\textbf{2SLS vs Panel IV DML.} 
Table~\ref{tab:moriconi2019_tab4} reports our estimates obtained from 2SLS regressions with FD (Columns~1 and~5) and panel IV DML regressions using the FD approach (Columns~2-4, and~6-8), based on the aggregated panel dataset. The first-stage regression results in Panels~A and~B are identical for the same skill group, as they use the same subsample. In both panels, the first-stage F-statistics for HS and LS immigrants are substantially smaller in the panel IV DML specifications than in the corresponding 2SLS regressions. %\blue{This difference reflects the fact that, under panel IV DML, the instrument equations are estimated separately and allow for more flexible control of a rich set of confounding variables. As a result, the instruments' exogeneity is strengthened, making them closer to being \emph{as good as randomly assigned}.}

%%%%%%%%%%%%%%%%%%%%%%%%%%%%%%%%%%%%%%%%%%%%%%%%%%%%%%%%%%%%%%%%%%%%%%%%%%%%
%%%%% Tab D2 - col (4) 
%%%%%%%%%%%%%%%%%%%%%%%%%%%%%%%%%%%%%%%%%%%%%%%%%%%%%%%%%%%%%%%%%%%%%%%%%%%%
\begin{table}[t!]
\centering
 \caption{Political preferences over 2007–2016 -- Aggregated individual voters}\label{tab:moriconi2019_tab4}
 \vspace{-2mm}
\scalebox{.52}{
\begin{threeparttable}	%use it for notes
  \begin{tabular}{lcccccccc}   
\vspace{-3mm}\\   
\hline\hline
\vspace{-3mm}\\	
\emph{Sample: }&\multicolumn{4}{c}{HS immigrants}&\multicolumn{4}{c}{LS immigrants}\\
\cmidrule(l{.35cm}r{.25cm}){2-5}\cmidrule(l{.35cm}r{.25cm}){6-9}
&2SLS&  DML-Lasso&DML-NNet& DML-Boosting&2SLS&  DML-Lasso&DML-NNet& DML-Boosting\\
&(1)&(2)&(3)&(4)&(5)&(6)&(7)&(8)\\
\vspace{-3mm}\\
\hline
\vspace{-3mm}\\
\multicolumn{9}{c}{\textbf{Panel A:} \emph{Net Welfare State}}\\	
\multicolumn{9}{l}{\emph{Second-stage results}}\\
Fr. Immigrants	&	0.054***	&	0.368*	&	0.59**	&	0.34	&	0.05	&	0.057	&	0.327	&	-0.777	\\
	&	(0.014)	&	(0.189)	&	(0.285)	&	(0.276)	&	(0.033) 	&	(0.09)	&	(0.526)	&	(1.082)	\\
%p-value theta	&		&	0.051	&	0.038	&	0.219	&		&	0.529	&	0.534	&	0.473	\\
\vspace{-3mm}\\
\multicolumn{9}{l}{\emph{Robust Weak IV Tests}}\\
AR $\chi^2$ stat	&	  15.36***	&	13.114***	&	15.398***	&	7.692***	&	3.6*	&	1.091	&	1.743	&	7.036***	\\
%p-value AR $\chi^2$	&	0	&	0	&	0	&	0.006	&	0.058	&	0.296	&	0.187	&	0.008	\\
AR 95\% CS	&	[0.029, 0.081] 	&	$( -\infty, -0.154 ] \cup [-0.008 , +\infty ) $	&	$( -\infty, -0.288 ] \cup [-0.019 , +\infty )$ 	&	$( -\infty, +\infty )$ 	&	[0.003,  0.159]	&	$( -\infty, +\infty )$ 	&	$( -\infty, +\infty )$ 	&	$( -\infty, +\infty )$ 	\\

\vspace{-3mm}\\
\multicolumn{9}{l}{\emph{Quality of learners}}\\
Model RMSE	&	0.135	&	0.374	&	0.518	&	0.323	&	0.145	&	0.213	&	0.608	&	1.444	\\
MSE of l	&		&	0.145	&	0.147	&	0.139	&		&	0.145	&	0.147	&	0.139	\\
MSE of r	&		&	1.009	&	0.768	&	0.743	&		&	1.369	&	1.779	&	1.711	\\
MSE of m	&		&	0.3	&	0.295	&	0.338	&		&	0.674	&	0.645	&	0.695	\\

\vspace{-3mm}\\
\multicolumn{9}{c}{\textbf{Panel B:} \emph{Net Public Education}}\\	
\multicolumn{9}{l}{\emph{Second-stage results}}\\
Fr. Immigrants	&	0.024	&	0.265**	&	0.333**	&	0.19	&	 -0.088*	&	-0.125	&	-0.279	&	-0.07	\\
	&	(0.015)	&	(0.118)	&	(0.133)	&	(0.145)	&	(0.046)	&	(0.135)	&	(0.52)	&	(0.134)	\\
%p-value theta	&		&	0.024	&	0.012	&	0.192	&		&	0.351	&	0.591	&	0.6	\\
\vspace{-3mm}\\
\multicolumn{9}{l}{\emph{Robust Weak IV Tests}}\\
AR $\chi^2$ stat	&	2.78*	&	8.414***	&	7.194***	&	2.364	&	6.65***	&	3.003*	&	0.608	&	0.024	\\
%p-value AR $\chi^2$	&	0.096	&	0.004	&	0.007	&	0.124	&	0.01	&	0.083	&	0.435	&	0.878	\\
AR 95\% CS	&	[-0.004, 0.055]	&	$( -\infty, +\infty )$ 	&	$( -\infty, +\infty )$ 	&	$( -\infty, +\infty )$ 	&	[-0.25, -0.022]	&	$( -\infty, +\infty )$ 	&	$( -\infty, +\infty )$ 	&	$( -\infty, +\infty )$ 	\\

\vspace{-3mm}\\
\multicolumn{9}{l}{\emph{Quality of learners}}\\
Model RMSE	&	0.139	&	0.306	&	0.314	&	0.227	&	0.181	&	0.276	&	0.571	&	0.264	\\
MSE of l	&		&	0.149	&	0.14	&	0.146	&		&	0.149	&	0.14	&	0.146	\\
MSE of r	&		&	1.009	&	0.768	&	0.743	&		&	1.369	&	1.779	&	1.711	\\
MSE of m	&		&	0.3	&	0.295	&	0.338	&		&	0.674	&	0.645	&	0.695	\\

\vspace{-3mm}\\
\multicolumn{9}{c}{\textbf{Panels A and B}}\\
\multicolumn{9}{l}{\emph{First-stage results} }\\
Shift Share IV	&	1.655***	&	0.487**	&	0.491**	&	0.341	&	0.522***	&	0.259	&	0.209	&	-0.204	\\
	&	(0.263) 	&	(0.199)	&	(0.215)	&	(0.236)	&	(0.162)	&	(0.198)	&	(0.335)	&	(0.293)	\\
%p-value pi	&		&	0.014	&	0.022	&	0.148	&		&	0.191	&	0.532	&	0.485	\\
F stat	&	39.51	&	5.904	&	5.127	&	2.056	&	10.37	&	1.681	&	0.383	&	0.478	\\

\vspace{-3mm}\\
\hline
\vspace{-3mm}\\
Observations	&	146	&	146	&	146	&	146	&	146	&	146	&	146	&	146	\\
No. clusters	&	113	&	113	&	113	&	113	&	113	&	113	&	113	&	113	\\
%No. variables	&	9	&	297	&	18	&	18	&	9	&	297	&	18	&	18	\\

\hline
\end{tabular} 	
\footnotesize
\textbf{Note:} The table displays our estimates based on Specifications~(2), (4) and (6) of Table~4 (Panels A and B) in \cite{moriconi2019} obtained from conventional 2SLS regression with FD transformation (Columns 1 and 5), and our panel IV DML estimation with different base learners (Columns 2-4 and 6-8).
%%%%%%%
The sample is aggregated sample at regional (NUTS2) level to construct an unbalanced panel data set. The treatment and instrumental variables in Columns (1)-(2) and (5)-(6) refer to the fraction of high-skilled workers, and in Columns (3)-(4) and (7)-(8) of low-skilled workers. The dependent variable in Panel A is `Net Welfare State, and in Panel B `Net Public Education'. 
%%%%%%%
Raw control variables in all panels are: the share of women, average age, share of tertiary/post-tertiary education, average GDP per capita (in log), share of tertiary sector (in log), average unemployment rate, and election year dummies. The set of control variables in the panel IV DML estimation with NNet and Boosting does not include interaction terms because these base learners are designed to capture nonlinearities in the data. Panel IV DML estimation with Lasso (Columns 2 and 6) uses an \emph{extended dictionary} of the raw variables, including polynomials up to order three and interaction terms between all the covariates, to satisfy Lasso's weak sparsity assumption. The number of covariates used for panel IV DML estimation doubles due to the inclusion of the lags of all included covariates, following the FD (exact) approach.
%%%%%%%
Panel IV DML technical note: 2 folds, cross-fitting, hyperparameters are tuned as per Table~\ref{tab:hyperpara}. 
%%%%%%
Standard errors in parenthesis are clustered at the regional level.
Significance levels: * p $<$ 0.10, ** p $<$ 0.05, *** p $<$ 0.01. 
\end{threeparttable}
}
\end{table}

Focusing on HS immigrants (Panels~A and~B of Columns~1-4 in Table~\ref{tab:moriconi2019_tab4}), the first-stage F-statistics from panel IV DML regressions fall well below the conventional rule-of-thumb value of 10, whereas the corresponding F-statistics of 2SLS is above the \citet{stock2005} threshold ($F = 39.5$). Therefore, under panel IV DML, the shift-share instrument for HS immigrants appears extremely weak, unlike with conventional 2SLS.
The AR test statistics, robust to weak instruments by construction, strongly reject the null hypothesis of no effect at 1\% level with panel IV DML in both panels (with the exception of gradient boosting in Panel B), indicating the presence of an effect of HS immigration on both outcomes. By contrast, the AR test statistic from 2SLS regressions rejects the null hypothesis at 1\%  in Panel~A only, and at 10\% level in Panel~B.
The corresponding AR CS from the panel IV DML regressions are either disjoint sets that include zero or unbounded, reflecting limited information to identify the effect due to weak instruments, even when a causal effect may be present. 
%Second-stage coefficients from the panel IV DML regressions are statistically significant in most cases, with the exception of the boosting specification. Given the known instability of boosting in small samples, this result should be interpreted with caution. 
More generally, all second-stage estimates of the treatment parameter in these specifications should be interpreted cautiously with both estimators, as the weakness of the shift-share instrument makes standard inference unreliable.

For LS immigrants (Columns 5-8 of Panels~A and~B in Table~\ref{tab:moriconi2019_tab4}), the first-stage F-statistics from panel IV DML estimators are well below the rule-of-thumb threshold of 10 while F-statistic from 2SLS barely exceeds 10. This raises serious concerns about the relevance of the shift-share instrument for LS immigrations. Consistent with this, the AR test statistics generally fail to reject the null hypothesis with panel IV DML in both panels, with a few exceptions. In Panel~A, the AR test statistic rejects the null hypothesis at the 1\% level for boosting specification; however, given the known instability of the boosting in small samples, the latter result should be interpreted with caution. In Panel~B, the AR test statistic is borderline significant (at 10\% level) for the Lasso specification, but broadly consistent with the absence of a treatment effect with the other learners. The AR CS obtained from panel IV DML are unbounded (entire real line) in all cases, indicating insufficient  information to identifying the range of causal effects, if any. By contrast, the AR confidence sets from the 2SLS specifications are always bounded and contain the corresponding second-stage point estimates, a pattern consistent with the tendency of 2SLS to inflate identification strength under weak instruments.

Overall, our panel IV DML does not fully support 2SLS findings mainly due concerns about the strength of the shift-share instruments. While the instruments appear moderately strong under standard 2SLS specifications, they are found to be weak once high-dimensional and potentially nonlinear confounding is flexibly accounted for using the panel IV DML estimator. The AR diagnostics further indicate that, although some effects may be present, weak IV prevents the identification of the causal effect. While panel IV DML makes the identifying assumptions more plausible under the included controls, in this setting this comes at the cost of reduced effective relevance, limiting the reliability of second-stage point estimates.

%% PARA BELOW REMOVED BECAUSE NO POINT IN COMMENTING 2ND STAGE!
%The second-stage estimates obtained with panel IV DML suggest positive effects of HS immigration for individual voters, with magnitudes generally larger than those from conventional 2SLS, which is consistent with panel IV DML capturing nonlinearities in the confounders more flexibly. We also find a positive and significant effect of HS immigration on political parties' preferences for welfare state expansion when using NNet and Boosting, whereas LS immigration shows no significant effect for either outcome with panel IV DML. However, because the shift-share instrument appears weak in these specifications, all second-stage results should be interpreted with caution.

%%%%%%%%%%%%%%%%%%%%%%
%\red{\textbf{2SLS, DML vs POLS with FE.} Overall, our conventional 2SLS results with aggregated data support original \citet{moriconi2019}'s findings on the pro-redistribution effect of HS immigrants on natives’ votes and anti-redistribution effect of LS immigrants on party platforms. We also find a strong negative effect of LS immigration on public education expansion among individual voters. The first-stage F statistic is larger than the critical value of 16.30 suggested by \citet{stock2005} for individual voters (except for Panel~B, where the effect of LS skill immigration is tested), but not for parties preferences.}

\subsubsection{Empirical Application:  \citet{moriconi2022}}\label{sec:moriconi2022}

In line with the article discussed in the previous section, \citet{moriconi2022} examine the impact of high-skilled (HS) and low-skilled (LS) immigration on individual voting patterns towards parties with a nationalist agenda, and perceived attitudes towards politics and immigration. The authors find that a higher share of HS immigrants decreases the intensity of nationalist preferences of native voters, and an opposite effect for LS immigrants. %In addition, the authors observe that immigration affects nationalistic voting through a change in individual attitudes towards politics and immigrants. 

In this third empirical application, we revisit the baseline specifications of \citet{moriconi2022} on the effect of HS and LS immigrants on nationalism (Columns 2-3 from their Table~6), and on the change in attitudes towards politics and immigrants  (Columns 1, 3, 4 and~6 from their Table~10).  The dependent variables of interest for the reanalysis are:  an index of nationalism intensity of parties (\emph{Nationalism}), constructed by matching individual-level party votes in national elections with the content of each party’s political manifesto; an index of trust in country parliament as political attitude; and a measure for a better place to live as attitude towards migrants.
%\footnote{Table~\ref{tab:moriconi2022_tab10_other} in the Appendix reports our estimates of  Table~10 (Columns 2 and 5) in \citet{moriconi2022} for the other dependent variables on political and migration attitudes present in the original article, such as measures for \emph{Trust in EU Parliament} (Panel A), \emph{More EU} (Panel B), \emph{Good economy} (Panel C), and \emph{Enrich culture} (Panel D).}
The set of control variables employed in analysis are the same as those in Table~\ref{tab:moriconi2019_tab4}.

We first discuss the key insights from the comparison between conventional 2SLS estimates with FE and with FD transformation (see the Online Appendix~\ref{sec:app_moriconi2022} for more details). First-stage strength is consistently higher under FE than FD, mainly because FD drops the first time-period and subjects without consecutive observations. The shift-share instruments in both FE and FD regressions are not always strong, in contrast with the original study, where $16.30 < F < 104.70$ in all first-stage specifications. While the instruments can be considered moderately strong in all specifications using the sample of HS immigrants, this is not always the case for LS immigrants samples. Specifically, for the nationalism outcome, the instrument is essentially irrelevant ($F \approx 0$) for LS immigration under both estimators, undermining the credibility of the relative estimates. The second-stage results seem to be estimator-sensitive. The effect of HS immigration on political attitudes (originally only marginally significant at 10\% level) survives only in FE, and the LS immigration effect on immigration attitudes disappears entirely. 
%\blue{The first-stage F-statistics are consistently larger in FE than in FD, which is estimated on a smaller sample by construction; observations at $T=1$ are removed after the FD transformation. The shift-share instruments are moderately strong for HS immigration across specifications and outcomes, and for LS immigration only in the political and immigration-attitude specifications, though FD F-statistics only marginally exceed the \citet{stock2005}'s critical value. By contrast, the instrument is  irrelevant ($F \approx 0$) for LS immigration in the nationalism specification under both estimators. In the original study, all instruments are always moderately strong ($16.30 < F < 104.70$), instead. The second-stage effect on political attitudes for HS immigration (significant at 10\% in the original study) remains only in FE models, while the LS immigration effect on immigration attitudes disappears.}

Moving to the panel IV DML results, Table~\ref{tab:moriconi2022_tabs6-10} displays conventional 2SLS with FD estimates (Columns~1 and~5) and our panel IV DML estimation results (Columns~2-4, and~6-8) by dependent variable. 
%%%%%%%%%%%%%
For HS immigrants (specifications in Columns~1-4 Panels A-C), the first-stage F-statistics obtained with panel IV DML never exceed even the rule-of-thumb threshold of 10 and are always substantially smaller than those from 2SLS (always $F>16.30$). This provides initial evidence of the presence of weak instruments in these specifications for HS immigrants with panel IV DML. % indicating that the inclusion of many variables in a flexible way in the additional instrument regression equation has made the instrument \emph{as good as randomly assigned}. 
In all panels, the AR test  with 2SLS and panel IV DML fails to reject the null hypothesis of no reduced-form effect when assuming that the treatment has no effect. The associated AR CS are either bounded with zero included, suggesting no second-stage effect, or unbounded (real line), indicating that the causal parameter cannot be identified with the available variation in the data due to weak IV. Both estimators therefore point to the absence of a causal effect, although only panel IV DML explicitly reveals the weak-identification problem. %Without AR diagnostics, conventional 2SLS would misleadingly suggest strong instruments and no effect, whereas the instruments are weak and inference is uninformative.

%%%%%%%%%%%%%%%%%%%%%%%%%%%%%%%%%%%%%%%%%%%%%%%%%%%%%%%%%%%%%%%%%%%%%%%%%%%%
%%%%% Tables 6 (cols 2-3) and 10 (cols 1 and 3, 4 and 6)Nationalismintensityandimmigrant share.
%%%%%%%%%%%%%%%%%%%%%%%%%%%%%%%%%%%%%%%%%%%%%%%%%%%%%%%%%%%%%%%%%%%%%%%%%%%%
\begin{table}[t!]
\centering
 \caption{Nationalism intensity, attitudes towards politics and immigration}\label{tab:moriconi2022_tabs6-10}
 \vspace{-2mm}
\scalebox{.535}{
\begin{threeparttable}	%use it for notes
  \begin{tabular}{lcccccccc}   
\vspace{-3mm}\\   
\hline\hline
\vspace{-3mm}\\	
\emph{Sample: }&\multicolumn{4}{c}{HS immigrants}&\multicolumn{4}{c}{LS Immigrants}\\
\cmidrule(l{.35cm}r{.25cm}){2-5}\cmidrule(l{.35cm}r{.25cm}){6-9}
& 2SLS &  DML-Lasso&DML-NNet& DML-Boosting&2SLS&  DML-Lasso&DML-NNet& DML-Boosting\\
&(1)&(2)&(3)&(4)&(5)&(6)&(7)&(8)\\
\vspace{-3mm}\\
\hline
\vspace{-3mm}\\
\multicolumn{9}{c}{\textbf{Panel A:} \emph{Nationalism intensity of parties}}\\	
\multicolumn{9}{l}{\emph{Second-stage results}}\\
Fr. Immigrants	&	-0.049	&	-0.100	&	-0.054	&	-0.017	&	0.084	&	-2.561	&	-0.086	&	1.41	\\
	&	(0.044)	&	(0.089)	&	(0.079)	&	(0.159)	&	 (0.067)	&	(3.186)	&	(0.297)	&	(8.882)	\\
%p-value theta	&		&	0.261	&	0.5	&	0.916	&		&	0.421	&	0.773	&	0.874	\\
AR 95\% CS	&	[-0.140, 0.025]	&	$( -\infty, +\infty ) $	&	$( -\infty, +\infty ) $	&	$( -\infty, +\infty ) $	&	[-0.028, 0.316]	&	$( -\infty, +\infty ) $	&	$( -\infty, +\infty ) $	&	$( -\infty, +\infty ) $	\\
 \vspace{-3mm}\\
\multicolumn{9}{l}{\emph{First-stage results}}\\
Shift Share IV	&	1.476***	&	0.601**	&	0.742**	&	0.406*	&	0.602***	&	0.167	&	0.189	&	-0.051	\\
	&	(0.241)   	&	(0.257)	&	(0.308)	&	(0.217)	&	(0.189) 	&	(0.208)	&	(0.356)	&	(0.262)	\\
%p-value pi	&		&	0.019	&	0.016	&	0.062	&		&	0.424	&	0.595	&	0.846	\\
F stat	&	37.54	&	5.381	&	5.699	&	3.433	&	8.31	&	0.629	&	0.277	&	0.037	\\
AR $\chi^2 $ stat	&	1.39	&	0.307	&	0.401	&	0.142	&	2.13	&	11.632***	&	0.066	&	3.456*	\\
%p-value AR $\chi^2 $	&	0.239	&	0.579	&	0.527	&	0.706	&	0.144	&	0.001	&	0.797	&	0.063	\\
 \vspace{-3mm}\\
\multicolumn{9}{l}{\emph{Quality of learners}}\\
Model RMSE	&	0.244	&	0.365	&	0.297	&	0.337	&	0.274	&	4.536	&	0.353	&	4.714	\\
MSE of l	&		&	0.27	&	0.248	&	0.285	&		&	0.27	&	0.248	&	0.285	\\
MSE of r	&		&	0.961	&	0.872	&	0.813	&		&	1.365	&	1.521	&	1.85	\\
MSE of m	&		&	0.28	&	0.278	&	0.318	&		&	0.713	&	0.72	&	0.671	\\
Observations	&	147	&	147	&	147	&	147	&	147	&	147	&	147	&	147	\\
No. clusters	&	114	&	114	&	114	&	114	&	114	&	114	&	114	&	114	\\
%No. variables	&	9	&	207	&	18	&	18	&	9	&	207	&	18	&	18	\\

%%%%%%%%%%%%%%%%%%%%%%%%%%%%%%%%%%%%%%%%%%%%%%%%%%%%%%%%%%%%%%%%%%%%%%%%%%%%%%%%%%%%%%%%%%%%%%%%%%%%
\vspace{-3mm}\\
\multicolumn{9}{c}{\textbf{Panel B:} \emph{Political attitudes -- Trust in country parliament}}\\			
\multicolumn{9}{l}{\emph{Second-stage results}}\\
Fr. Immigrants	&	0.067	&	0.014	&	-0.035	&	0.045	&	0.05	&	0.137	&	0.344	&	0.205	\\
	&	(0.044)	&	(0.066)	&	(0.053)	&	(0.058)	&	 (0.051)	&	(0.142)	&	(0.5)	&	(0.179)	\\
%p-value theta	&		&	0.833	&	0.508	&	0.438	&		&	0.333	&	0.491	&	0.251	\\
AR 95\% CS	&	[-0.015,0.174]	&	[-0.139, 0.146] 	&	[-0.154, 0.077] 	&	[-0.082, 0.151] 	&	[-0.030,0.181]	&	[-0.016, 0.513] 	&	$( -\infty, +\infty ) $	&	$( -\infty, +\infty ) $	\\
 \vspace{-3mm}\\
\multicolumn{9}{l}{\emph{First-stage results}}\\
Shift Share IV	&	1.174***	&	1.062***	&	1.268***	&	1.15***	&	0.647***	&	0.559**	&	0.166	&	0.291	\\
	&	(0.277)	&	(0.397)	&	(0.444)	&	(0.394)	&	 (0.155)	&	(0.244)	&	(0.206)	&	(0.207)	\\
%p-value pi	&		&	0.008	&	0.004	&	0.004	&		&	0.022	&	0.42	&	0.16	\\
F stat	&	19.09	&	7.016	&	8.002	&	8.353	&	16.39	&	5.139	&	0.639	&	1.941	\\
AR $\chi^2 $ stat	&	3.3*	&	0.001	&	0.507	&	0.449	&	1.17	&	1.936	&	2.992*	&	1.558	\\
%p-value AR $\chi^2 $	&	0.069	&	0.982	&	0.476	&	0.503	&	0.279	&	0.164	&	0.084	&	0.212	\\

 \vspace{-3mm}\\
\multicolumn{9}{l}{\emph{Quality of learners}}\\
Model RMSE	&		&	0.423	&	0.383	&	0.413	&		&	0.506	&	0.782	&	0.63	\\
MSE of l	&		&	0.249	&	0.228	&	0.241	&		&	0.249	&	0.228	&	0.241	\\
MSE of r	&		&	0.739	&	0.851	&	0.811	&		&	1.172	&	1.345	&	1.248	\\
MSE of m	&		&	0.239	&	0.227	&	0.255	&		&	0.409	&	0.398	&	0.408	\\
Observations	&	327	&	327	&	327	&	327	&	327	&	327	&	327	&	327	\\
No. clusters	&	114	&	114	&	114	&	114	&	114	&	114	&	114	&	114	\\
%No. variables	&		&	250	&	20	&	20	&		&	250	&	20	&	20	\\

%%%%%%%%%%%%%%%%%%%%%%%%%%%%%%%%%%%%%%%%%%%%%%%%%%%%%%%%%%%%%%%%%%%%%%%%%%%%%%%%%%%%%%%%%%%%%%%%%%%%
\vspace{-3mm}\\
\multicolumn{9}{c}{\textbf{Panel C:} \emph{Migration attitudes -- Better place to live}}\\			
\multicolumn{9}{l}{\emph{Second-stage results}}\\													
Fr. Immigrants	&	-0.023	&	-0.027	&	-0.109	&	-0.074	&	-0.100*	&	0.065	&	-0.311	&	-0.254	\\
	&	(0.048) 	&	(0.07)	&	(0.089)	&	(0.071)	&	(0.051)	&	(0.082)	&	(0.471)	&	(0.227)	\\
%p-value theta	&		&	0.703	&	0.218	&	0.302	&		&	0.429	&	0.509	&	0.265	\\
AR 95\% CS	&	[-0.168, 0 .041]	&	[-0.193, 0.099] 	&	[-0.232, 0.008] 	&	[-0.208, 0.033] 	&	[-0.233,-0.026]	&	[-0.106, 0.339] 	&	$( -\infty, +\infty ) $	&	$( -\infty, 0.106] \cup [0.434, +\infty )$ 	\\
 \vspace{-3mm}\\
\multicolumn{9}{l}{\emph{First-stage results}}\\
Shift Share IV	&	1.174***	&	1.062***	&	1.268***	&	1.15***	&	0.647***	&	0.559**	&	0.166	&	0.291	\\
	&	(0.277)	&	(0.397)	&	(0.444)	&	(0.394)	&	 (0.155)	&	(0.244)	&	(0.206)	&	(0.207)	\\
%p-value pi	&		&	0.008	&	0.004	&	0.004	&		&	0.022	&	0.42	&	0.16	\\
F stat	&	19.09	&	7.016	&	8.002	&	8.353	&	16.39	&	5.139	&	0.639	&	1.941	\\
AR $\chi^2 $ stat	&	0.28	&	0.381	&	2.683	&	1.745	&	7.99***	&	0.78	&	3.514*	&	3.307*	\\
%p-value AR $\chi^2 $	&	0.595	&	0.537	&	0.101	&	0.186	&	0.005	&	0.377	&	0.061	&	0.069	\\
 \vspace{-3mm}\\
\multicolumn{9}{l}{\emph{Quality of learners}}\\
Model RMSE	&		&	0.434	&	0.403	&	0.424	&		&	0.44	&	0.819	&	0.739	\\
MSE of l	&		&	0.253	&	0.22	&	0.241	&		&	0.253	&	0.22	&	0.241	\\
MSE of r	&		&	0.739	&	0.851	&	0.811	&		&	1.172	&	1.345	&	1.248	\\
MSE of m	&		&	0.239	&	0.227	&	0.255	&		&	0.409	&	0.398	&	0.408	\\
Observations	&	327	&	327	&	327	&	327	&	327	&	327	&	327	&	327	\\
No. clusters	&	114	&	114	&	114	&	114	&	114	&	114	&	114	&	114	\\
%No. variables	&	11	&	250	&	20	&	20	&	11	&	250	&	20	&	20	\\

\hline
\end{tabular} 	
\footnotesize
\textbf{Note:} The table reports our estimates based on Specifications~(2) and (3) of Table~6 in \cite{moriconi2022} (our Panel A), and Specifications~(3) and (5) (Panels A and B) of Table~10 in \cite{moriconi2022} (our Panels B-C). 
%%%%%%%%%%%%%
The table displays our estimates from conventional 2SLS regression with FD transformation (Columns 1 and 5), and our panel IV DML estimation with different base learners (Columns 2-4 and 6-8).
%%%%%%%
The sample is aggregated sample at regional (NUTS2) level to construct an unbalanced panel data set. The treatment and instrumental variables in Columns (1)-(2) and (5)-(6) refer to the fraction of high-skilled workers, and in Columns (3)-(4) and (7)-(8) of low-skilled workers.
%%%%%%%%%%%%%%%%%%%%
Each panel uses a different depended variable. The raw control variables in all panels are: the share of women, average age, share of tertiary/post-tertiary education, average GDP per capita (in log), share of tertiary sector (in log), average unemployment rate, and year dummies. 
%%%%%%%%
Raw variables in the panel IV DML estimation with NNet and Boosting does not include interaction terms because these base learners are designed to capture nonlinearities in the data.
Panel IV DML with Lasso (Columns 2 and 6) uses an \emph{extended dictionary} of the raw variables, including polynomials up to order three and interaction terms between all the covariates, to satisfy Lasso's weak sparsity assumption. The number of covariates used for panel IV DML estimation doubles due to the inclusion of the lags of all included covariates, following the FD (exact) approach.
%%%%%
%Panel A: the number of observations ($NT$) is 147, and the number of groups is 114. Panels B-C: the number of observations ($NT$) is 327, and the number of groups is 114.
%%%%%%%
Panel IV DML technical note: 2 folds, cross-fitting, hyperparameters are tuned as per Table~\ref{tab:hyperpara}.
%%%%%%%%%
Standard errors in parenthesis are clustered at the regional level.
Significance levels: * p $<$ 0.10, ** p $<$ 0.05, \mbox{*** p $<$ 0.01.}
\end{threeparttable}
}
\end{table}
%x_ln_gdpcap x_tertiary_perc x_unemp_rate (tab6)

%The first-stage F-statistics obtained with panel IV DML for HS immigrants fall below the conventional threshold of 16.30, providing initial evidence of weak identification in these specifications. This reflects the fact that flexibly conditioning on a rich set of covariates—required to render the instrument plausibly exogenous and \emph{as good as randomly assigned}—substantially reduces its predictive power. Across all panels, the Anderson-Rubin (AR) tests from both 2SLS and panel IV DML fail to reject the null hypothesis of no reduced-form effect when no causal effect is present. The corresponding AR confidence sets are either bounded but include zero, or unbounded (the real line), indicating that the causal parameter cannot be precisely identified due to weak instruments. Both estimators therefore point to the absence of a causal effect, although only panel IV DML explicitly reveals the weak-identification problem. Absent AR diagnostics, conventional 2SLS would misleadingly suggest strong instruments and no effect, whereas the correct conclusion is that the instruments are weak and inference is uninformative.

For LS immigrants we comment on the results of the three outcomes (panels) separately. In Columns~5-8 of Panel A, the first-stage F-statistics are well below 10 for both 2SLS and panel IV DML, clearly indicating the presence of weak instruments. Although AR test statistics from panel IV DML suggest the possible presence of a second-stage effect across all three learners, not detected by 2SLS, the associated AR confidence sets span the entire real line. This reflects the lack of information in the instrument that prevents identification of the causal parameter. %In this case,  without AR diagnostics, the researcher might conclude from 2SLS alone that the instrument is weak and there is no second-stage effect, whereas panel IV DML suggests that weak identification precludes meaningful inference about the second-stage effect. 
In this case, without AR diagnostics, the researcher might draw the same conclusion from 2SLS and panel IV DML, i.e. that the instrument is weak and no inference on the second stage can be made. AR diagnostics can inform us if there is a second stage effect regardless of the strength of the instrument. In this case, only the AR diagnostics from panel IV DML allow us to conclude that there might be an effect, which cannot be identified due to the weakness of the instrument.

In Columns~5-8 of Panels~B-C, where the first-stage regressions are identical across outcomes, panel IV DML again produces substantially smaller F-statistics than 2SLS, whose values now only marginally exceed the \citet{stock2005}'s threshold. In Panel B, AR tests fail to reject the null hypothesis with both estimators (and only at 10\% level with neural network), indicating no evidence of a causal effect on political attitudes. In Panel~C on migration attitudes, results are mixed: AR tests from panel IV DML with Lasso fails to reject the null of no effect while neural networks and boosting reject at 10\% level only. In contrast, the result of the 2SLS AR test conveys the message that there is a strong second-stage effect. %However, the latter should be treated with caution given the small sample size and the known instability of boosting in this setting. 
Consistent with weak identification, AR confidence sets from panel IV DML either cover the entire real line (NNet) or disjoint segments (boosting), whereas those from 2SLS are bounded and exclude zero. In contrast, AR CS from Lasso is bounded but includes zero, which is aligned with the AR test result. In general, given the weakness of the shift-share instrument, statistical inference based on conventional 2SLS estimator is unreliable, and the second-stage coefficient cannot be interpreted as a credible causal effect.

Overall, panel IV DML reveals that once instrument exogeneity is strengthened through flexible control for confounding, in this case the shift-share instruments lose substantial relevance. As a result, the available variation is insufficient to support reliable inference on second-stage effects, remarking the importance of weak-identification diagnostics in IV applications.

\section{Monte Carlo Simulations}\label{sec:mcsimul}
For the Monte Carlo simulations, we consider a data generating process (DGP) inspired by the estimating equations of the  empirical applications discussed in Section~\ref{sec:empirical}. %with many controls but only subset of which are relevant and enter the model both linearly and through nonlinear interaction terms. This design allows us to investigate the performance of our panel IV DML estimator over the conventional 2SLS when the functional form of the covariates is flexible. 
The data are generated from a static panel data model with high-dimensional covariates and individual fixed effects, which are strongly correlated with the included variables. The outcome depends on an endogenous treatment; treatment endogeneity arises from two sources: (a) the correlation between the structural and first-stage error term, and (b) the presence of the fixed effect in both structural and treatment equations.\footnote{The first source of endogeneity can be addressed with an IV approach, while the second with a panel data estimation approach (i.e., FE, FD and correlated random effects).} The instrument is not randomly assigned, being affected by covariates, but is exogenous to the structural error. The nuisance functions linking controls to the outcome, treatment, and instrument are sparse (only few variables  out of thirty are relevant) and nonlinear (interactions are included). We consider both a strong-instrument design and a weak-instrument design. More details on the DGP are provided in the Online Appendix~\ref{sec:mc_dgp}.

As in the empirical applications, we employ our Panel IV DML estimator with different learners -- i.e., Lasso, a single-layer neural network (NNet), and gradient boosting with 100 trees (Boosting) -- to predict the nuissance functions of the covariates in a flexible way, and use the conventional 2SLS estimator as benchmark to assess the finite sample properties of the proposed estimator.
The set of confounding variables used in conventional 2SLS and panel IV DML with NNet and Boosting regressions includes all raw variables and no nonlinear terms because, in practice, the analyst is agnostic on the true functional form of the covariates.\footnote{Neural network and gradient boosting should be able to capture those nonlinearities, by construction, even if unspecified, unlike 2SLS.} By contrast, panel IV DML with Lasso requires the analyst to specify a rich set of variables, including polynomials and interactions of the raw covariates (extended dictionary) to satisfy the sparsity assumption.\footnote{The constructed extended dictionary does not include the interaction terms present in Equations~\eqref{eqn:l0}-\eqref{eqn:m0} of the DGP, allowing us to assess how effectively Lasso can flexibly recover similar covariate complexity.} The hyperparameters of the base learners are tuned via grid search \citep{bergstra2012} as explained in the Online Appendix~\ref{sec:tuning}. 

%The population of cross-sectional units from which each Monte Carlo subsample is drawn consists of $10,000$ units observed over a number of fixed $T=10$ periods, but the study is conducting by subsampling $N=\{100,500, 1000, 5000\}$ to compare finite-sample performance for small, medium and large sample sizes.
%We draw $100$ bootstrapped samples for each combination of cross-sectional sample size $N$ and estimation approach.

%We now discuss the Monte Carlo simulation results from estimating the model above via panel IV DML and conventional 2SLS with FD transformation approach.\footnote{Monte Carlo simulation results for panel IV DML estimated with the within-group transformation (or fixed effects) are reported in Tables~\ref{tab:mc_wg_strong} and~\ref{tab:mc_wg_weak} of the printed Appendix~\ref{sec:app_tabs}. The estimated target parameter is always upward biased regardless of the learner, and test statistics are always unable to detect weak instrumental variables. These results can be explained by the highly nonlinear nature of the nuisance functions (polynomials and interactions), which is not well-approximated by transformed variables.}  

Tables~\ref{tab:mc_strong} and~\ref{tab:mc_weak} report the average results across 100 bootstrap samples for the strong- and weak-instrument cases, respectively. Each table summarizes the averages (for numerical quantities) or proportions (for binary indicators) over all Monte Carlo replications. The metrics reported include: (a) the bias and root mean squared error (RMSE) of the estimated target parameter $\thetano_n$, together with the ratio of its estimated standard error to its Monte Carlo standard deviation (SE/SD); (b) the RMSE of the three nuisance functions; (c) the first-stage F-statistic and threshold rules for assessing instrument strength; and (d) the p-value of the Anderson-Rubin (AR) test statistic together with the corresponding type of AR confidence set.
Each panel displays the results by cross-sectional sample size, \mbox{$N=\{100,500, 1000, 5000\}$,} with $T=10$ fixed, where each row represents a different estimator. 

\begin{table}[t!]
\centering
 \caption{MC Results, FD Approach with Strong IV }\label{tab:mc_strong}
 \vspace{-3mm}
\scalebox{.55}{
\begin{threeparttable}	%use it for notes
\begin{tabular}{lcccccccccccccc}  
\vspace{-3mm}\\   
\hline\hline
\vspace{-3mm}\\
 &\multicolumn{3}{c}{Target parameter $\widehat{\theta}$}&\multicolumn{3}{c}{Nuisance parameters}&\multicolumn{3}{c}{First-stage F statistic}& AR $\chi^2$&\multicolumn{4}{c}{Anderson-Rubin Confidence Set}\\  
 \cmidrule(l{.35cm}r{.25cm}){2-4}\cmidrule(l{.35cm}r{.25cm}){5-7}\cmidrule(l{.35cm}r{.25cm}){8-10}\cmidrule(l{.35cm}r{.25cm}){11-11}\cmidrule(l{.35cm}r{.25cm}){12-15}
 &	Bias&	RMSE	&	SE/SD&	RMSE $l$	&RMSE $r$	&	RMSE $m$	&$F$	&	$F>16.3$	&	$F>104.7$	&	$p<0.05$	&	Bounded & Real Line& Disjoint& Includes 0	\\
% 	&(1)&(2)&(3)&(4)&(5)&(6)&(7)&(8)&(9)&(10)&(11)&(12)&(13)&(14)\\
\vspace{-2mm}\\
\hline 
\vspace{-3mm}\\
\multicolumn{14}{c}{\textbf{Panel A:} N=100, T=10}\\
\vspace{-3mm}\\										
2SLS	&	0.508	&	0.260	&	1.213	&	--&	--&	--&	747.6	&	1.00	&	1.00	&	1.00	&	1.00	&	0	&	0	&	0	\\
\multicolumn{12}{l}{\emph{Panel IV DML with:}}\\
Lasso	&	0.005	&	0.331	&	3.420	&	1.982	&	1.480	&	0.389	&	44.6	&	0.99	&	0	&	0.9	&	1.00	&	0	&	0	&	0.07	\\
NNet	&	0.097	&	0.132	&	1.711	&	2.296	&	1.727	&	0.424	&	33.8	&	0.87	&	0.01	&	0.74	&	1.00	&	0	&	0	&	0.23	\\
Boosting	&	0.448	&	0.209	&	1.232	&	2.415	&	1.903	&	0.661	&	136.3	&	1.00	&	0.68	&	1.00	&	1.00	&	0	&	0	&	0	\\

\vspace{-3mm}\\
\multicolumn{14}{c}{\textbf{Panel B:} N=500, T=10}\\																\vspace{-3mm}\\												
2SLS	&	0.505	&	0.255	&	1.112	&	--&	--&	--&	3760.9	&	1.00	&	1.00	&	1.00	&	1.00	&	0	&	0	&	0	\\
\multicolumn{12}{l}{\emph{Panel IV DML with:}}\\
Lasso	&	0.055	&	0.008	&	1.037	&	1.953	&	1.456	&	0.374	&	179.7	&	1.00	&	1.00	&	1.00	&	1.00	&	0	&	0	&	0	\\
NNet	&	0.062	&	0.013	&	1.244	&	2.021	&	1.529	&	0.387	&	158.8	&	0.99	&	0.96	&	1.00	&	1.00	&	0	&	0	&	0	\\
Boosting	&	0.280	&	0.084	&	1.330	&	2.125	&	1.625	&	0.491	&	243.1	&	1.00	&	1.00	&	1.00	&	1.00	&	0	&	0	&	0	\\

\vspace{-3mm}\\
\multicolumn{14}{c}{\textbf{Panel C:} N=1000, T=10}\\														\vspace{-3mm}\\										
2SLS	&	0.505	&	0.256	&	1.177	&	--&	--&	--&	7561.6	&	1.00	&	1.00	&	1.00	&	1.00	&	0	&	0	&	0	\\
\multicolumn{12}{l}{\emph{Panel IV DML with:}}\\
Lasso	&	0.048	&	0.005	&	1.072	&	1.932	&	1.443	&	0.372	&	352.5	&	1.00	&	1.00	&	1.00	&	1.00	&	0	&	0	&	0	\\
NNet	&	0.052	&	0.007	&	1.206	&	1.985	&	1.498	&	0.385	&	309.1	&	0.99	&	0.97	&	1.00	&	1.00	&	0	&	0	&	0	\\
Boosting	&	0.193	&	0.041	&	1.258	&	2.055	&	1.562	&	0.455	&	394.8	&	1.00	&	1.00	&	1.00	&	1.00	&	0	&	0	&	0	\\

\vspace{-3mm}\\
\multicolumn{14}{c}{\textbf{Panel D:} N=5000, T=10}\\														\vspace{-3mm}\\											
2SLS	&	0.505	&	0.255	&	0.835	&--&--&--&	37612.8	&	1.00	&	1.00	&	1.00	&	1.00	&	0	&	0	&	0	\\
\multicolumn{12}{l}{\emph{Panel IV DML with:}}\\
Lasso	&	0.065	&	0.004	&	0.697	&	1.880	&	1.406	&	0.367	&	1808.4	&	1.00	&	1.00	&	1.00	&	1.00	&	0	&	0	&	0	\\
NNet	&	0.073	&	0.008	&	1.979	&	1.954	&	1.462	&	0.393	&	1509.3	&	1.00	&	0.98	&	1.00	&	1.00	&	0	&	0	&	0	\\
Boosting	&	0.050	&	0.004	&	1.215	&	1.986	&	1.497	&	0.415	&	1390.3	&	1.00	&	1.00	&	1.00	&	1.00	&	0	&	0	&	0	\\

\hline
\end{tabular} 	
\begin{tablenotes}[para,flushleft]	
\textbf{Note:} The figures in the table are average values and frequencies over 100 bootstrapped replications by estimation method (conventional 2SLS and panel IV DML) and learner (Lasso, neural network, and gradient boosting). The true structural parameter $\theta$ is 0.50, and the true first-stage coefficient is $\pi=0.8$. Panel IV DML estimation details: cross-fitting with 3 folds.
\end{tablenotes}
\end{threeparttable}}
\end{table}

%\textbf{Strong IV.} 
We begin the discussion of the Monte Carlo results with Table~\ref{tab:mc_strong}, which corresponds to the strong-instrument design. In this setting, inference is reliable and, hence, the focus is primarily on the performance of the panel IV DML estimator for the target parameter.
Panel IV DML consistently outperforms conventional 2SLS across all sample sizes, reflecting its ability to capture nonlinearities in the covariates that 2SLS cannot accommodate, by construction. In particular, the panel IV DML estimator recovers the structural parameter with high accuracy and precision when using Lasso and neural networks: the bias remains small (below 0.10) even in small samples ($N=100$). Gradient boosting exhibits larger bias in small samples, likely due to overfitting when the training sample is small, but still performs slightly better than 2SLS even in small samples. As the sample size increases, its bias converges toward that of Lasso and neural networks. The RMSE of the panel IV DML estimator declines remarkably with sample size, being consistent with root-N convergence of the panel IV DML estimator, whereas the 2SLS estimator does not display the same improvement. Consistent with the design, first-stage F-statistics are large and frequently exceed the \citet{lee2022} threshold of 104.70, indicating strong identification. The main exceptions are for Lasso and neural network in very small samples ($N=100$), where the average F-statistics exceed the \citet{stock2005} critical value of 16.30 but remain below 104.70. 
This does not indicate genuine weak identification, but rather reflects the limited information available in small samples. The Anderson-Rubin (AR) diagnostics support this interpretation: both the AR test statistic and the AR confidence set (CS) provide evidence of the existence of a treatment effect under this DGP across estimators and sample sizes. This pattern mirrors our empirical reanalysis of \citet{tabellini2020}, where F-statistics fell similarly below 104.70, but the AR test statistic rejected the null hypothesis of no effect and the AR CS remained bounded.
Exceptions arise again for Lasso and neural networks in very small samples ($N=100$), where limited information implies that the absence of a second-stage effect cannot be fully ruled out.

%test statistic strongly rejects the joint null hypothesis of no effect, and the corresponding AR confidence sets are bounded and exclude zero for all but the smallest samples with Lasso and NNet, confirming the existence of a causal effect under the DGP.

\begin{table}[t!]
\centering
 \caption{MC Results, FD Approach with Weak IV}\label{tab:mc_weak}
 \vspace{-2mm}
\scalebox{.55}{
\begin{threeparttable}	%use it for notes
\begin{tabular}{lcccccccccccccc}  
\vspace{-3mm}\\   
\hline\hline
\vspace{-3mm}\\
 &\multicolumn{3}{c}{Target parameter $\widehat{\theta}$}&\multicolumn{3}{c}{Nuisance parameters}&\multicolumn{3}{c}{First-stage F statistic}& AR $\chi^2$&\multicolumn{4}{c}{Anderson-Rubin Confidence Set }\\  
 \cmidrule(l{.35cm}r{.25cm}){2-4}\cmidrule(l{.35cm}r{.25cm}){5-7}\cmidrule(l{.35cm}r{.25cm}){8-10}\cmidrule(l{.35cm}r{.25cm}){11-11}\cmidrule(l{.35cm}r{.25cm}){12-15}
 &	Bias&	RMSE	&	SE/SD&	RMSE $l$	&RMSE $r$	&	RMSE $m$	&$F$	&	$F>16.3$	&	$F>104.7$	&	$p<0.05$	&	Bounded & Real Line& Disjoint& Includes 0	\\
% 	&(1)&(2)&(3)&(4)&(5)&(6)&(7)&(8)&(9)&(10)&(11)&(12)&(13)&(14)\\
\vspace{-2mm}\\
\hline 
\vspace{-3mm}\\
\multicolumn{14}{c}{\textbf{Panel A:} N=100, T=10}\\																	\vspace{-3mm}\\										
2SLS	&	1.007	&	1.020	&	1.174	&	--&	--&	--&	192.9	&	1.00	&	0.99	&	1.00	&	1.00	&	0	&	0	&	0	\\
\multicolumn{12}{l}{\emph{Panel IV DML with:}}\\
Lasso	&	1.249	&	20.434	&	1.240	&	1.969	&	1.429	&	0.389	&	2.8	&	0.01	&	0	&	0.26	&	0.36	&	0.53	&	0.11	&	0.71	\\
NNet	&	0.495	&	12.131	&	0.378	&	2.304	&	1.666	&	0.424	&	2.1	&	0.01	&	0	&	0.2	&	0.25	&	0.6	&	0.15	&	0.78	\\
Boosting	&	1.136	&	2.094	&	4.201	&	2.322	&	1.682	&	0.661	&	37.1	&	0.93	&	0	&	1.00	&	1.00	&	0	&	0	&	0	\\

\vspace{-3mm}\\
\multicolumn{14}{c}{\textbf{Panel B:} N=500, T=10}\\																	\vspace{-3mm}\\											
2SLS	&	1.001	&	1.002	&	1.108	&	--&	--&	--&	958.5	&	1.00	&	1.00	&	1.00	&	1.00	&	0	&	0	&	0	\\
\multicolumn{12}{l}{\emph{Panel IV DML with:}}\\
Lasso	&	1.334	&	106.700	&	2.848	&	1.942	&	1.415	&	0.374	&	4.2	&	0	&	0	&	0.27	&	0.53	&	0.32	&	0.15	&	0.63	\\
NNet	&	0.417	&	147.068	&	2.659	&	2.004	&	1.457	&	0.387	&	3.9	&	0	&	0	&	0.31	&	0.46	&	0.4	&	0.14	&	0.64	\\
Boosting	&	1.364	&	7.888	&	8.154	&	2.086	&	1.519	&	0.491	&	26.5	&	0.82	&	0	&	1.00	&	1.00	&	0	&	0	&	0	\\

\vspace{-3mm}\\
\multicolumn{14}{c}{\textbf{Panel C:} N=1000, T=10}\\																	
\vspace{-3mm}\\
2SLS	&	1.001	&	1.003	&	1.082	&	--&	--&	--&	1929.5	&	1.00	&	1.00	&	1.00	&	1.00	&	0	&	0	&	0	\\
\multicolumn{12}{l}{\emph{Panel IV DML with:}}\\
Lasso	&	0.651	&	19.555	&	4.002	&	1.923	&	1.403	&	0.372	&	6.9	&	0.04	&	0	&	0.46	&	0.79	&	0.14	&	0.07	&	0.51	\\
NNet	&	0.408	&	23.187	&	2.721	&	1.972	&	1.435	&	0.385	&	7.1	&	0.08	&	0	&	0.46	&	0.7	&	0.2	&	0.1	&	0.48	\\
Boosting	&	0.792	&	1.078	&	3.339	&	2.025	&	1.473	&	0.455	&	27.8	&	0.84	&	0	&	1.00	&	1.00	&	0	&	0	&	0	\\

\vspace{-3mm}\\
\multicolumn{14}{c}{\textbf{Panel D:} N=5000, T=10}\\																
\vspace{-3mm}\\
2SLS	&	1.002	&	1.004	&	0.826	&	--&	--&	--&	9575.5	&	1.00	&	1.00	&	1.00	&	1.00	&	0	&	0	&	0	\\
\multicolumn{12}{l}{\emph{Panel IV DML with:}}\\
Lasso	&	0.506	&	0.333	&	1.932	&	1.869	&	1.361	&	0.367	&	34.3	&	1.00	&	0	&	1.00	&	1.00	&	0	&	0	&	0	\\
NNet	&	0.096	&	14.171	&	11.232	&	1.931	&	1.400	&	0.393	&	32.1	&	0.85	&	0.03	&	0.96	&	1.00	&	0	&	0	&	0.03	\\
Boosting	&	-5.328	&	3370.120	&	31.939	&	1.964	&	1.428	&	0.415	&	21.5	&	0.72	&	0	&	0.95	&	1.00	&	0	&	0	&	0.03	\\
\hline
\end{tabular} 	
\begin{tablenotes}[para,flushleft]	
\textbf{Note:} The figures in the table are average values and frequencies over 100 bootstrapped replications by estimation method (conventional 2SLS and panel IV DML) and learner (Lasso, neural network, and gradient boosting). The true structural parameter $\theta$ is 0.50, and the true first-stage coefficient is $\pi=0.001$. Panel IV DML estimation details: cross-fitting with 3 folds.
\end{tablenotes}
\end{threeparttable}}
\end{table}

%\textbf{Weak IV.} 
When the instrument is designed to be weak (Table~\ref{tab:mc_weak}), the second-stage estimates and statistical inference are unreliable for all estimators, so the Monte Carlo metrics for $\thetano_n$ (bias, RMSE, SE/SD) are largely uninformative.\footnote{As expected, both conventional 2SLS and panel IV DML produce biased estimates of the target parameter across all sample sizes. Surprisingly, neural network exhibits comparatively smaller average bias in large samples ($N=5,000$) though at the cost of reduced precision ($SE/SD\gg1$).} %The large bias and RMSE of $\thetano_n$ with gradient boosting in large samples are a consequence of weak identification rather than learner instability.}
The key consideration, therefore, is whether the weak-identification diagnostics correctly detect the lack of instrument relevance when using conventional 2SLS and panel IV DML estimators.  

Conventional 2SLS systematically fails to detect weak instruments. That is, the first-stage F-statistics always exceed 104.70, the AR test always rejects the null hypothesis of no effect at 5\% level in all samples, and AR 95\% CS are always bounded and exclude zero. These diagnostics incorrectly suggest strong identification and identification of the treatment effect, despite the DGP being designed with weak instruments. 
By contrast, panel IV DML with Lasso and NNet  correctly detects weak identification in most cases, especially in small samples. For these learners, the average first-stage F-statistic does not exceed the rule-of-thumb threshold of 10 for $N\le1000$, is  always below the \citet{stock2005}'s critical value of 16.30 in small samples, and never approaches the \citet{lee2022} threshold of 104.70 with exceptions in large samples ($N=5,000$). Boosting never reaches the \citet{lee2022}'s  threshold, although its F-statistics are systematically larger (around 25 on average) than the other learners.
The AR test further highlights the contrast between panel IV DML and conventional 2SLS. For Lasso and NNet, the AR rejection rates are low in small samples (on average around 30\% at $N\le500$ and 46\% at $N=1,000$), appropriately reflecting weak identification, and increase to about 95\% at $N=5,000$ since larger samples provide more information. The corresponding AR CS are often unbounded (either disjoint or real line) in small samples, signalling weak instruments. As $N$ increases, the AR CS become more often bounded, but still include zero, correctly indicating that the causal effect cannot be precisely identified under weak instruments. Boosting behaves similarly to 2SLS in small samples, but converges toward the neural-network patterns as the sample size grows.
In this regard, panel IV DML provides inference that is more reliable in finite samples than 2SLS, following the patterns observed in the empirical applications of \citet{moriconi2019,moriconi2022}, where the instrument was found weak by panel IV DML but not by 2SLS and the AR CS were mostly bounded with zero excluded. % Weak-identification diagnostics reinforce these findings.} 

In conclusion, the Monte Carlo simulations indicate that conventional 2SLS produces biased and inconsistent estimates under both weak and strong instruments, with the magnitude of the bias remaining similar across sample sizes and failing to exhibit root-N convergence. Conversely,  panel IV DML not only improves estimation and statistical inference of the treatment effect under strong identification, regardless of the sample size, but it is also able to provide more robust and informative inference when identification is weak and conventional 2SLS can be severely misleading.

\section{Conclusion}\label{sec:conclusion}
% This paper introduces novel DML estimation procedures for partially linear panel regression model with fixed effects when the treatment variable is endogenous and the functional form of the covariates possibly nonlinear. %, hence, an instrumental variable strategy is required for valid inference. We focus on the homogeneous case because this is a common empirical set up. 
% We show that the panel IV DML method is a powerful tool for the estimation of treatment effects, being able to capture the nonlinearities in the data and deliver inference that is more precise in finite samples and substantially more reliable than 2SLS. The availability of the panel IV DML toolkit allows researchers to complement traditional estimation techniques by providing robustness for their analyses.
% Moreover, we provide applied researchers with a flexible, theoretically grounded, and empirically practical tool for causal estimation in panel settings where traditional methods are increasingly strained.
This paper introduces novel DML estimation procedures for partially linear panel regression models with fixed effects under endogenous treatment and potentially nonlinear covariate effects. We show that panel IV DML is a powerful alternative to 2SLS, capturing nonlinearities in the data while delivering more precise and reliable finite-sample inference. The panel IV DML toolkit allows researchers to complement traditional estimation techniques, providing a flexible, theoretically grounded, and empirically practical approach to causal estimation in panel settings where conventional methods are falling short.

Another key contribution of this paper is the integration of weak-instrument diagnostics (i.e., the first-stage F-statistic and the Anderson-Rubin test and confidence sets) into the panel IV DML framework. Our results show that AR inference is essential for reliable conclusions, as it remains valid under weak identification and can uncover treatment effects even when the F-statistic suggests limited instrument strength. Therefore, we advocate reporting AR diagnostics alongside conventional measures in panel IV DML applications.

%A key contribution is the integration of weak-instrument diagnostics, i.e., the first-stage F-statistic and the Anderson-Rubin (AR) robust test statistic and confidence set, into the panel IV DML framework. Our empirical applications and Monte Carlo simulations demonstrate that reporting AR diagnostics alongside the F-statistic is essential for drawing reliable inferential conclusions in IV settings. In practice, which diagnostic a researcher relies on can determine the conclusions they draw. The F-statistic informs only about instrument strength, and when it signals weakness, a researcher relying on it alone would conclude that no inference on the treatment effect is possible. The AR test, by contrast, is robust to weak identification by construction: it can detect evidence of a second-stage effect regardless of instrument strength, providing inferential content precisely where the F-statistic offers none. This can determine whether a substantive conclusion can be reached or the analysis yields no conclusion. We, therefore, recommend the use of AR robust diagnostics as a standard complement to any IV-based analysis, not only within the panel IV DML framework but more broadly. Relying exclusively on the F-statistic risks discarding valid evidence of treatment effects, especially when identification is weak.

%%%%%%%%%%%%%%%%%%%%%%%%%%%
%%References 
\bibliographystyle{apalike}
\bibliography{biblio}
%%%%%%%%%%%%%%%%%%%%%%%%%%%
\appendix
\newpage
%\section*{Appendix}
\section{Proofs}
\renewcommand{\theequation}{A.\arabic{equation}}
\setcounter{equation}{0}

\subsection{Proof of Proposition~\ref{prop:score}} %and Derivation of the Panel IV DML Score Function}
\label{sec:proof_score}
The derivation of the Neyman orthogonal (NO) score function for IV panel is based on the following conditional moment restrictions
\begin{equation*}
    \E[\uit|\vector \zit,{\vector X_{it}}, \xi_i]=\E[\rit|\vector \zit,{\vector X_{it}},\xi_i]= \E[\uit^*|\vector \zit,{\vector X_{it}}, \xi_i]=0
\end{equation*}
where  $\uit^*= \Yit - l_0(\vector X_{it}) - \vector V_{it}' \bm{\delta} - \alpha_i$ is the reduced-form regression residual, which induce the conditional moment restrictions 
\begin{equation*}
    \E[{\widetilde U}_{it}|{\widetilde{\vector Z}}_{it},{\vector X_{it}},{\vector X}_{it-1}, \xi_i]=%
    \E[{\widetilde R}_{it}|{\widetilde{\vector Z}}_{it},{\vector X_{it}},{\vector X}_{it-1},\xi_i]=%
    \E[{\widetilde U}_{it}^*|\vector {\widetilde{\vector Z}}_{it},{\vector X_{it}},{\vector X}_{it-1}, \xi_i]=0
\end{equation*}
under Assumptions \ref{item:asm_feedback} and \ref{item:asm_nolags}.

From \citet[Lemma 2.6]{chernozhukov2018}, the semiparametrically efficient NO score for ${\vector b}_0=(\vector{\pi}_0,\theta_0)$ or ${\vector b}^{\rm{rf}}_0 = ({\vector\delta}_0,\vector{\pi}_0)$ based on the moment restrictions above is ${\psi}^{\perp}({\widetilde{\vector Y}}_i,{\widetilde{\vector D}}_i,{\widetilde{\matrix Z}}_i,{\matrix X}_i)={\vector\mu}(\widetilde{\vector Z}_i,{\vector X}_i){\vector r}_i$ where
\[
{\vector\mu}(\widetilde{\vector Z}_i,{\vector X}_i ) = A^\prime (\widetilde{\vector Z}_i,{\vector X}_i)\Omega^{-1}(\widetilde{\vector Z}_i,{\vector X}_i)-G(X)\Gamma(\widetilde{\vector Z}_i,{\vector X}_i)\Omega^{-1}(\widetilde{\vector Z}_i,{\vector X}_i)
\]
and the components of ${\vector\mu}(\widetilde{\vector Z}_i,{\vector X}_i )$ are given as follows:
%%%%%%%%%%%%%%%%%
\begin{align}
    &{\vector r}_{i}^\prime  =\pmat{ \widetilde{{\bm R}}_{i}^\prime & \widetilde{{\bm U}}_{i}^\prime {\rm ~~or~~ } \widetilde{{\bm U}}_{i}^{*\prime}} \label{eqn:r}\\
    &A({\widetilde V},X)  \equiv \E\big[ \partial_{\bm{\pi}, \theta {\rm ~or~} \bm{\delta}} \,\, {\vector r}_{i} | \vector {\widetilde V}_{i}, \vector X_{i}\big]
    = - \pmat{\widetilde{\vector V}_{i}' & {\vector 0}  \\ {\vector 0} & \widetilde{\vector V}_{i}'\bm{\pi}_0 \\ \hline {\vector 0} & \widetilde{\vector V}_{i}'} \label{eqn:A} \\
   &\Omega({\widetilde V},X)\equiv\E\left[
        \begin{array}{cc|c}
        \widetilde{{\vector R}}_{i}\widetilde{{\vector R}}_i'  &         \widetilde{{\vector R}}_{i}\widetilde{{\vector U}}_i' & \widetilde{{\vector R}}_{i}\widetilde{{\vector U}}_i^{*\prime}\\
        \widetilde{{\vector U}}_{i}\widetilde{{\vector R}}_i'  & \widetilde{{\vector U}}_{i}\widetilde{{\vector U}}_i^\prime  & \\ \hline
        \widetilde{{\vector U}}_i\widetilde{{\vector R}}_{i}^{*\prime} & & \widetilde{{\vector U}}_{i}^*\widetilde{{\vector U}}_i^{*\prime}
        \end{array}
        \,\Bigg|\, {\widetilde{\vector V}}_{i}, \vector X_{i}
        \right] 
        =\left(
        \begin{array}{cc|c}
        \Omega_{\pi\pi} & \Omega_{\pi\theta} & \Omega_{\pi\delta} \\
        \Omega_{\pi\theta}^\prime & \Omega_{\theta\theta} & \\ \hline
        \Omega_{\pi\delta}^\prime & & \Omega_{\delta\delta}
        \end{array} 
        \right) \\
        \label{eqn:omega}
        &\Gamma({\widetilde V},X)  \equiv \E\big[ \partial_{\bmeta} \,\, {\vector r}_{i} | \vector {\widetilde V}_{i}, \vector X_{i}\big] =
        \left( 
        \begin{array}{cc|c}
 \zero & -{\vector 1}_{T-1} & {\vector\pi}_0^\prime\bigotimes{\vector 1}_{T-1} \\
-{\vector 1}_{T-1} & {\vector 1}_{T-1}\theta_0 & {\matrix 0} \\ \hline
        & {\vector 0} & {\vector\delta}_0^\prime\bigotimes{\vector 1}_{T-1}
        \end{array} 
        \right), 
\end{align}
\noindent where ${\vector 0}$ indicates the conformable matrix or vector of zeros, and the within-matrix lines indicate (above and left) the entries associated with $\theta_0$ and the others (below and right) associated with  $\bm{\delta}_0$. The lower-diagonal element of~\eqref{eqn:A} follows under \eqref{eqn:yit2}-\eqref{eqn:dit2}: $\E^*[\widetilde{D}_{it}- \widetilde{r}_0(\vector X_{it})] = \E^*[{\widetilde{\vector V}}_{it}' \bm{\pi}_0 + \widetilde{r}_0(\vector X_{it}) + \widetilde{R}_{it} - {\widetilde{r}_0(\vector X}_{it})] = \widetilde{\vector V}_{it}' \bm{\pi}_0$, where $\E^*[.]=\E[ .\mid {\widetilde{\vector V}}_{it}, \vector X_{it-1}, \vector X_{it}]$. Finally,
\[
G(X) \equiv \E[ A^\prime (\widetilde{\vector Z}_i,{\vector X}_i)\Omega^{-1}(\widetilde{\vector Z}_i,{\vector X}_i) \Gamma(\widetilde{\vector Z}_i,{\vector X}_i)|{\matrix X}_i]={\bf 0}
\]
under the locally efficient condition that $\Omega(\widetilde{\vector Z}_i,{\vector X}_i)=\Omega({\vector X}_i)$ and because $\Gamma({\widetilde{\vector Z}}_i,{\vector X}_i)=\Gamma$ and $\E[ A({\widetilde{\vector Z}}_i,{\vector X}_i)|{\matrix X}_i]={\bf 0}$ under both models. The final step in the derivation of (\ref{eqn:prop_score1}) (and that in the footnote for it) is the locally efficient condition that $\E[\widetilde{\vector U}_i{\widetilde{\vector R}}_i^\prime\mid {\matrix X}_i]=\E[\widetilde{\vector U}_i^*{\widetilde{\vector R}}_i^\prime\mid{\matrix X}_i]={\bf 0}$.

It then remains to verify that the existence of finite moments for the above expressions to be valid as per \citet[Lemma 2.6]{chernozhukov2018}:\ these conditions are that $\E[\|\Gamma(R)\|^4]$, $\E[\|A(R)\|^4]$, $\E[\|G(X)\|^4]$ and $\E[\|\Omega(R)\|^{-2}]$ are finite (noting $R={\widetilde{\matrix Z}}_i,{\matrix X}_i$).  Clearly, these conditions are satisfied by $\Gamma$ and $G$ if $\|\theta_0\|$, $\|{\vector\pi}_0\|$ and $\|{\vector\delta}_0\|=O(1)$ (see proof of Proposition \ref{prop:dml} below), and also by $\Omega$ if $\Omega_{\theta\theta}$ (or $\Omega_{\delta\delta}$) and $\Omega_{\pi\pi}$ have positive and finite singular values (implying both are positive-definite and non-singular, see also the Proposition \ref{prop:dml} proof).  For the structural model, $\|A(R)\|=\| {\widetilde{\matrix V}}^\prime_i {\vector\pi}_0\|+\|{\widetilde{\matrix V}}_i\|\le (C+1)\|{\widetilde{\matrix Z}}_i-{\widetilde{\matrix M}}_{0i}\| \le (C+1)\big(\|{\widetilde{\matrix Z}_i}\|+\|{\widetilde{\matrix M}}_{0i}\|\big) \le 2(C+1)\|{\widetilde{\matrix Z}_i}\|$ first because ${\vector{\pi}}=O(1)$ (see Proposition \ref{prop:dml} proof) and then because Jensen's inequality gives $\|{\widetilde{\matrix Z}_i}\|\geq \|{\widetilde{\matrix M}}_{0i}\|$. Similar arguments for the reduced-form model give $\|A(R)\|=\| {\widetilde{\matrix V}}_i\|+\|{\widetilde{\matrix V}}_i\|\le 2\|{\widetilde{\matrix Z}}_i-{\widetilde{\matrix M}}_{0i}\| \le 2\big(\|{\widetilde{\matrix Z}_i}\|+\|{\widetilde{\matrix M}}_{0i}\|\big) \le 4\|{\widetilde{\matrix Z}_i}\|$

Hence, all of the necessary conditions are satisfied. 

\subsection{Proof of Proposition \ref{prop:dml}}%: The Panel IV DML Estimator}
\label{sec:dml}

\noindent \textbf{Preliminaries.}  For vectors $v=(v_1,\ldots,v_{r+1})^\prime$, we use the vector norm $\|v\|_q=\left(\sum_{i=1}^{r+1}|v_i|^q\right)^{1/q}$ where the norm satisfies the H\"older inequality such that $\|ab\|_s\le\|a\|_p\|b\|_q$ for $1⁄p+1⁄q = 1/s$ for any conformable $a$ and $b$ whose product is a vector.
For real square matrices $M=(m_1,\ldots,m_p)$, where $m_k=(m_{1k},\ldots,m_{rk})'$ is column $k$ of $M$, we use the Schatten norm $\|M\|_q=\left(\sum_{k=1}^p|\sigma_k(M)|^q\right)^{1/q}$, that is, the vector norm of the singular values of $M$ $\sigma_1(M),\ldots,\sigma_p(M)$.  The H\"older inequality for conformable real matrices $A$ and $B$ is $\|AB\|_2^2\le\|A\|_p \|B\|_q$ for $1⁄p+1⁄q=1$.  
Generalising to $L_q(P)$ for non-counting measure $P\in P$, the norm of measurable $f(W)$ for random vector $W$ is $\|f(W)\|_{P,q}=\left(\int|f(w)|^q \partial P(w) \right)^{1/q}$, and for vector functional $f(W)=(f_1 (W),\ldots,f_l (W))$ it is $\|f(W)\|_{P,q}=\max_k(\|f_k (W)\|_{P,q})$, and H\"older’s inequality is $\|f_i (W) f_k (W)\|_{P,s}\le\|f_i (W)\|_{P,p} \|f_k (W)\|_{P,q}$ if $1⁄p+1⁄q=1/s$.
Regularity Condition~\ref{item:e} sets ${\cal T}_N$ to be defined by the following conditions:
\begin{itemize}
    \item	$\|\eta-\eta_0\|_{P,q}\le C$ for $q>4$,
	\item $\|\eta-\eta_0\|_{P,2}\le\delta_N$,
	\item $\|\widetilde{M}-\widetilde{M}_0\|_{P,2}\left(\|\widetilde{l}-\widetilde{l}_0\|_{P,2}+\|\widetilde{r}-\widetilde{r}_0\|_{P,2}+\|\widetilde{M}-\widetilde{M}_0\|_{P,2}\right)\le\delta_N N^{-1/2}$,
\end{itemize}
Crudely put, these conditions are that, overall, $\cal{T}_N$ can shrink towards $\eta_0$ in MSE at a rate slower than $\sqrt{N}$, but $\widetilde{M}\widetilde{l}$, $\widetilde{M}\widetilde{r}$ and $\widetilde{M}\widetilde{M}'$ must converge at a rate at least as fast as $\sqrt{N}$.  Note that throughout the labelling of bounding constants is informal but this does not affect the conclusions because these are arbitrary.  

Throughout, note that we use italicised {\em Theorem}, {\em Assumption}~etc. to refer to the associated results in \citet{chernozhukov2018}.  The justification follows the same outline as their proof of {\em Theorem~4.2} and, as such, ignores contributions from the residual moments $\Omega$ for $\psi(W;\theta_0,\pi_0,\eta)$ defined in Proposition \ref{prop:score}.
\\
\\
\noindent \textbf{Step 1: Verify {\em Assumptions~3.1(a)-3.1(c)}.} The locally efficient linear score from Proposition \ref{prop:dml} can be written

\begin{equation}
\psi(W; \theta_0, \pi_0, \eta)= -
\pmat{v_0^{*} (\widetilde{D} - \widetilde{r}) & 0 \\
0_r' & \widetilde{V}' \widetilde{V}}
\pmat{\theta_0 \\\pi_0} + \pmat{v_0^{*} (\widetilde{Y} - \widetilde{l}) \\
\widetilde{V}' (\widetilde{D} - \widetilde{r})}
\equiv 
\psi^{a}(W; \eta)
\pmat{\theta_0 \\ \pi_0} +  \psi^{b}(W; \eta),
\label{eqn:lin_score}
\end{equation}
%$\psi(W;\theta_0,\pi_0,\eta)=-\begin{matrix} v_0^*(\widetilde{D} -\tild{r}) & 0 ̃)&0_r^'@0_r&V ̃'V ̃ ))(■(θ_0@π_0 ))\end{matrix}+(■(v_0^* (Y ̃-l ̃ )@V ̃'(D ̃-r ̃ ) ))≡\psi^a (W;η)(■(θ_0@π_0 ))+\psi^b (W;η),     (A.1)
where row vector $v_0^*=\pi_0^\prime\widetilde{V}^\prime=\pi_0(\widetilde{Z}-\widetilde{M})^\prime$ is the combined effect of the $r$ instrumental variables on $\widetilde{D}$.  This is linear, satisfies $\E_P\left[ \psi(W;\theta_0,\pi_0,\eta_0)\right]=0$ and, if it exists, is twice continuously Gateaux differentiable.
\\
\\
\textbf{Step 2: Verify {\em Assumption~3.1(d)}.} That Neyman orthogonality of~\eqref{eqn:r} with $\lambda_N=0$ follows from {\em Lemma 2.6} has already been verified by Proposition \ref{prop:score}.
\\
\\
\textbf{Step 3: Verify {\em Assumptions~3.1(e)} and {\em 3.2(d)}}. 
\\ \noindent First, consider decomposing the norm of $\psi_a^0=\psi^a(W;\eta_0)$ in terms of the singular values of $\E_P\left(\psi_0^A(W;\eta_0)\right)$:
\[
\|\E_P \left(\psi_0^a\right)\|=\Big|\E_P\left(\pi_0^\prime(\widetilde{Z}-\widetilde{M}_0)(\widetilde{D}-\widetilde{r}_0)\right)\Big|+\sum_{k=1}^r\sigma_k\left(\E_P(\widetilde{V}_0^\prime\widetilde{V}_0)\right),
\]
where $\sigma_k\left(\E_P(\widetilde{V}_0^\prime\widetilde{V}_0)\right)$ is singular value $k$ of $\E_P\left[\psi_0^A(W;\eta_0)\right]$.  From Regularity Condition~\ref{item:c}, all $r+1$ components are greater than some positive constant and so $\|\E_P(\psi^a)\|\geq c$ follows.  Similarly,
\[
\|\E_P(\psi^a)\|\le \big|\E_P\{\pi_0^\prime(\widetilde{Z}-\widetilde{M}_0)(\widetilde{D}-\widetilde{r}_0)\}\big|+r\big|\rho\left(\E_P(\widetilde{V}_0^\prime\widetilde{V}_0)\right)\big|\le (1+r)C,
\]
where $\rho\Big(\E_P(\widetilde{V}_0^\prime\widetilde{V}_0)\Big)=\max_k\sigma_k \left(\E_P(\widetilde{V}_0^\prime\widetilde{V}_0)\right)$ is the maximum singular value of $\E_P(\widetilde{V}_0^\prime\widetilde{V}_0)$.  Regularity Condition (c) also ensures that $\|\E_P(\psi^a)\|$ is bounded above, as required.

Second, because  $\E_P(\psi_0)=0$ it follows under (\ref{eqn:lin_score}) that $\E_P(\psi_0^b)=\E_P (\psi_0^a)b_0$, where $b_0=(\theta_0,\pi_0)'$, and so $\psi_0^b=\psi_0^a b_0+\epsilon$, where $b_0=\E_P^{-1} (\psi_0^a)\E_P(\psi_0^b)$ is the coefficient of the linear projection of $\psi_0^a$ onto $\psi_0^b$ for the just-identified case, so $\E_P(\epsilon)=\E_P(\psi_0^a\epsilon)=0$.  Then, using the singular value decomposition,
\[
\E_P(\psi_0^b\psi_0^{b\prime})=\E_P\big(\psi_0^a b_0 b_0^\prime \psi_0^a+\epsilon\epsilon^\prime\big)=\E_P\big(A\Sigma A^\prime b_0 b_0^\prime A^\prime\Sigma A+\epsilon\epsilon^\prime\big),
\]
where $\Sigma$ is a diagonal matrix with diagonal elements the singular values of normal matrix $\psi_0^a$, and $A$ is a unitary matrix satisfying $A^\prime A=AA^\prime=I$ so that $A^\prime b_0 b_0^\prime A=b_0^\prime b_0 I$.  Furthermore,
\[
\E_P \big[A\Sigma A^\prime b_0 b_0^\prime A\Sigma A^\prime\big]=\E_P\big[A(b_0^\prime b_0 \Sigma^2)A^\prime\big],
\]
which shows $E_P(\psi_0^b\psi_0^{b\prime})$ to be a matrix with all-positive singular values because $b_0^\prime b_0>0$.  Finally,
\[
\E_P(\psi_0\psi_0^\prime)=2\E_P (\psi_0^a b_0 b_0^\prime \psi_0^a)+\E_P (\epsilon\epsilon^\prime)
\]
has positive singular values because it is the sum of two matrices with positive singular values; noting that positive-definite $\E_P (\epsilon\epsilon^\prime)={\rm Var}_P(\epsilon)$ has positive eigenvalues. %\red{can we take the positive-definite residual variance-covariance matrix for granted, or state it as part of Regularity condition (c)?  Positive definite => positive singular values}.
\\
\\
\textbf{Step 4: {\em Assumption~3.2(a)}.} That the following results hold with the following limits lying in ${\cal T}_N$ with probability at least $1-\Delta_N$ is asserted to be true by Regularity Condition~\ref{item:e}.
\\
\\
\textbf{Step 5: Verify {\em Assumption~3.2(b)} that $m_N^* = \sup \big(\E_P(\|\psi^a\|^q)^{1/q}\big) \le c_2$.} From the block-diagonal structure of $\psi^a$,  $\|\psi^a\| = |v_0^*(\widetilde{D}-\widetilde{r})| + {\rm Tr}(\widetilde{V}^\prime \widetilde{V})$.  Now we follow {\em Theorem~4.2} by specifying $q/2$ and requiring that $q > 4$. Then applying the Minkowski inequality (for the sum of two scalars) gives
\[
\|\psi^a\| = |v_0^*(\widetilde{D}-\widetilde{r})| + {\rm Tr}(\widetilde{V}^\prime \widetilde{V}).
\]
Now we follow {\em Theorem~4.2} by specifying $q/2$ and requiring that $q > 4$. Then, applying the Minkowski inequality (for the sum of two scalars) gives
\[
\E_P \left(\|\psi^a\|^{q/2}\right)^{2/q} = \big\||v_0^*(\widetilde{D}-\widetilde{r})| + {\rm Tr}(\widetilde{V}^\prime \widetilde{V})\big\|_{P,q/2} \le \|v_0^*(\widetilde{D}-\widetilde{r})\|_{P,q/2} + \|{\rm Tr}(\widetilde{V}^\prime \widetilde{V})\|_{P,q/2},
\]

\noindent and a further application of Minkowski gives
\[
\|{\rm Tr}(\widetilde{V}^\prime \widetilde{V})\|_{P,q/2}
= \Big\|\sum_{k=1}^r \sigma_k(\widetilde{V}^\prime \widetilde{V})\Big\|_{P,q/2}
\le \sum_{k=1}^r \|\sigma_k(\widetilde{V}^\prime \widetilde{V})\|_{P,q/2}
\le r \|\rho(\widetilde{V}^\prime \widetilde{V})\|_{P,q/2}
= r \|\rho(\widetilde{V})\|_{P,q}^2,
\]
where $\rho(\widetilde{V}) = \max \sigma_k(\widetilde{V})$ is the maximum singular value of $\widetilde{V}$ and, being a linear combination of all $r$ mean-centred instrumental variables, respects the assumptions satisfied by the other instrumental variables so that $\rho(\widetilde{V}) = \widetilde{Z}_\rho - m^\rho(X)$ and $\rho_0(\widetilde{V}) = \widetilde{Z}_\rho - m^\rho_0$ is its true mean-centred value.  Successive applications of Minkowski’s inequality to $\|\rho(\widetilde{V})\|_{P,q}$ gives
\[
\|\rho(\widetilde{V})\|_{P,q}=\|\widetilde{Z}_\rho-\widetilde{m}_0^\rho-(\widetilde{m}^\rho-\widetilde{m}_0^\rho)\|_{P,q}\le\|\rho_0 (\widetilde{V})\|_{P,q}+\|\widetilde{m}^\rho-\widetilde{m}_0^\rho\|_{P,q}\le\|\widetilde{Z}_\rho\|_{P,q}+\|\widetilde{m}_0^\rho\|_{P,q}+\|\widetilde{m}^\rho-\widetilde{m}_0^\rho\|_{P,q},
\]
and Jensen’s inequality gives $\|\widetilde{m}_0^\rho\|_{P,q}\le\|\widetilde{Z}_\rho\|_{P,q}$ so that
\[
\|\widetilde{Z}_\rho\|_{P,q}+\|\widetilde{m}_0^\rho\|_{P,q}+\|\widetilde{m}_0^\rho-\widetilde{m}_0^\rho\|_{P,q}\le2\|\widetilde{Z}_\rho\|_{P,q}+\|\widetilde{m}_\rho-\widetilde{m}_0^\rho\|_{P,q}\le2C+C.
\]

\noindent Before moving on to the second component, note that $\pi_0$ and $\theta_0$ can be bounded empirically as follows:
\[
\|\pi_0\|=\|\widetilde{V}_0^\prime(\widetilde{D}-\widetilde{r}_0)\|_{P,1}\|\widetilde{V}_0^\prime\widetilde{V}_0\|_{P,1} \le r^{-1} c_0^{-1} \|\widetilde{V}_0^\prime(\widetilde{D}-\widetilde{r}_0)\|_{P,1},
\]
because from Regularity Condition~\ref{item:c} it follows that $\|\widetilde{V}_0^\prime\widetilde{V}_0\|_{P,1}\geq rc_0$.  Then H\"older’s inequality followed by Minkowski’s and then Jensen’s inequalities further gives
\begin{align}
\|\pi_0\|\le r^{-1} c_0^{-1} \|\widetilde{V}_0\|_{P,2} \|\widetilde{D} -\widetilde{r}_0\|_{P,2}&
\le r^{-1} c_0^{-1} \left(\|\widetilde{Z}\|_{P,2}+\|\widetilde{M}_0\|_{P,2}\right)\left(\|\widetilde{D}\|_{P,2}+\|\widetilde{r}_0\|_{P,2}\right) \notag \\ 
&\le r^{-1} c_0^{-1}\left(\|\widetilde{Z}\|_{P,2}+\|\widetilde{Z}\|_{P,2}\right)\left(\|\widetilde{D}\|_{P,2}+\|\widetilde{D}\|_{P,2}\right)\notag \\
&\le 4r^{-1} c_0^{-1} C^2 \equiv \alpha. \notag
\end{align}
with the final inequality following from Regularity Condition~\ref{item:b} that $\|\widetilde{Z}\|_{P,2}$ and $\|\widetilde{D}\|_{P,2}\le C$. Similarly, $\theta_0=\E_P \big[v_0^* (\widetilde{Y}-\widetilde{l}_0)\big]⁄\E_P \big[v_0^* (\widetilde{D}-\widetilde{r}_0)\big]$ and so applying Minkowski’s inequality
\[
|\theta_0|=\|v_0^*(\widetilde{Y}-\widetilde{l}_0)\|_{P,1}/\|v_0^*(\widetilde{D}-\widetilde{r}_0)\|_{P,1} \le c_0^{-1}\|\pi_0\| \, \|\widetilde{Z}-\widetilde{M}_0\|_{P,2} \|\widetilde{Y}-\widetilde{l}_0\|_{P,2}\le r\alpha^2,
\]
with the first inequality following from the lower bound on $\pi_0$, the second from H\"older’s inequality combined with the norm property $\|\pi_0\|_2\le\|\pi_0\|$, and the final one from Regularity Condition~\ref{item:b}. Now returning to the second component $|v_0^*(\widetilde{D}-\widetilde{r})|$, successively applying Minkowski’s, H\"older’s, Minkowski’s and Jensen’s inequalities gives
\begin{align}
\big[\E_P(|v_0^*(\widetilde{D}-\widetilde{r})|^{q/2})\big]^{2/q}&=\|v_0^* (\widetilde{D} - \widetilde{r}_0)-v_0^*(\widetilde{r}-\widetilde{r}_0)\|_{P,q/2}\notag\\
&\le |v_0^*(\widetilde{D}-\widetilde{r}_0)\|_{P,q/2}+\|v_0^*(\widetilde{r}-\widetilde{r}_0)\|_{P,q/2}\notag\\
&\le \|\pi_0\|_q \|\widetilde{Z}-\widetilde{M}_0-(\widetilde{M}-\widetilde{M}_0)\|_{P,q} \left(\|\widetilde{D}-\widetilde{r}_0\|_{P,q}+\|\widetilde{r}-\widetilde{r}_0\|_{P,q}\right)\notag\\
&\le\|\pi_0\|_q (\|\widetilde{Z}\|_{P,q}+\|\widetilde{M}_0\|_{P,q}+\|\widetilde{M}_0\|_{P,q})\left(\|\widetilde{D}\|_{P,q}+\|\widetilde{r}_0\|_{P,q}+\|\widetilde{r}-\widetilde{r}_0\|_{P,q}\right)\notag\\
&\le\|\pi_0\|_q (2\|\widetilde{Z}\|_{P,q}+\|\widetilde{M}-\widetilde{M}_0\|_{P,q})\left( 2\|\widetilde{D}\|_{P,q}+\|\widetilde{r}-\widetilde{r}_0\|_{P,q}\right)\notag\\
& \le \alpha(3C)(3C)\notag\\
&=\alpha\beta^2, \notag
\end{align}
where the final inequality follows from Regularity Conditions~\ref{item:c} and~\ref{item:e} together with Jensen’s inequality to give $\|\widetilde{Z}\|_{P,q}+\|\widetilde{M}_0\|_{P,q}\le 2\|\widetilde{Z}\|_{P,q}\le 2C$ and $\|\widetilde{D}\|_{P,q}+\|\widetilde{r}_0\|_{P,q}\le 2\|\widetilde{D}\|_{P,q}\le 2C$ and $3C\equiv \beta$.  Therefore, because $r$ is fixed,
\[
\left(\E_P\|\Psi^a\|^{q/2}\right)^{2/q}\le\beta^2(r+\alpha)<c_2,
\]
noting that both $\alpha$ and $\beta^2$ are $O(C^2)$.  Now we show that $m_N = \sup \left(\E_P\|\psi\|^q\right)^{1/q} \le c$, where
\[
\psi 
= \pmat{v_0^* \widetilde{U} \\
\widetilde{V}^\prime \widetilde{R}}
= \pmat{v_0^* \widetilde{U}_0 \\
\widetilde{V}^\prime \widetilde{R}_0}
-
\pmat{v_0^* \big(\widetilde{l} - \widetilde{l}_0 + (\widetilde{r} - \widetilde{r}_0)\theta_0\big) \\
\widetilde{V}^\prime \big(\widetilde{r} - \widetilde{r}_0 + (\widetilde{V} - \widetilde{V}_0)\pi_0\big)}
\equiv \psi_0 - \psi_1.
\]
\noindent $\widetilde{U}_0=\widetilde{Y}-\widetilde{l}_0-(\widetilde{D}-\widetilde{r}_0)\theta_0$, $\widetilde{R}_0=\widetilde{D}-\widetilde{r}_0-\widetilde{V}_0\pi_0$, and $\widetilde{V}=\widetilde{Z}-\widetilde{M}_0$ are the true model residuals, and $\widetilde{U}=\widetilde{Y}-\widetilde{l}_0-(\widetilde{D}-\widetilde{r}_0)\theta_0$, $\widetilde{R}=\widetilde{D}-\widetilde{r}-\widetilde{V}\pi_0$ and $\widetilde{V}=\widetilde{Z}-\widetilde{M}$.  Applying the Minkowski inequality to
\[
(\E_P\left(\|\psi\|^{q/2}\right)^{2/q}=\|\psi\|_{P,q/2}\le\|\psi_0\|_{P,q/2}+\|\psi_1\|_{P,q/2},
\]
and then Minkowski’s followed by H\"older’s inequalities followed by Regularity Condition~\ref{item:c} to
\begin{align}
\|\psi_0\|_{P,q/2}=\left\{\E_P \left(|v_0^*\widetilde{U} _0|^{q/2}+\sum_{k=1}^r|\widetilde{v}_k^\prime\widetilde{R}_0|^{q/2}\right)\right\}^{2/q}&\le\|v_0^*\widetilde{U}_0\|_{P,q/2}+r\|\breve{v}^\prime\widetilde{R}_0\|_{P,q/2}\notag\\
&\le\|\pi_0\|_q\|\widetilde{U}_0\|_{P,q}+r\|\breve{v}\|_{P,q}\|\widetilde{R}_0\|_{P,q}\notag\\
&=\alpha\beta\|\widetilde{U}_0\|_{P,q}+r\beta\|\widetilde{R}_0\|_{P,q}, \notag
\end{align}
where $\|\pi_0\|_q\le\alpha$,  $\breve{v} =\arg \max_{\widetilde{v}_k}\|\widetilde{v}_k\|_{P,q}$, $\|\breve{V}\|_{P,q}\le\beta$ and $\|\widetilde{v}_0^*\|_{P,q}\le\|\pi_0\|\|\widetilde{Z}-\widetilde{M}_0-(\widetilde{M}-\widetilde{M}_0)\|_{P,q}\le\alpha\beta$.  Successive applications of these inequalities along the lines above further gives
\begin{align*}
\|\psi_1\|_{P,q/2}&\le\|v_0^* (\widetilde{l}-\widetilde{l}_0+(\widetilde{r}-\widetilde{r}_0)\theta_0)\|_{P,q/2}+r\|\breve{v}^\prime (\widetilde{r}-\widetilde{r}_0)+(\widetilde{V}-\widetilde{V}_0)\pi_0\|_{P,q/2}\\
&\le\alpha\beta\|\widetilde{l}-\widetilde{l}_0\|_{P,q} +\alpha\|\widetilde{r}-\widetilde{r}_0\|_{P,q} r\alpha^2+r\beta\|\widetilde{r}-\widetilde{r}_0\|_{P,q}+r\beta\|\widetilde{r}-\widetilde{r}_0\|_{P,q}+r\beta\|\widetilde{V}-\widetilde{V}_0\|_{P,q}\alpha\\
&\le C\alpha\beta+rC\alpha^3+2rC\beta+rC\alpha\beta,
\end{align*}
because $\|\theta_0\|_{P,q}\le r^{-1}\alpha^2$ so that
\begin{equation*}
\|\psi\|_{P,q/2}\le\alpha\beta\|\widetilde{U}_0\|_{P,q}+r\beta\|\widetilde{R}_0\|_{P,q}+C\alpha\beta+rC\alpha^3+2rC\beta+rC\alpha\beta.
\end{equation*}
Finally,
\[
\|\widetilde{U}_0\|_{P,q}\le\|\widetilde{Y}-\widetilde{l}_0\|_{P,q}+\|(\widetilde{D}-\widetilde{r}_0)\theta_0\|_{P,q}\le2C+2C\|\theta_0\|_q\le2C(1+r\alpha^2),
 \]
and
\[
\|\widetilde{R}_0\|_{P,q}=\|\widetilde{D}-\widetilde{r}_0\|_{P,q}+\|(\widetilde{Z}-\widetilde{M}_0)\pi_0\|_{P,q}\le 2C+2C\|\pi_0\|_q\le2C(1+\alpha).
\]
Therefore,
\begin{align*}
\|\psi\|_{P,q/2}&\le\alpha\beta 2C(1+r\alpha^2)+r\beta2C(1+\alpha)+C\alpha\beta+rC\alpha^3+2rC\beta+rC\alpha\beta\\
&=\alpha\beta2C(1+r\alpha^2 )+r\beta2C(1+\alpha)+C\alpha\beta+rC\alpha^3+2rC\beta+rC\alpha\beta\\
&=C\beta(2\alpha+2r\alpha^3+2r(1+\alpha)+(1+r)\alpha+(r\alpha^3)⁄\beta+2r)<c_2.
\end{align*}
This implies that $c_2=O(C^7)$ because $\alpha^3⁄\beta=O(C^5)$.
\\
\\
\noindent \textbf{Step 6: Verify {\em Assumption~3.2(c)}.}
Now we show that $\|\E_P(\Psi^a)-\E_P (\Psi_0^a)\|$ converges at rate $\delta_N$, where
\begin{align}
\psi(W; \theta_0, \pi_0, \eta) &= 
-\pmat{v_0^*(\widetilde{D} - \widetilde{r}) & 0_{\,\widetilde{}} & 0_r' \\
0_r & \widetilde{V}'\widetilde{V}}
\pmat{\theta_0 \\ \pi_0} +
\pmat{v_0^*(\widetilde{Y} - \widetilde{l}) \\
\widetilde{V}'(\widetilde{D} - \widetilde{r})} \nonumber \\
&\equiv 
\psi^a(W; \eta)
\pmat{\theta_0 \\\pi_0}
+ 
\psi^b(W; \eta) \nonumber,
\end{align}
%\[
%\|\E_P (\Psi^a)-\E_P(\Psi_0^a)\|=\|(■(E_P [v_0^* (D ̃-r ̃ )]&0'@0&E_p [(Z ̃-M ̃ )'(Z ̃-M ̃ )] ))-(■(E_P [π_0^' (Z ̃-M ̃_0 )'(D ̃-r ̃ )]&0'@0&E_p [(Z ̃-M ̃_0 )'(Z ̃-M ̃_0 )] ))‖,
%\]
\noindent and, taking the top-left and bottom-right diagonals in turn,
\begin{align*}
\big\|\E_P &\big[v_0^*(\widetilde{D}-\widetilde{r})\big]-\E_P \big[\pi_0^\prime(\widetilde{Z}-\widetilde{M}_0)^\prime(\widetilde{D}-\widetilde{r})\big]\big\|\\
&=\|\pi_0^\prime(\widetilde{M}-\widetilde{M}_0)^\prime(\widetilde{D}-{\widetilde r}_0)+\pi_0^\prime (\widetilde{Z}-\widetilde{M}_0)^\prime(\widetilde{r}-\widetilde{r}_0)+\pi_0^\prime (\widetilde{M}-\widetilde{M}_0)^\prime(\widetilde{r}-\widetilde{r}_0)\|_{P,1}\\
&\le\|\pi_0^\prime (\widetilde{M}-\widetilde{M}_0)^\prime(\widetilde{R}-\widetilde{r}_0)\|_{P,1}+\|\pi_0^\prime (\widetilde{Z}-\widetilde{M}_0)^\prime(\widetilde{r}-\widetilde{r}_0)\|_{P,1}+\|\pi_0^\prime (\widetilde{M}-\widetilde{M}_0)^\prime(\widetilde{r}-\widetilde{r}_0)\|_{P,1}\\
&\le|\pi_0\|_2 (\|\widetilde{M}_0\|_{P,2} \|\widetilde{D}-\widetilde{r}_0\|_{P,2}+\|\widetilde{Z}-\widetilde{M}_0\|_{P,2} |\widetilde{r}-\widetilde{r}_0\|_{P,2}+\|\widetilde{M}-\widetilde{M}_0\|_{P,2} \|\widetilde{r}-\widetilde{r}_0\|_{P,2})\\
&\le \alpha(2C\delta_N+2C\delta_N+\delta_N N^{-1/2} )\\
&\le\alpha(2C\delta_N+2C\delta_N+\delta_N )=\delta_N (4C+1)\alpha,
\end{align*}
and
%‖E_P [(Z ̃-M ̃ )'(Z ̃-M ̃ )]-E_P [(Z ̃-M ̃_0 )'(Z ̃-M ̃_0 )]‖=‖-2(M ̃-M ̃_0 )'(Z ̃-M ̃_0 )+(M ̃-M ̃_0 )'(M ̃-M ̃_0 )‖_(P,1)\le2‖(M ̃-M ̃_0 )'(Z ̃-M ̃_0 )‖_(P,1)+‖(M ̃-M ̃_0 )'(M ̃-M ̃_0 )‖_(P,1)\le2‖M ̃-M ̃_0 ‖_(P,2) ‖Z ̃-M ̃_0 ‖_(P,2)+‖M ̃-M ̃_0 ‖_(P,2)^2\le\delta_N (4C+N^(-1/2) )\le\delta_N (4C+1)\le\delta_N (4C+1)\alpha,
\begin{align*}
\big\| \E_P \big[(\widetilde{Z} - \widetilde{M})'(\widetilde{Z} - \widetilde{M})\big]&- \E_P \big[(\widetilde{Z} - \widetilde{M}_0)'(\widetilde{Z} - \widetilde{M}_0)\big]\big\|\\
& = \big\| -2(\widetilde{M} - \widetilde{M}_0)'(\widetilde{Z} - \widetilde{M}_0)
+ (\widetilde{M} - \widetilde{M}_0)'(\widetilde{M} - \widetilde{M}_0)
\big\|_{P,1}\\
&\le 2\big\|(\widetilde{M} - \widetilde{M}_0)'(\widetilde{Z} - \widetilde{M}_0)\big\|_{P,1}
+ \big\|(\widetilde{M} - \widetilde{M}_0)'(\widetilde{M} - \widetilde{M}_0)\big\|_{P,1}\\
&\le 2\|\widetilde{M} - \widetilde{M}_0\|_{P,2} \, \|\widetilde{Z} - \widetilde{M}_0\|_{P,2}
+ \|\widetilde{M} - \widetilde{M}_0\|_{P,2}^2\\
&\le \delta_N (4C + N^{-1/2})
\le \delta_N (4C + 1)
\le \delta_N (4C + 1)\alpha,
\end{align*}
by successive applications of Minkowski’s and H\"older’s inequalities and Regularity Conditions~\ref{item:c} and~\ref{item:e}.  Now we show $\big[\E_P\|\psi-\psi_0\|^2\big]^{1⁄2}$ converges at the same rate:
\begin{align*}
\|\psi-\psi_0\|_{P,2}^2 &= \left\|\pmat{v_0^* \widetilde{U}\pi_0^\prime -\widetilde{V}_0^\prime\widetilde{U}_0 \\ \widetilde{V}^\prime\widetilde{R}-\widetilde{V}_0\prime\widetilde{R}_0}\right\|_{P,2}^2 \\
&=\|z_0^*\widetilde{U} -\pi_0^\prime \widetilde{V}_0^\prime\widetilde{U}_0\|_{P,2}^2+\sum_{k=1}^r\Big\|(\widetilde{z}_k-\widetilde{m}_k)\prime\widetilde{R}-(\widetilde{z}_k-\widetilde{m}_{k0})^\prime\widetilde{R}_0\Big\|_{P,2}^2 \\
&=\left\|\pi_0^\prime\left(\widetilde{V}_0-(\widetilde{M}-\widetilde{M}_0)\right)^\prime\left(\widetilde{U}_0-(\widetilde{l}-\widetilde{l}_0)+(\widetilde{r}-\widetilde{r}_0)\theta_0\right)-\pi_0^\prime\widetilde{V}_0^\prime\widetilde{U}_0\right\|_{P,2}^2 \\
&+ \sum_{k=1}^r\left\| \big(\widetilde{v}_{k0}-(\widetilde{m}_k-\widetilde{m}_{k0})\big)^\prime\left(\widetilde{R}_0-(\widetilde{r}-\widetilde{r}_0)+ (\widetilde{M}-\widetilde{M}_0)\pi_0\right)-\widetilde{v}_{0k}^\prime\widetilde{R}_0 \right\|_{P,2}^2 \\
&=\left\|\pi_0^\prime \left\{ (\widetilde{M}-\widetilde{M}_0)^\prime\widetilde{U}_0+\left(\widetilde{V}_0-(\widetilde{M}-\widetilde{M}_0)\right)^\prime\left(\widetilde{l}-\widetilde{l}_0-(\widetilde{r}-\widetilde{r}_0)\theta_0\right)\right\} \right\|_{P,2}^2 \\
&+ \sum_{k=1}^r\left\| (\widetilde{m}_k-\widetilde{m}_{k0})^\prime\widetilde{R}_0-\big(\widetilde{v}_{k0}+(\widetilde{m}_k-\widetilde{m}_{k0})\big)^\prime\left( (\widetilde{r}-\widetilde{r}_0)-(\widetilde{M}-\widetilde{M}_0)\pi_0\right)\right\|_{P,2}^2 \\
&\leq \left\|\pi_0^\prime\left\{(\widetilde{M}-\widetilde{M}_0)^\prime\widetilde{U}_0+\left(\widetilde{V}_0-(\widetilde{M}-\widetilde{M}_0)\right)^\prime\left(\widetilde{l}-\widetilde{l}_0-(\widetilde{r}-\widetilde{r}_0)\theta_0\right)\right\}\right\|_{P,2}^2 \\
&+ r\max_k\left\|(\widetilde{m}_k-\widetilde{m}_{k0})^\prime\widetilde{R}_0-\big(\widetilde{v}_{k0}+(\widetilde{m}_k-\widetilde{m}_{k0})\big)^\prime\big((\widetilde{r}-\widetilde{r}_0)-(\widetilde{M}-\widetilde{M}_0)\pi_0\big)\right\|_{P,2}^2.
\end{align*}
%where $\breve{m} + \breve{v}$ is the instrumental variable with the largest norm, and $\breve{v}_0$ and $\breve{m}_0$ are the true values of its residual and conditional mean under $P$. 
Then, applying H\"older’s inequality, gives an expression of the form
\[
\|\psi-\psi_0\|_{P,2}^2\le \|\pi_0\|_1^2 (\|A_1\|_{P,2}^2+\ldots+\|A_6\|_{P,2}^2)+r(\|B_1\|_{P,2}^2+\ldots+\|B_6\|_{P,2}^2),
\]
based on the six cross-products of $\widetilde{M}-\widetilde{M}_0$, $\widetilde{V}_0$, $\widetilde{l}-\widetilde{l}_0$ and $\widetilde{r}-\widetilde{r}_0$ in the first expression and similarly for the second.  For example, the following cross-products involving two nuisance functions satisfies 
\[
\|A_1\|_{P,2}^2\leq \|\widetilde{M}-\widetilde{M}_0\|_{P,2}^2 \, \|\widetilde{U}_0\|_{P,2}^2 \, \|\widetilde{l}-\widetilde{l}_0\|_{P,2}^2\leq (C\delta_N N^{-1/2})^2,
\]
as do the other seven, and the cross-product involving only one nuisance parameter satisfies
\[
\|B_5\|_{P,2}^2\leq\|\widetilde{v}_0\|_{P,2}^2 \, \|\widetilde{r}-\widetilde{r}_0\|_{P,2}^2 \, \leq(2C\delta_N )^2
\]
as do the other three (these follow from H\"older’s inequality and Regularity Conditions~\ref{item:c} and~\ref{item:e} and Regularity Condition~\ref{item:d}) gives $\|\widetilde{U}_0\|_{P,2}=\left(\E_P(\widetilde{U}_0^2)\right)^{1/2}\leq\sqrt{C}$ (and ditto for $\widetilde{R}_0$, $\widetilde{V}_0$ and $\breve{v}_0$) and so the convergence rate of the product of the two nuisance parameters is $\delta_N N^{-1/2}$. Hence,
\[
\|\psi-\psi_0\|_{P,2}^2 \leq \sqrt{ O(C^2 ) \Big(O(C^2 )\big(1+o(N^{-1})\big) \delta_N^2\Big)\Big(O(C^2 )\big(1+o(N^{-1})\big)\delta_N^2\Big)} = O(C^2 \delta_N )
\]
because $\|\pi_0\|_1^2\leq\alpha^2=O(C^4)$. Finally, it remains to show
\[
\sup_{d \in (0,1), \eta \in \mathcal{T}_N} \big\| \partial_r^2 \mathbb{E}_P \ell\big[\psi\big(W; \theta_0, \eta_0 + (\eta - \eta_0)d\big) \big]\big\| \leq \delta_N N^{-1/2}.
\]
Here,
\[
f(d) = \mathbb{E}_P \left[
\pmat{\pi_0' \{ \widetilde{V}_0 - d(\widetilde{M} - \widetilde{M}_0) \}' \{ \widetilde{U}_0 - d(\widetilde{l} - \widetilde{l}_0) d - d(\widetilde{r} - \widetilde{r}_0) \theta_0 \} \\
\{ \widetilde{V}_0 - d(\widetilde{M} - \widetilde{M}_0) \}' \{ \widetilde{R}_0 - (\widetilde{r} - \widetilde{r}_0)d - d(\widetilde{M} - \widetilde{M}_0)\pi_0 \}}
\right].
\]
and its second derivative is
\[
\partial_d^2 f(d) = \mathbb{E}_P \left[
\begin{pmatrix}
2 \pi_0' (\widetilde{M} - \widetilde{M}_0)' \{ (\widetilde{l} - \widetilde{l}_0) - (\widetilde{r} - \widetilde{r}_0)\theta_0 \} \\
2 (\widetilde{M} - \widetilde{M}_0)' \{ (\widetilde{r} - \widetilde{r}_0) - (\widetilde{M} - \widetilde{M}_0)\pi_0 \}
\end{pmatrix}
\right].
\]
Straightforward application of Minkowski’s and H\"older’s inequalities and Regularity Condition~\ref{item:e} gives
%|E_P [2π_0^' (M ̃-M ̃_0 )'{(l ̃-l ̃_0 )-(r ̃-r ̃_0 ) θ_0 }]|\le2\alpha(1+r\alpha) \delta_N N^(-1/2),
\[
\left| \mathbb{E}_P \left[ 2 \pi_0' (\widetilde{M} - \widetilde{M}_0)' \{ (\widetilde{l} - \widetilde{l}_0) - (\widetilde{r} - \widetilde{r}_0)\theta_0 \} \right] \right|
\le 2 \alpha (1 + r \alpha) \delta_N N^{-1/2},
\]
and
\[
\Big\lVert 
\mathbb{E}_P \!\left[ 2(\widetilde{M} - \widetilde{M}_0)' \bigl\{ (\widetilde{r} - \widetilde{r}_0) - (\widetilde{M} - \widetilde{M}_0)\pi_0 \bigr\} \right] \Big\rVert
\le 2(1+\alpha)\,\delta_N N^{-1/2}
\le 2\alpha(1+r\alpha)\,\delta_N N^{-1/2},
\]
so that $\delta_N \equiv 2\alpha(1+r\alpha)\,\delta_N N^{-1/2}.$

%‖E_P [2(M ̃-M ̃_0 )'{(r ̃-r ̃_0 )-(M ̃-M ̃_0 ) π_0 }]‖\le2(1+\alpha) \delta_N N^(-1/2)\le2\alpha(1+r\alpha) \delta_N N^(-1/2)
%so that \delta_N≡2\alpha(1+r\alpha) \delta_N N^(-1/2).
All the conditions set out in {\em Assumption~3.1} and {\em Assumption~3.2} are all satisfied, and so {\em Theorem~3.1} holds for the DML estimator.
\newpage
%%%%%%%%%%%%%%%%%%%%%%%%%%%%%%%%%%%%%%%%%%
\section{Algorithm for Panel IV DML}\label{sec:algorithm}
The panel IV DML algorithm for the estimation and statistical inference of $\mathbf{b}_0=(\theta_0,\bm{\pi}_0)'$ and for $\mathbf{b}^{\rm rf}_0=(\bm{\delta}_0,\bm{\pi}_0)'$, follows the steps below. A summary is provided in Algorithm~\ref{alg:xtdml}. 

\

\noindent \textbf{\textsc{Stage 1 (Panel dataset)}} Consider a dataset with repeated cross-sectional units $N$ (e.g., households, firms, countries, regions) and multiple time periods such that $T\ge2$. Apply the FD (exact) approach to remove the unobserved individual heterogeneity from the model equations. This consists in taking the first difference of the following variables $\{\Yit,\Dit, \vector Z_{it}\}$, and generating the first-order lag of $\vector X_{it}$.

\
    
%%algorithm
%\begin{enumerate}[label=\textsc{Step} \arabic*]
    \noindent \textbf{\textsc{Stage 2 (Block-k sample splitting and cross-fitting)}} Randomly partition the cross-sectional units $N$ into $K$ folds of the same size. For each fold \mbox{$k=1,\ldots,K$} denote $\Wk\subset\mathcal{W}$ as the estimation sample, and $\Wk^c$ as its complement, where $N_k\equiv |\Wk| = N/K$, $|\Wk^c|=N-N_k$. The folds are mutually exclusive and exhaustive such that $\Wk\cap\mathcal{W}_j=\Wk\cap\Wk^c=\varnothing$ and $\Wk\cup\Wk^c=\mathcal{W}_1\cup\ldots\cup\mathcal{W}_K=\mathcal{W}$. For \mbox{$K>2$}, the larger complementary sample $\Wk^c$ is used to learn the potentially complex nuisance parameters $\bmeta_0$, and $\Wk$ for the relatively simple task of estimating the parameters $\mathbf{b}_0$ and $\mathbf{b}_0^{\rm rf}$. 
    %Because each unit $i$ and its whole time series is assigned to the same fold, the algorithm performs a \emph{block-k-fold} cross-fitting.  Sample splitting with cross-fitting (when $K\ge2$ and the folds swap roles) reduces the overfitting bias and restores efficiency.

\ 

    \noindent \textbf{\textsc{Stage 3 (Learning the nuisance parameters)}} For each fold $k$, use ML algorithms to predict the vector of nuisance parameters $\bmeta$ from the complementary data $\{W_i:i\in \Wk^c\}$. 

     \ 
     
    %%%%%%%%%%%%%%%%%%%%%%%%%%%%%%%%%%%%%%%%%%
    \noindent \textbf{Stage 4 \textsc{(Construction of Neyman-orthogonal score)}} Use the predicted nuisances $\widehat{\bmeta}_k$ from \textsc{Stage 3} to construct Neyman orthogonal scores $\mathbf{s}(W_k, \mathbf{b}_k;\widehat{\bmeta}_k)$ for the structural model and $\mathbf{s}^{\rm rf}(W_k, \mathbf{b}_k^{\rm rf};\widehat{\bmeta}_k)$ for the reduced form model using the estimation sample in fold $k$ ($\W_k$).

\

    \noindent \textbf{Stage 5 \textsc{(Estimation and inference on the parameters)}} The Panel IV DML estimators $\widehat{\mathbf{b}}_k$ $\widehat{\mathbf{b}}_k^{\rm rf}$ are the unique solutions to $(K N_k)\inv\sum_{k=1}^K\sum_{i\in \Wk}\sum_{t\in \mathcal{T}} \mathbf{s}(W_{it}, \mathbf{b}_k;\widehat{\bmeta}_k)  = \zero$ and $(K N_k)\inv\sum_{k=1}^K\sum_{i\in \Wk}\sum_{t\in \mathcal{T}} \mathbf{s}^{\rm rf}(W_{it}, \mathbf{b}_k^{\rm rf};\widehat{\bmeta}_k)  = \zero$. 
    A finite-sample correction of $(\widehat{\mathbf{b}}_k-K\inv\sum_k\widehat{\mathbf{b}}_k)^2$ is applied to the average variance across the $k$-folds, $\widehat{\matrix \Sigma}^2$, weighted by the number of units in the cluster to account for the variation introduced by sampling splitting \citep[p.~C30]{chernozhukov2018}. 
    
    \ 
    
    \noindent \textbf{\textsc{Stage 6 (Iteration)}} Repeat \textsc{Steps}~3-5 for each of the $k$ folds and average the intermediate results.
    
    \ 
    
    \noindent \textbf{\textsc{Stage 7 (Robust weak-IV diagnostics)}} 
    Calculate the first-stage {\em F}-statistic $F^{DML}$, the AR test statistic $\mathrm{AR}(\thetano)^{DML}$, and the AR confidence set $CS_{AR}(\thetano)$. Compare the test statistics with the appropriate critical values at the desired levels of significance to make inferences about the causal parameter of interest. 

\begin{algorithm}[t!]
\SetAlgoLined
\SetKwData{Left}{left}
\SetKwData{This}{this}
\SetKwData{Up}{up}
\SetKwFunction{Union}{Union}
\SetKwFunction{FindCompress}{FindCompress}
\SetKwInOut{Input}{Input}
\SetKwInOut{Output}{Output}
\SetKwInOut{Require}{Require}
\SetKwInOut{State}{State}
\SetKwInOut{Initialize}{Initialize}

\caption{Panel IV DML Algorithm}\label{alg:xtdml}
\vspace{2mm}
\Require{Panel dataset with $N$ subjects, and $T\ge2$ time periods.}
\Input{Data $\{Y_{it},D_{it},\vector Z_{it},\vector X_{it}\}$, panel and time identifiers, and cluster variable if different from panel identifier.}
\Output{$\widehat{\mathbf{b}}=(\thetano,\widehat{\bm{\pi}})'$, $\widehat{\mathbf{b}}_k^{\rm rf}=(\widehat{\bm{\delta}},\widehat{\bm{\pi}})'$, $\widehat{\bm{\Sigma}}$, $\widehat{\bm{\Sigma}}^{\rm rf}$, $F^{DML}$, ${AR}(\thetano)^{DML}$, $CS_{AR}(\thetano)$, model RMSE, RMSE~of~$l,r,\vector m$.}

\Initialize{Set number of folds $k\ge2$; $i$ randomly assigned to fold $k$. Assign learners for nuisance parameters $\bmeta = \{l,r,\vector m\}$.}

Divide the sample into $K$ folds, and randomly assign unit $i$ and its time series to fold $k$. Define the estimation sample in fold $k$ ($W_k$) with size $|N_k|$, and the complementary sample in folds $-k$ ($W_k^c$).

\For{$k\leftarrow 1$ \KwTo $K$}{
%\tcc{Data in fold $k$ is the estimation sample ($\W_k$); data in $-k$ is the complementary sample ($\W_k^c$)}
    Predict $\widehat{\bmeta}_k$ using base learners on data $\mathcal{W}_k^c$ .\\
    Construct $\mathbf{s}(W_k, \mathbf{b}_k; \widehat{\bmeta}_k)$ and $\mathbf{s}^{\rm rf}(W_k, \mathbf{b}_k^{\rm rf}; \widehat{\bmeta}_k)$.\\
    Solve $\frac{1}{|N_k|}\sum_{i\in \mathcal{W}_k}\mathbf{s}(W_k, \mathbf{b}_k; \widehat{\bmeta}_k)=0$ and $\frac{1}{|N_k|}\sum_{i\in \mathcal{W}_k}\mathbf{s}^{\rm rf}(W_k, \mathbf{b}_k^{\rm rf}; \widehat{\bmeta}_k)=0$ for $\widehat{\mathbf{b}}_k$ and $\widehat{\mathbf{b}}_k^{\rm rf}$, respectively.
}
Average $\{\widehat{\mathbf{b}}_k,\widehat{\mathbf{b}}_k^{\rm rf}\}_{k=1}^K$ and model RMSE, compute finite-sample variance-covariance matrix $\widehat{\bm{\Sigma}}$,$\widehat{\bm{\Sigma}}^{\rm rf}$ and  RMSE~of~$l,r,\vector m$. \\
Compute $F^{DML}$, ${AR}(\thetano)^{DML}$, $CS_{AR}(\thetano)$.
\end{algorithm}

\newpage
\section{Additional Tables for \citet{tabellini2020}}\label{sec:app_tabellini}
\renewcommand{\thetable}{C.\arabic{table}}
\setcounter{table}{0}

In this section, we discuss additional results based on conventional 2SLS estimators using fixed effects (FE) and first-differences (FD) transformations in the reanalysis of \citet{tabellini2020}. This comparison, briefly commented in Section~\ref{sec:tabellini}, is a necessary robustness step to verify that any possible differences found with panel IV DML estimates mainly reflect methodological rather than specification choices.   %Therefore, we verify that conventional 2SLS estimates based on FE and FD are broadly comparable in terms of sign, magnitude, and statistical significance. 
%This provides complementary benchmarks and robustness checks for the panel IV DML analysis discussed in the main text. 
%This serves as robustness check of panel IV DML analysis presented in the main text.
Table~\ref{tab:tabellini2020_fefd} presents the results of this supplementary check. Panel A reports baseline specifications while Panel B, not originally implemented, displays the augmented specifications with additional control variables included simultaneously. The dependent variables are: \emph{DW Nominate Score} (Columns 1-2) for the political outcome, and \emph{Log Occupational Score} (Columns 3-4) for economic outcome.

Panel A of Table~\ref{tab:tabellini2020_fefd} reports the original 2SLS estimates with FE in Columns~1 and~3 \citep[originally from Table~3, Column~4, Panel~B, and Table~5, Column~1, Panel~B of][]{tabellini2020} alongside our 2SLS estimates with FD in Columns~2 and~4. 
The first-stage F-statistics from the 2SLS specifications with FD for both outcomes, although somewhat smaller than those from FE, always exceed \citet{stock2005}'s critical value of 16.30, even if slightly above this threshold for the political outcome with the FD estimator.\footnote{Using a more conservative threshold of 104.70  for the strength of the instrument \citep{lee2022}, only the specification for the economic outcome estimated with 2SLS with FE (Column 3 in Panel A) passes the test for instrumental relevance.} The results from Anderson-Rubin (AR) test statistics and AR confidence sets (bounded without zero included), not computed in the original study, provide support about the existence of a strong positive effect of immigration on both outcome variables with either 2SLS estimators. %The second-stage estimates are broadly similar across the two estimators in terms of magnitude, sign, and statistical significance. 
Given the strength of the instruments, the second-stage coefficients (close in magnitudes for both panel estimators) suggest that higher immigration shares led to the election of more conservative representatives, while employment gains for natives, induced by European immigrants, were accompanied by occupational or skill upgrading.

%%%%%%%%%%%%%%%%%%%%%%%%%%%%%%%%%%%%%%%%%%%%%%%%%%%%%%%%%%%%%%%%%%%%%%%%%%%%
%%%%% Tab D2 - col (4) 
%%%%%%%%%%%%%%%%%%%%%%%%%%%%%%%%%%%%%%%%%%%%%%%%%%%%%%%%%%%%%%%%%%%%%%%%%%%%
\begin{table}[t!]
\centering
 \caption{The political and economic effects of immigration}\label{tab:tabellini2020_fefd}
 \vspace{-2mm}
\scalebox{.75}{
\begin{threeparttable}	%use it for notes
  \begin{tabular}{lcccc}   
\vspace{-3mm}\\   
\hline\hline
\vspace{-3mm}\\	
\emph{Dependent variable: }&\multicolumn{2}{c}{DW Nominate Score}&\multicolumn{2}{c}{Log Occupational Score}\\
\cmidrule(l{.35cm}r{.25cm}){2-3}\cmidrule(l{.35cm}r{.25cm}){4-5}
2SLS with: &FE& FD &FE& FD\\
&(1)&(2)&(3)&(4)\\
\vspace{-3mm}\\
\hline
\vspace{-3mm}\\
\multicolumn{5}{c}{\textbf{Panel A:} \emph{Baseline specification with fixed-effects interactions }}\\	
\vspace{-3mm}\\
\multicolumn{5}{l}{\emph{Second-stage results}}\\
Fr. Immigrants	&	 1.658**&1.772**	&	 0.097***&0.095***\\
 	&	 (0.808)  &(0.829)	&(0.036)  	 & (0.042)	\\
AR CS 95\%&[0.298, 3.499]  &[0.049, 3.495]   &[0.030, 0.162]  &[0.017,0.173]  \\
\vspace{-3mm}\\
\multicolumn{5}{l}{\emph{First-stage results}}\\
Shift-Share IV	&	  1.007***&0.965***	&0.993***	 &  0.933***\\
 	& (0.209) 	&(0.226) 	& (0.061) 	 & (0.094) 	\\
F stat	&23.11	&18.23	&  251.31 	& 99.45	\\
AR $\chi^2$  stat&4.71	&	3.84	&	8.27&	5.40\\
AR $\chi^2$ p-value& 0.030  & 0.050&   0.004    &  0.020\\

Model RMSE& 0.015	& 0.232&	   0.013&  0.015	  \\
\vspace{-3mm}\\
Observations&  460&     303   &     538   &     342    \\
No. clusters& 157&     157   &       127  &     125  \\

%%%%%%%%%%%%%%%%%%
\vspace{-3mm}\\
\multicolumn{5}{c}{\textbf{Panel B:} \emph{ Specification with additional controls (not originally implemented)}}\\
\vspace{-3mm}\\
\multicolumn{5}{l}{\emph{Second-stage results}}\\
Fr. Immigrants	&	 1.736 &0.928	&	0.091* 	& 0.094 	 \\
 	&	(1.902)  &(2.103)   	&(0.049) 	&	(0.060) 			\\
 AR CS 95\% & [-2.143, 6.070]  &[-4.276, 4.883]  &	[-0.004, 0.181] &[-0.032, 0.207] \\
 \vspace{-3mm}\\
\multicolumn{5}{l}{\emph{First-stage results}}\\
Shift-Share IV&	0.544***&0.541***	&0.794***	&	 0.745***			\\
 	& (0.143)	&(0.178)	&	 (0.083)  &	(0.117)  			\\
F stat	&  14.62  	&9.26	&	85.79	&	40.89			\\
AR $\chi^2$ stat& 0.82 	&0.18	&3.38	&	 2.20		\\
AR $\chi^2$ p-value& 0.366   &    0.668       &   0.066   & 0.138  \\

Model RMSE& 0.013	&	0.218	&   0.011& 0.019		\\
\vspace{-3mm}\\
Observations&   451 &     297   &    526   &     338    \\
No. clusters&  154  &     154   &       126   &     125  \\

\hline
\end{tabular} 	
\footnotesize
\textbf{Note:}
The table displays the estimates of baseline specifications from Table 3 (column 4, Panel B) and Table 5 (column 2, Panel B) in \cite{tabellini2020}, obtained with conventional 2SLS with fixed effects, and our estimates  with conventional 2SLS regression with FD transformation (Columns 2 and 4). The raw control variables only include interactions between state and year fixed effects, for a total of 77 variables in Panel A and 96 in Panel B; some variables are omitted due to multicollinearity.
The number of clusters differs between Columns 3 and 4 because the FD estimator requires at least two consecutive non-missing observations per group, while the FE estimator only requires two non-missing observations, not necessarily consecutive. In this case, two groups lack consecutive data and, hence, excluded from the estimation.
The Anderson-Rubin test statistic and confidence sets are not originally implemented in the analysis.
%%%%%
Standard errors in parenthesis are clustered at the metropolitan area in Columns 1-4 and at the city code level in Columns 5-8.  Significance levels: * p $<$ 0.10, ** p $<$ 0.05, *** p $<$ 0.01.
\end{threeparttable}
}
\end{table}

% While occupational Score can be used to proxy for natives’ income, they do not capture within occupation changes in earnings. 

%% p=77 for 1-2, p=79 for 3-5

In Panel B of Table~\ref{tab:tabellini2020_fefd} with the augmented specifications, we find much smaller first-stage F-statistics from FD than those from FE regressions. The instrument appears weak with both estimators in the political outcome specification (Columns 1-2). Specifically, the FE F-statistic is below \citet{stock2005}'s threshold, while the FD F-statistic does not even exceed the conventional rule-of-thumb value of 10. By contrast, the shift-share instrument in the economic outcome specification (Columns 3-4) can be considered strong, as both first-stage F-statistics exceed \citet{stock2005}'s threshold, but never \citet{lee2022}'s threshold of 104.70. The AR test statistics either fail to reject the null hypothesis of no reduced-form effect, when assuming no second-stage effect, or only at 10\% level (in Column 3). The AR CS always include zero as a possible value of the treatment effect, suggesting no second-stage effect in both outcomes. Given the moderate strength of the instruments, we can comment on the second-stage estimates. The second-stage coefficients are statistically insignificant (or borderline significant in Column~3) with both panel estimators and in both specifications, confirming the AR test results. %There is an exception for Specification (3 ) in Panel B, where both AR test and second-stage coefficients are is only significant at 10\% level.

In conclusion,  FE and FD estimators yield very similar results in many dimensions across both outcomes and, therefore, they may identify the same underlying parameter under standard assumptions. The baseline findings of the original study are confirmed with both FE and FD estimators. However,  after controlling for additional confounders in the regression,  both estimators provide no evidence of any impact of immigration on both political and economic outcomes, with the exception of the FE regression of log occupational score.
%when additional controls are included, both estimators produce statistically insignificant effects (Panel~B of Table~\ref{tab:tabellini2020_fefd}), which do not supporting the main results of the original article. 

\section{Additional Tables for \citet{moriconi2019,moriconi2022}}\label{sec:app_moriconi}
\renewcommand{\thetable}{D.\arabic{table}}
\setcounter{table}{0}

In the following sections, we discuss the results on individual voters' behaviour from conventional 2SLS estimators using FE and FD transformations for \citet{moriconi2019,moriconi2022}, which were briefly commented in Section~\ref{sec:moriconi2019_2022}. In our re-analysis, the estimates are obtained from a conventional unbalanced panel dataset, where we aggregated the individual voters' information at the regional (either NUTS2 or country) and year levels such that the same regional units from each of the twelve European countries are observed over consecutive election years. For completeness of the analysis, we also report the original estimates obtained with the individual level data without any aggregation.

In revisiting the study conducted by \citet{moriconi2019}, we also discuss the results of the political parties to complement the analysis. The party platforms data are constructed as a conventional balanced panel dataset, where information of the same political parties from twelve European countries is collected over up to three election years. Unlike the voters' data, no aggregation was required.  The set of covariates used in the political party analysis differs from the aggregated individual voters' analysis and includes: the average GDP per capita (in log), share of tertiary sector (in log), average unemployment rate, and election year fixed effects.

%For \citet{moriconi2022}, we complement the analysis in Section~\ref{sec:moriconi2022} by reporting additional results for political and migration attitude outcomes, which were omitted from the main text for brevity, estimated using both conventional 2SLS and panel IV DML methods.
%%%%%%%%%%%%%%%%%%%%%%%%%%%%%%%%%%%%%%%%%%%%%%%%%%%%%%%%%%%%
%% MORICONI 2019
%%%%%%%%%%%%%%%%%%%%%%%%%%%%%%%%%%%%%%%%%%%%%%%%%%%%%%%%%%%%
\subsection{\citet{moriconi2019}}\label{sec:app_moriconi2019}

Table~\ref{tab:moriconi2019_tab4_fefd} reports our estimates obtained from conventional 2SLS with FE and 2SLS with FD  using aggregated individual data (Columns 1-4) and political party data (Columns 5-8). The results are reported by skill-group, as in the original analysis: HS immigrants (Columns 1-2, and 5-6) and LS immigrants (3-4, and 7-8). The dependent variables are: \emph{Net Welfare State} in Panel~A, and \emph{Net Public Education} in Panel~B. 
For completeness,  Table~\ref{tab:moriconi2019_tab4_original} displays the corresponding original results from \cite{moriconi2019}'s Table~4 (individual voters' data) and Table~5 (parties' data) relative to their Specifications~(2), (4) and (6) reported in their Panels~A and~B. 

%%%%%%%%%%%%%%%%%%%%%%%%%%%%%%%%%%%%%%%%%%%%%%%%%%%%%%%%%%%%%%%%%%%%%%%%%%%%
%%%%% Tab D2 - col (4) 
%%%%%%%%%%%%%%%%%%%%%%%%%%%%%%%%%%%%%%%%%%%%%%%%%%%%%%%%%%%%%%%%%%%%%%%%%%%%
\begin{table}[t!]
\centering
 \caption{Political preferences over 2007–2016: 2SLS with FD and with FE}\label{tab:moriconi2019_tab4_fefd}
 \vspace{-2mm}
\scalebox{.6}{
\begin{threeparttable}	%use it for notes
  \begin{tabular}{lcccccccc}   
\vspace{-3mm}\\   
\hline\hline
\vspace{-3mm}\\	
Sample: &\multicolumn{4}{c}{Aggregated Individual voters}&\multicolumn{4}{c}{Political parties}\\
\cmidrule(l{.35cm}r{.25cm}){2-5}\cmidrule(l{.35cm}r{.25cm}){6-9}
&\multicolumn{2}{c}{HS immigrants}&\multicolumn{2}{c}{LS immigrants}&\multicolumn{2}{c}{HS immigrants}&\multicolumn{2}{c}{LS immigrants}\\
\cmidrule(l{.35cm}r{.25cm}){2-3}\cmidrule(l{.35cm}r{.25cm}){4-5}\cmidrule(l{.35cm}r{.25cm}){6-7}\cmidrule(l{.35cm}r{.25cm}){8-9}
2SLS with:&FE&FD &FE &FD&FE&FD &FE &FD  \\
&(1)&(2)&(3)&(4)&(5)&(6)&(7)&(8)\\
\vspace{-3mm}\\
\hline
\vspace{-3mm}\\
\multicolumn{9}{c}{\textbf{Panel A:} \emph{Net Welfare State}}\\			
\multicolumn{9}{l}{\emph{Second-stage results}}\\
Share&   0.062***&  0.054***& 0.040* &0.050&0.157& 0.076 &-0.331*&-0.448*\\
          & (0.011)   & (0.014)   &(0.024)&  (0.033)&(0.477) &(0.350)& (0.168)&(0.237)\\
AR CS 95\% 	&[0.043, 0.087]   &[0.029, 0.080]  &[-0.004, 0.101]   &[0.003, 0.159]&$[-0.940,+\infty)$  &[-0.471, 1.329]  &$(-\infty,-0.121]$   &$(-\infty,-0.209]$ \\
\vspace{-3mm}\\
\multicolumn{9}{l}{\emph{Robust Weak IV Tests}}\\
% Shift-share IV 	&	 1.856***&1.655***& 0.600***&  0.522***\\
% & (0.285)   &(0.263)   & (0.166)   & (0.162)   \\
% F stat	&	 42.90   &   39.51   &  14.61   &   10.37  \\
AR $\chi^2$ stat&   27.93   &   15.36   &    3.24   &    3.60&0.14   &    0.06   &   11.38   &   20.06\\
AR $\chi^2$ p-value &   0.000   &   0.000     &   0.072   &   0.058  &   0.710   &   0.810 &   0.001   &   0.000 \\

\vspace{-3mm}\\

%%%%%%%%%%%%%%%%%%%%%%%
\vspace{-3mm}\\
\multicolumn{9}{c}{\textbf{Panel B:} \emph{Net Public Education}}\\	
\multicolumn{9}{l}{\emph{Second-stage results}}\\
Share&   0.026** &  0.024 &-0.070**&-0.088* &0.356& 0.229&  0.239&0.333\\
          & (0.013)   &(0.015)&(0.032)& (0.046) &(0.743) &(0.551)&(0.195)& (0.284)\\
AR CS 95\%              &  [0.005, 0.055]   &[-0.004, 0.055]   &[-0.160, -0.017]   &[-0.250, -0.022]  &$[  -2.493,+\infty)$  &[-0.731, 1.998]   &$[0.007,+\infty)$  &$[-0.005,+\infty)$\\
\vspace{-3mm}\\
\multicolumn{9}{l}{\emph{Robust Weak IV Tests}}\\
% Shift-share IV 	&	 1.856***&1.655***& 0.600***&  0.522***\\
% & (0.285)   &(0.263)   & (0.166)   & (0.162)   \\
% K-P rk Wald F-stat       &   42.90   &   39.51   &     14.61   &   10.37 \\
AR $\chi^2$                   &    5.24   &    2.78   &   6.70   &    6.65 &0.28   &    0.22 &   3.09   &    3.66   \\
AR $\chi^2$ p-value           &   0.022   &   0.096   &  0.010   &   0.010  &0.598   &   0.639   &   0.079   &   0.056  \\

\vspace{-3mm}\\
\multicolumn{9}{c}{\textbf{Panels A and B}}\\
\multicolumn{9}{l}{\emph{First-stage results}}\\
Shift-share IV 	&  1.995***&1.655***& 0.636***&  0.522***& 0.739*  & 0.915**&0.779* & 0.731*   \\
& (0.305)   &(0.263)   & (0.167)   & (0.162) &(0.402)& (0.340)&(0.366) & (0.401) \\
F-stat       &   42.90   &   39.51   &     14.61   &   10.37 &  3.30   &    7.25   &  4.60   &    3.32  \\

\vspace{-3mm}\\
\hline
\vspace{-3mm}\\
% Region FE&Yes&Yes&Yes&Yes&Yes&Yes&Yes&Yes\\
% Election FE&Yes&Yes&Yes&Yes&Yes&Yes&Yes&Yes\\
%No. variables	&	9	&	297	&	18	&	18&	9	&	297	&	18	&	18	\\
Observations&259&146&259&146& 179&97&179&97 \\
No. clusters&113&113&113&113& 12&12& 12&12\\
\hline
\end{tabular} 	
\footnotesize
\textbf{Note:} 
The table displays our estimates of Specifications~(2), (4) and (6) (Panels A and B) in \cite{moriconi2019}'s Tables~4 (individual voters' data) and 5 (parties' data), obtained from conventional 2SLS regression with FE (Columns 1, 3, 5 and 7) and FD transformation (Columns 2, 4, 6 and 8). The original sample consists of different individual voters from twelve European countries sampled each year, which we aggregate at regional (NUTS2) level to obtain an unbalanced panel data set. The treatment and instrumental variables in Columns (1)-(2) and (5)-(6) refer to the fraction of high-skilled workers, and in Columns (3)-(4) and (7)-(8) of low-skilled workers. The number of observations differs from 2SLS with FE and with FD regressions because the first time period is removed after the first-difference transformation. Raw control variables in all panels are: the share of women, average age, share of tertiary/post-tertiary education, average GDP per capita (in log), share of tertiary sector (in log), average unemployment rate, and election year dummies. Dependent variables are `Net Welfare State' in Panel A, and `Net Public Education in Panel B.
% %%%%%%%
Standard errors in parenthesis are clustered at the regional level in Columns 1-4 and at the country level in Column 5-8.
Significance levels: \mbox{* p $<$ 0.10,} \mbox{** p $<$ 0.05,} *** p $<$ 0.01. 
\end{threeparttable}
}
\end{table}

We begin by discussing the results for individual voters based on the aggregated sample (Columns 1-4 of Table~\ref{tab:moriconi2019_tab4_fefd}) for both panels, as the first-stage regressions are identical across specifications. In Columns 1-2, the shift-share instruments for HS immigration appear strong under both estimators. The corresponding F-statistics are commonly viewed as sufficiently large based on \citet[][]{stock2005}'s critical value of 16.30 and, hence, strong. However, their strength may be considered quite moderate relative to more conservative thresholds proposed in the literature \citep[e.g., 104.700 in][]{lee2022}. The same consideration applies to the original study, where the reported F-statistic ($F=32.87$ in Columns 1-2 of Table~\ref{tab:moriconi2019_tab4_original}) is also below \citet{lee2022}'s threshold and smaller than those from the panel 2SLS specifications.
The AR test statistics, not computed in the original study,  reject the null hypothesis of no reduced-form effect with both estimators (albeit only at the 10\% level for FD in Panel B), providing evidence of a second-stage effect in both Panels~A and~B. This conclusion is supported by the corresponding 95\% AR confidence sets, which are always bounded and exclude zero, with the exception of Column~2 in Panel~B. Overall, the FE estimates align closely with the original cross-sectional findings about the effect of HS immigration on both welfare and public education expansion, whereas the FD results only support the presence of an effect for welfare expansion.

%%%%%%%%%%%%%%%%%%%%%%%%%%%%%%%%%%%%%%%%%%%%%%%%%%%%%%%%%%%%%%%%%%%%%%%%%%%%
%%%%% Tab D2 - col (4) 
%%%%%%%%%%%%%%%%%%%%%%%%%%%%%%%%%%%%%%%%%%%%%%%%%%%%%%%%%%%%%%%%%%%%%%%%%%%%
\begin{table}[t!]
\centering
 \caption{Political preferences over 2007–2016 (\citet{moriconi2019}'s results)}\label{tab:moriconi2019_tab4_original}
 \vspace{-2mm}
\scalebox{.95}{
\begin{threeparttable}	%use it for notes
  \begin{tabular}{lcccccc}   
\vspace{-3mm}\\   
\hline\hline
\vspace{-3mm}\\	
Sample: &\multicolumn{2}{c}{Individual voters}&\multicolumn{2}{c}{Political parties}\\
\cmidrule(l{.35cm}r{.25cm}){2-3}\cmidrule(l{.35cm}r{.25cm}){4-5}
&HS&LS &HS &LS  \\
&(1)&(2)&(3)&(4)\\
\vspace{-3mm}\\
\hline
\vspace{-3mm}\\
\multicolumn{5}{c}{\textbf{Panel A:} \emph{Net Welfare State}}\\			
\vspace{-3mm}\\
Share of immigrants&  0.049$^{**}$& 0.009&0.062 &-0.357$^{*}$ \\
      & (0.022)&(0.012)&(0.494)&(0.183)  \\

Observations&50,304&50,304&177 &177 \\
K-P rk Wald F-stat &32.87 & 32.58& 21.47 & 17.39 \\
Adj. R-Square & 0.70&0.70&0.57 &0.54 \\

\vspace{-3mm}\\
\multicolumn{5}{c}{\textbf{Panel B:} \emph{Net Public Education}}\\			
\vspace{-3mm}\\
Share of immigrants& 0.058$^{***}$& -0.028&0.378& 0.261\\
          &(0.020)& (0.022)& (0.458)& (0.181) \\
Observations& 50,304& 50,304& 177& 177 \\

\vspace{-3mm}\\
\hline
\vspace{-3mm}\\
K-P rk Wald F-stat & 32.87 & 32.58&21.47 &17.39 \\
Adj. R-Square& 0.56& 0.56 & 0.62 &  0.59\\
\vspace{-3mm}\\
\hline
\vspace{-3mm}\\
Election FE & Yes& Yes& Yes& Yes\\
NUTS2 Controls & Yes& Yes& Yes& Yes\\
Individual Controls & Yes& Yes& Yes& Yes\\
\hline
\end{tabular} 	
\footnotesize
\textbf{Note:} The table displays the original estimates for Specifications~(2), (4) and (6) in Panels A and B from \cite{moriconi2019}'s Table~4 (individual voters' data) and Table~5 (parties' data), obtained from conventional IV estimation. 
% %%%%%%%
Standard errors in parenthesis are clustered at the regional level.
Significance levels: \mbox{* p $<$ 0.10,} \mbox{** p $<$ 0.05,} ***~p~$<$~0.01. 
\end{threeparttable}
}
\end{table}

In Columns~3-4, the instruments appear much weaker than in the original study under both FE and FD estimators. The corresponding first-stage F-statistics lie between the conventional cut-off values of 10 and 16.30, raising  concerns about the strength of the shift-share instrument in these specifications. 
By contrast, the original cross-sectional analysis reports considerably larger values ($F=32.58$ in Column~2 of Table~\ref{tab:moriconi2019_tab4_original}), showing no evidence of weak instruments under pooled 2SLS.  The lower F-statistics obtained from FE and FD regressions may partly reflect a loss of statistical power after the region-year aggregation, which substantially reduced the sample size (from around 50,000 observations to roughly 200 units in the estimation sample), thereby attenuating the apparent relevance of the instrument. The AR test statistics suggest the presence of some effect of LS immigration on both outcomes among individual voters (at 10\% level for \emph{Net Welfare State}, and at 5\% level for \emph{Net Public Education}). The corresponding AR CS are often bounded and do not include zero (except for Column~3 of Panel~A). Given the apparent weakness of the shift-share instruments in panel data regressions, second-stage results from both estimators should be interpreted with caution. Nevertheless, the AR diagnostic results indicate that both panel estimators agree on a negative effect of LS immigration on public education, whereas only the FD specification points to a potential positive effect on welfare expansion.The original study, by contrast, finds no significant effect for LS immigrants.
 
 %While the original study finds no significant effect of LS immigration on \emph{Net Public Education} using individual data, our results suggest that the estimated relationship is sensitive to instrument strength.

Looking at the political parties in Columns~5-8 of Table~\ref{tab:moriconi2019_tab4_fefd}, The shift-share instruments for both HS and LS immigrants are weak in all panel regressions, with F-statistics below 10, compared to $F=21.47$ for HS and $F=17.39$ for LS in the original study. %Therefore, corresponding second-stage results should be interpreted with caution. 
The AR test statistics indicate the absence of any effect in HS specifications (Columns 5-6) and the presence of some second-stage effect only in LS specifications  (Columns 7-8) on both outcomes (positive on welfare and negative on education). The corresponding AR confidence sets are either unbounded (Columns 5, 7 and 8), indicating that the data contain limited information to precisely identify the causal effect, or bounded with  zero included (Column 6), indicating no second-stage effect. Unlike us, the original study finds only a borderline anti-redistribution (negative) effect of LS immigration on welfare expansion (Column 4, Panel~A of Table~\ref{tab:moriconi2019_tab4_original}). Although AR results show that in some cases an effect may exist, the weakness of the instrument prevents precise identification when looking at the AR CS. On the basis of the F-statistics and AR diagnostics, the second-stage coefficients from the FE and FD estimators cannot be regarded as consistent causal estimates. 

%%% POLITICAL PARTIES %%%
%%%%%%%%%%%%%%%%%%%%%%%%%%%%%%%%%%%%%%%%%%%%%%%%%%%%%%%%%%%%%%%%%%%%%%%%%%%%
%%%%% Tab 5 - cols 2-6 
%%%%%%%%%%%%%%%%%%%%%%%%%%%%%%%%%%%%%%%%%%%%%%%%%%%%%%%%%%%%%%%%%%%%%%%%%%%%
\begin{table}[t!]
\centering
 \caption{Political preferences -- Political Parties}\label{tab:moriconi2019_tab5}
 \vspace{-2mm}
\scalebox{.6}{
\begin{threeparttable}	%use it for notes
  \begin{tabular}{lcccccccc}   
\vspace{-3mm}\\   
\hline\hline
\vspace{-3mm}\\	
\emph{Sample of: }&\multicolumn{4}{c}{HS immigrants}&\multicolumn{4}{c}{LS Immigrants}\\
\cmidrule(l{.35cm}r{.25cm}){2-5}\cmidrule(l{.35cm}r{.25cm}){6-9}
&2SLS&  DML-Lasso&DML-NNet& DML-Boosting&2SLS&  DML-Lasso&DML-NNet& DML-Boosting\\
&(1)&(2)&(3)&(4)&(5)&(6)&(7)&(8)\\
\vspace{-3mm}\\
\hline
\vspace{-2mm}\\
\multicolumn{9}{c}{\textbf{Panel A:} \emph{Net Welfare State }}\\			
\multicolumn{9}{l}{\emph{Second-stage results}}\\
Share	&	0.076	&	-0.359	&	0.951***	&	1.497	&	-0.448*	&	0.189	&	1.264	&	1.782	\\
	&	-0.35	&	(0.912)	&	(0.352)	&	(0.99)	&	-0.237	&	(0.288)	&	(1.066)	&	(3.762)	\\
%P-value theta	&		&	0.694	&	0.007	&	0.13	&		&	0.512	&	0.236	&	0.636	\\
AR 95\% CS	&	[-0.471,1.329]	&	[-2.844, 0.627]	&	[0.322, 0.87]	&	[0.357, 8.611]	&	$(-\infty, -0.209]$	&	[-0.212, -0.028]	&	[-0.177, 0.14]	&	$( -\infty, +\infty ) $\\

\vspace{-3mm}\\
\multicolumn{9}{l}{\emph{First-stage results}}\\
Shift-Share IV	&	0.915**	&	0.754**	&	1.256***	&	0.626	&	0.731*	&	0.165	&	0.465	&	-0.098	\\
	&	-0.34	&	(0.367)	&	(0.369)	&	(0.527)	&	-0.401	&	(1.073)	&	(0.611)	&	(0.49)	\\
%P-value pi	&		&	0.04	&	0.001	&	0.235	&		&	0.877	&	0.446	&	0.842	\\
F stat	&	7.25	&	3.525	&	9.677	&	1.177	&	3.32	&	0.02	&	0.483	&	0.033	\\
AR $\chi^2$ stat	&	0.06	&	1.034	&	3.059*	&	6.457**	&	20.06***	&	4.546**	&	0.875	&	0.14	\\
%P-value AR $\chi^2$	&	0.81	&	0.309	&	0.08	&	0.011	&	0	&	0.033	&	0.35	&	0.709	\\

\vspace{-3mm}\\
\multicolumn{9}{l}{\emph{Quality of learners}}\\
Model RMSE	&	0.964	&	3.691	&	4.018	&	3.844	&	1.019	&	3.298	&	8.41	&	6.825	\\
MSE of l	&		&	1.152	&	1.279	&	1.142	&		&	1.152	&	1.279	&	1.142	\\
MSE of r	&		&	0.801	&	1.428	&	0.754	&		&	3.977	&	3.09	&	1.353	\\
MSE of m	&		&	1.076	&	1.078	&	0.322	&		&	2.582	&	1.853	&	0.675	\\

\vspace{-3mm}\\
\multicolumn{9}{c}{\textbf{Panel B:} \emph{Net Public Education}}\\	
\multicolumn{9}{l}{\emph{Second-stage results}}\\
Share	&	0.229	&	-0.491	&	0.025	&	1.043	&	0.333	&	-0.269	&	0.031	&	-0.75	\\
	&	-0.551	&	(0.473)	&	(0.588)	&	(1.361)	&	-0.284	&	(0.285)	&	(0.306)	&	(2.557)	\\
%P-value theta	&		&	0.3	&	0.967	&	0.443	&		&	0.345	&	0.92	&	0.769	\\
AR 95\% CS	&	[-0.731,1.998]	&	[-1.343, 1.068]	&	[0.261, 0.979]	&	[-0.417, 4.898]	&	$[0.005, +\infty)$	&	[-0.034, 0.159]	&	[0.281, 0.646]	&	$( -\infty, +\infty ) $\\

\vspace{-3mm}\\
\multicolumn{9}{l}{\emph{First-stage results}}\\
Shift-Share IV	&	0.915**	&	0.754**	&	1.256***	&	0.626	&	0.731*	&	0.165	&	0.465	&	-0.098	\\
	&	-0.34	&	(0.367)	&	(0.369)	&	(0.527)	&	-0.401	&	(1.073)	&	(0.611)	&	(0.49)	\\
%P-value pi	&		&	0.04	&	0.001	&	0.235	&		&	0.877	&	0.446	&	0.842	\\
F stat	&	7.25	&	3.525	&	9.677	&	1.177	&	3.32	&	0.02	&	0.483	&	0.033	\\
AR $\chi^2$ stat	&	0.22	&	2.464	&	0.059	&	0.319	&	3.66*	&	1.499	&	4.292**	&	0.991	\\
%P-value AR $\chi^2$	&	0.639	&	0.117	&	0.809	&	0.572	&	0.056	&	0.221	&	0.038	&	0.319	\\

\vspace{-3mm}\\
\multicolumn{9}{l}{\emph{Quality of learners}}\\
Model RMSE	&	0.895	&	2.698	&	3.386	&	4.973	&	0.959	&	2.687	&	3.647	&	6.035	\\
MSE of l	&		&	1.128	&	1.321	&	0.982	&		&	1.128	&	1.321	&	0.982	\\
MSE of r	&		&	0.801	&	1.428	&	0.754	&		&	3.977	&	3.09	&	1.353	\\
MSE of m	&		&	1.076	&	1.078	&	0.322	&		&	2.582	&	1.853	&	0.675	\\

\vspace{-3mm}\\
\hline
\vspace{-3mm}\\
Observations&	97	&	97	&	97	&	97	&	97	&	97	&	97	&	97	\\
No. clusters	&	12	&	12	&	12	&	12	&	12	&	12	&	12	&	12	\\

\hline
\end{tabular} 	
\footnotesize
\textbf{Note:} \textbf{Note:}  The table displays our estimates based on Specifications~(2), (4) and (6) of Table~5 (Panels A and B) in \cite{moriconi2019} obtained from conventional 2SLS regression with FD transformation (Columns 1 and 5), and our panel IV DML estimation with different base learners (Columns 2-4 and 6-8).
%%%%%%%
The sample is aggregated sample at regional (NUTS2) level to construct an unbalanced panel data set.The treatment and instrumental variables in Columns (1)-(2) and (5)-(6) refer to the fraction of high-skilled workers, and in Columns (3)-(4) and (7)-(8) of low-skilled workers. The dependent variable in Panel A is `Net Welfare State, and in Panel B `Net Public Education'. 
%%%%%%%
Raw control variables in all panels are: the share of tertiary sector (in log), average unemployment rate, and election year dummies. The set of controls variables in the panel IV DML estimation with NNet and Boosting does not include interaction terms because these base learners are designed to capture nonlinearities in the data. Panel IV DML estimation with Lasso (Columns 2 and 6) uses an \emph{extended dictionary} of the raw variables, including polynomials up to order three and interaction terms between all the covariates, to satisfy Lasso's weak sparsity assumption. The number of covariates used for panel IV DML estimation doubles due to the inclusion of the lags of all included covariates, following the FD (exact) approach.
%%%%%%%
Panel IV DML technical note: 2 folds, cross-fitting, hyperparameters are tuned as per Table~\ref{tab:hyperpara}. 
%%%%%%
Standard errors in parenthesis are clustered at the country level.
Significance levels: * p $<$ 0.10, ** p $<$ 0.05, *** p $<$ 0.01. 
\end{threeparttable}
}
\end{table}

%For the political parties' analysis, neither 2SLS with FE nor FD detects a causal effect of HS or LS immigration on public education expansion. For LS immigration, the effect on welfare expansion is negative and fairly significant, but this should be treated with caution as the instrument remains too weak ($F\ll10$) to identify the causal parameter reliably. This is consistent with the unbounded AR confidence sets. 

Extending the discussion on political parties,  Table~\ref{tab:moriconi2019_tab5} reports our estimates from 2SLS regressions with FD alongside panel IV DML regressions using the FD approach. 
For both HS and LS immigration, the shift-share instruments appear weak under both conventional 2SLS and panel IV DML. That is,  the corresponding first-stage F-statistics are well below the rule-of-thumb threshold of~10. The instruments are even weaker under panel IV DML, with first-stage F-statistics much lower than those from conventional 2SLS specifications. Consequently, the second-stage estimates should be interpreted with caution, as weak identification undermines the consistency of the estimator and the validity of statistical inference.
The AR diagnostic results with panel IV DML are mixed because the specification using Lasso-predictions supports the presence of some negative effect of LS immigrants on welfare expansion (Column~6 in Panel~A) while the specification using neural-network predictions provides evidence for some positive effect of LS immigrants on public education (Column~6 in Panel~A). However, in both cases the second-stage coefficients are outside the AR CS, reflecting the limited information in the data to identify the causal effect precisely and the unreliability of the estimates under weak identification.
%test statistics indicate no effect of immigration on either outcome (Panels~A and~B). In most specifications, AR CS are either disjoint or bounded and include zero, reflecting the limited information in the data to identify the causal effect precisely.}
%Two exceptions are worth noting. For HS immigration (Panel~A, Column~4), the panel IV DML estimate based on gradient boosting yields an AR test with p-value of 1\%, suggesting the presence of some effect (AR CS bounded but wide). However, given the instability of boosting in small samples, this finding should be treated cautiously (as documented in the Monte Carlo simulations in Section~\ref{sec:mcsimul}. For LS immigration (Panel~B, Column~7),  AR test statistic with neural-network rejects the null of no reduced-form effect, pointing to a possible underlying relationship. Nevertheless, the corresponding AR CS does not contain the second-stage point estimate, remarking the unreliability of that estimate under weak identification.

Overall, the negative effect of LS immigration on welfare expansion with political parties, found with conventional 2SLS, disappears in panel IV DML regressions. AR diagnostics from panel IV DML regression with Lasso reveal the possible presence of some negative effect of HS immigration on welfare expansion.

%%%%%%%%%%%%%%%%%%%%%%%%%%%%%%%%%%%%%%%%%%%%%%%%%%%%%%%%%%%%
%% MORICONI 2022
%%%%%%%%%%%%%%%%%%%%%%%%%%%%%%%%%%%%%%%%%%%%%%%%%%%%%%%%%%%%
\subsection{\citet{moriconi2022}}\label{sec:app_moriconi2022}

Table~\ref{tab:moriconi2022_tab6-10_fefd} reports our estimates obtained from conventional 2SLS regressions with FE and with FD using aggregated voters' data. The analysis is divided by skill-group: HS immigrants (Columns~1-2) and LS immigrants (Columns~3-4). The dependent variables are: \emph{Nationalism} in Panel~A, \emph{Trust in own country parliament} for political attitudes in Panel~B, and \emph{Better place} for immigration attitudes in Panel~C. 
For completeness, our Table~\ref{tab:moriconi2022_original} displays the original results for Specifications~(2) and (3) in \cite{moriconi2022}'s Table~6, and Specifications~(3) and (5) in their Table~10, estimated with pooled 2SLS.

We start comparing the effect of immigration on nationalism under FE and FD estimators (Panel~A of Table~\ref{tab:moriconi2022_tab6-10_fefd}). Regarding instrument strength, the shift-share instrument for HS immigrants (Columns 1-2) is strong with $F>16.30$,  while the instrument for LS immigrants (Columns 3-4)  is weak with $F<16.30$ under both panel estimators. %Specifically, the first-stage F-statistics obtained from the 2SLS with FE and FD specifications exceed the threshold value of 16.30 in Columns 1-2 of Panel~A, but are much lower than the threshold of 10 in Columns~3-4 of Panel~A (LS immigrants), raising concerns about weak instruments. 
Therefore, the regression results for LS immigration specification should be interpreted with extreme caution due to weak identification, unlike those for HS immigrants. By contrast, the original first-stage F-statistics (Columns 1-2 of Table~\ref{tab:moriconi2022_original}) are sufficiently large to consider both shift-share instruments strong \citep[according to][]{stock2005}. AR test statistics, not  implemented in the original study, show that there is no reduced form effect when the second-stage effect is assumed to be zero (the null hypothesis is not rejected) with both estimators, and AR 95\% CS are bounded with zero included. The original study finds a borderline significant effect (at 10\% level) for HS immigrants on nationalism (Column~1 of Table~\ref{tab:moriconi2022_original}) instead. 

%%%%%%%%%%%%%%%%%%%%%%%%%%%%%%%%%%%%%%%%%%%%%%%%%%%%%%%%%%%%%%%%%%%%%%%%%%%%
%%%%% Tab D2 - col (4) 
%%%%%%%%%%%%%%%%%%%%%%%%%%%%%%%%%%%%%%%%%%%%%%%%%%%%%%%%%%%%%%%%%%%%%%%%%%%%
\begin{table}[p]
\centering
 \caption{Nationalism intensity, immigration and attitudes towards politics and immigration -- 2SLS with FD and with FE}\label{tab:moriconi2022_tab6-10_fefd}
 \vspace{-2mm}
\scalebox{.75}{
\begin{threeparttable}	%use it for notes
  \begin{tabular}{lcccc}   
\vspace{-3mm}\\   
\hline\hline
\vspace{-3mm}\\	
&\multicolumn{2}{c}{HS immigrants}&\multicolumn{2}{c}{LS immigrants}\\
\cmidrule(l{.35cm}r{.25cm}){2-3}\cmidrule(l{.35cm}r{.25cm}){4-5}
2SLS with:&FE&FD &FE &FD\\
&(1)&(2)&(3)&(4)\\
\vspace{-3mm}\\
\hline
\vspace{-3mm}\\
\multicolumn{5}{c}{\textbf{Panel A:} \emph{Nationalism intensity of parties}}\\	
\vspace{-3mm}\\
\multicolumn{5}{l}{\emph{Second-stage results}}\\

Fr. Immigrants&	 0.003 &  -0.049&0.070  &   0.084\\
  & (0.037)   & (0.044)   &(0.048)   & (0.067) \\
AR 95\% CS   &[-0.08, 0.062]   &[-0.140, 0.025]  &[-0.014, 0.230]   &[-0.028, 0.316]   \\

\vspace{-3mm}\\
\multicolumn{5}{l}{\emph{First-stage results}}\\
Shift-Share IV	& 1.775*** &  1.476***&    0.721***   &    0.602***  \\
 &(0.267) &  (0.241)  &(0.193)   & (0.189)   \\
F stat	& 38.26   &   37.54     &    11.40   &    8.31     \\ 

\multicolumn{5}{l}{\emph{Robust Weak IV Tests}}\\
AR $\chi^2$ &    0.01   &    1.39   & 2.47   &    2.13   \\
AR $\chi^2$ p-value&  0.927   &   0.239   &    0.116   &   0.144  \\

\vspace{-3mm}\\
Observations &261&147&261&147\\
No. groups &114&114&114&114\\

\vspace{-3mm}\\
\multicolumn{5}{c}{\textbf{Panel B:} \emph{Political attitudes -- 
 Trust country parliament}}\\			
\vspace{-3mm}\\
\multicolumn{5}{l}{\emph{Second-stage results}}\\
Fr. Immigrants&	   0.052*  &  0.067   &   0.048   &  0.050   \\
&(0.031)   &(0.044)   &(0.042)   &(0.051)\\
AR CS 95\%& [0.000, 0.116]   & [-0.015, 0.174]   &[-0.030, 0.143]   &[-0.030, 0.181]   \\
\vspace{-3mm}\\

\multicolumn{5}{l}{\emph{Robust Weak IV Tests}}\\
AR $\chi^2$&    3.27   &    3.30   &   1.42   &   1.17\\
AR $\chi^2$ p-value&   0.071   &  0.069   &   0.234   &   0.279  \\

\vspace{-3mm}\\
\multicolumn{5}{c}{\textbf{Panel C:} \emph{Migration attitudes -- Better place to live}}\\	
\vspace{-3mm}\\
\multicolumn{5}{l}{\emph{Second-stage results}}\\
Fr. Immigrants&	  -0.036   &  -0.023  &  -0.040   &  -0.100*  \\
    & (0.028)   &     (0.048)   & (0.035)   &   (0.051) \\
AR CS 95\%& [-0.094, 0.016]   & [-0.168, 0.041] &[-0.125, 0.0186]   &[-0.233, -0.026]   \\
\vspace{-3mm}\\

\multicolumn{5}{l}{\emph{Robust Weak IV Tests}}\\
AR $\chi^2$&    1.73   &    0.28   &1.62   &7.99 \\
AR $\chi^2$ p-value&   0.189   &   0.595   &0.203   & 0.005 \\

\vspace{-3mm}\\
% \hline
% \vspace{-3mm}\\
\multicolumn{5}{c}{\textbf{Panels B and C}}\\
\multicolumn{5}{l}{\emph{First-stage results}}\\
Shift-Share IV	&  1.480***& 1.174***&  0.716*** &   0.647***\\  
  & (0.144)   & (0.277)   &  (0.125)   & (0.155)   \\
F stat	 &  106.48   &   19.09   &32.77    &   16.39  \\
 
\vspace{-3mm}\\
Observations  &441&327&441&327\\
No. clusters &114&114&114&114\\

\hline
\end{tabular} 	
\footnotesize
\textbf{Note:} The table reports our estimates based on Specifications~(2) and (3) of Table~6 in \cite{moriconi2022} (our Panel A), and Specifications~(3) and (5) (Panels A and B) of Table~10 in \cite{moriconi2022} (our Panels B-C). 
%%%%%%%%%%%%%
The estimates are obtained from conventional 2SLS regression with FE (Columns 1 and 3) and FD transformation (Columns 2 and 4). The original sample consists of different individual voters from twelve European countries sampled each year, which we aggregate at regional (NUTS2) level to obtain an unbalanced panel data set. The treatment and instrumental variables in Columns (1)-(2) refer to the fraction of high-skilled workers, and in Columns (3)-(4) of low-skilled workers. The number of observations differs from 2SLS with FE and with FD regressions because the first time period is removed after the first-difference transformation.
%%%%%%%%%%%%%%%%%%%%
Each panel uses a different depended variable. The raw control variables in all panels are: the share of women, average age, share of tertiary/post-tertiary education, average GDP per capita (in log), share of tertiary sector (in log), average unemployment rate, and year dummies. 
%%%%%%%%
Raw variables in the panel IV DML estimation with NNet and Boosting does not include interaction terms because these base learners are designed to capture nonlinearities in the data.
Panel IV DML with Lasso (Columns 2 and 6) uses an \emph{extended dictionary} of the raw variables, including polynomials up to order three and interaction terms between all the covariates, to satisfy Lasso's weak sparsity assumption. The number of covariates used for panel IV DML estimation doubles due to the inclusion of the lags of all included covariates, following the FD (exact) approach.
%%%%%
%Panel A: the number of observations ($NT$) is 147, and the number of groups is 114. Panels B-C: the number of observations ($NT$) is 327, and the number of groups is 114.
%%%%%%%%%
Standard errors in parenthesis are clustered at the regional level.
Significance levels: * p $<$ 0.10, ** p $<$ 0.05, \mbox{*** p $<$ 0.01.}
\end{threeparttable}
}
\end{table}

We now discuss the effect of immigration on political and immigration attitudes (Panels~B and~C of Table~\ref{tab:moriconi2022_tab6-10_fefd}). The first-stage regressions for both specifications together are the same because they use the same sample, endogenous treatment and instrumental variables. Both shift-share instruments appear strong across all specifications and panel estimators, with first-stage F-statistics exceeding the threshold of 16.30, although values are somewhat lower under the FD estimator which remain only marginally above \citet{stock2005}'s threshold. By contrast, the first-stage F-statistics in the original analysis are always above $F>35$ in both HS and LS specifications (Columns~3-8 of Table~\ref{tab:moriconi2022_original}).
In our re-analysis, there are only two cases when the AR diagnostics suggest the possible presence of a second-stage effect. First, in Columns 1-2 of Panel~B (HS specifications), the AR tests with FE and FD reject the null hypothesis of no effect at 10\% significance level; the corresponding AR CS are bounded, but include zero. Therefore, we can conclude that there is no effect of HS immigration on increased trust in own country parliament, unlike the original article which finds a positive but borderline effect. %The original result is borderline significant as well.
Second, in Column~4 of Panel~C (LS specification estimated with FD estimator), the AR test statistic is significant at 1\% level; the bounded AR confidence set identifies a negative effect that includes the estimated point estimate. The original study finds a significant negative effect, but their shift-share instrument is much stronger ($F=45.24$) than ours ($F=16.39$).
In all other cases, the AR tests fail to reject the null hypothesis and the AR CS are bounded with zero included, suggesting the absence of a statistically significant causal effect.

%%%%%%%%%%%%%%%%%%%%%%%%%%%%%%%%%%%%%%%%%%%%%%%%%%%%%%%%%%%%%%%%%%%%%%%%%%%%
%%%%% Tab D2 - col (4) 
%%%%%%%%%%%%%%%%%%%%%%%%%%%%%%%%%%%%%%%%%%%%%%%%%%%%%%%%%%%%%%%%%%%%%%%%%%%%
\begin{table}[t!]
\centering
 \caption{Outcomes and immigrant share (\citet{moriconi2022}'s results)}\label{tab:moriconi2022_original}
 \vspace{-2mm}
\scalebox{.9}{
\begin{threeparttable}	%use it for notes
  \begin{tabular}{lcccccc}   
\vspace{-3mm}\\   
\hline\hline
\vspace{-3mm}\\	
Dep. variable: & \multicolumn{2}{c}{Nationalism intensity}&\multicolumn{2}{c}{Trust country parliament}&\multicolumn{2}{c}{Better place to live}\\
\cmidrule(l{.35cm}r{.25cm}){2-3}\cmidrule(l{.35cm}r{.25cm}){4-5}\cmidrule(l{.35cm}r{.25cm}){6-7}
&(1)&(2)&(3)&(4)&(5)&(6)\\
\vspace{-3mm}\\
\hline
\vspace{-3mm}\\
\vspace{-3mm}\\
Share HS& -0.14$^{*}$& & 0.07$^*$&& -0.05&\\
      &  (0.07) & & (0.04)&&(0.03)&\\ 
Share LS& & 0.05& &0.04&&-0.05$^{**}$ \\
      & &(0.05)& & (0.04)&&(0.02)\\
\vspace{-3mm}\\
\hline
\vspace{-3mm}\\
Observations& 48,303& 48,303& 78,058& 78,058& 77,862& 77,862\\
K-P rk Wald F-stat & 32.24 & 38.72&35.77&45.07&35.77&45.24\\
Adj. R-Square &0.13&0.13&0.10&0.10&0.11&0.11\\

\vspace{-3mm}\\
\hline
\vspace{-3mm}\\
NUTS2 FE & Yes& Yes& Yes& Yes& Yes& Yes\\
Year FE & Yes& Yes& Yes& Yes& Yes& Yes\\
NUTS2 Controls & Yes& Yes& Yes& Yes& Yes& Yes\\
Individual Controls & Yes& Yes& Yes& Yes& Yes& Yes\\
\hline
\end{tabular} 	
\footnotesize
\textbf{Note:} The table displays the original results for Specifications~(2) and (3) from \cite{moriconi2022} 's Table~6, and Specifications~(3) and (5) from their Table~10. The estimation method the authors used is conventional IV estimation. 
% %%%%%%%
Standard errors in parenthesis are clustered at the regional level.
Significance levels: \mbox{* p $<$ 0.10,} \mbox{** p $<$ 0.05,} *** p $<$ 0.01. 
\end{threeparttable}
}
\end{table}

In general, FE and FD estimators widely agree on the results with the exception of LS immigrants in the specification for \emph{Better place to live} and, therefore, any difference observed in panel IV DML regressions is due to methodological rather than specification choices.

\newpage
\clearpage
\section{Monte Carlo Data Generating Process}\label{sec:mc_dgp}
\renewcommand{\theequation}{E.\arabic{equation}}
\setcounter{equation}{0}

We consider a data generating process (DGP) that resembles the specification of the re-analized empirical application.  We generate the following simulated regression model from Model~\eqref{eqn:yit}-\eqref{eqn:zit}
\begin{align}
  & Y_{it}  = D_{it}\theta + l_0(\vector X_{it}) + \alpha_i + U_{it} \\
  & D_{it}  = Z_{it}\pi + r_0(\vector X_{it}) + 0.5\alpha_i + R_{it}\\
  & Z_{it}  = m_0(\vector X_{it}) + \gamma_i+ V_{it}\\
  &\vector X_{it} = (X_{it,1}, \dots, X_{it,p})'\distas{}{}\Gamma_i+ N(0,1)\\
  & \alpha_i = \rho\Gamma_i + \sqrt{(1-\rho^2)} A_i,\\
  &  \Gamma_i\distas{}{}N(3,9), A_i\distas{}{}N(0,1), \gamma_i\distas{}{}N(0,25) \\
 &\begin{pmatrix}
    U_{it}\\
    R_{it}\\
    V_{it}
 \end{pmatrix}
 \distas{}{}N%
  \begin{bmatrix}
     \begin{pmatrix}
         0\\
         0\\
         0
     \end{pmatrix}
     , 
     \begin{pmatrix}
         1 & 0.6&0\\
         0.6 & 1&0\\
         0&0&0.25
     \end{pmatrix}
 \end{bmatrix},
\end{align}
where the target parameter is $\theta=0.5$; the parameter $\rho=0.9$ controls the extent of the influence of the fixed effect over the random effect; and $p=30$ is the number of control variables. We set up two Monte Carlo simulation designs: a setting in which the instrument is strong with $\pi= 0.8$, and one where the instrument is weak with $\pi= 0.001$. Treatment endogeneity is induced both by the correlation between the error terms $U_{it}$ and $R_{it}$, \mbox{$\sigma_{ur} = 0.6$}, and by the fixed effects $\alpha_i$ through $\Gamma_i$, which captures the dependence of unobserved heterogeneity on the included exogenous covariates.
%The vector of confounding variables is generated as $\vector X_{it} = (X_{it,1}, \dots, X_{it,p})'\distas{}{}\Gamma_i+ N(0,1)$ with $p=30$. 

The nuisance functions $(l_0, m_0, r_0)$ are modelled as follows:
%three DGPs: (1) linear in $\vector X_{it}$; (2) nonlinear in $\vector X_{it}$ resembling empirical specifications; (3) discontinuous (with interactions) in  $\vector X_{it}$. %The original dataset comprises N=5,000 cross-sections observed over $T=10$ periods.
\begin{align}
    l_0(\vector X_{it})  &= a_1  X_{it,1} + a_2  X_{it,3}  + a_3  X_{it,1}\cdot \one\{X_{it,1}>0\}  \label{eqn:l0}  \\
    r_0(\vector X_{it})  &=  b_1  X_{it,1} + b_2  X_{it,3}  + b_3  X_{it,1}\cdot \one\{X_{it,1}>0\} \label{eqn:r0}  \\
    m_0(\vector X_{it})  &=  c_1  X_{it,1} + c_2  X_{it,3}  + c_3  X_{it,1}\cdot \one\{X_{it,1}>0\} \label{eqn:m0}  
\end{align}
where $a_j=b_j=c_j=0.5$ for $j=\{1,2,3\}$, and $\one\{.\}$ is an indicator operator that transforms the continuous variable $X_{it,1}$ into a binary variable by assigning value of one to positive realisations of $X_{it,1}$ and zero otherwise.

In this design, nonlinearity enters the nuisance functions via the interaction term $ X_{it,1}\cdot \one\{X_{it,1}>0\}$. Practically, an example of the chosen form of nonlinearity is as follows:  suppose that $X_{it,1}$ is the continuous variable $Age_{it}$, and $\one\{X_{it,1}>0\}$ is a binary indicator of the variable when a specific a specific age threshold is surpassed, e.g. $\one\{Age_{it}>25\}$. In this example, the interaction the interaction term allows the slope of \emph{Age} to change when \emph{Age} is greater than 25, effectively modelling a kink (or slope shift) at the cut-off.

To allow for sparsity ($s\ll p$), we include many control variables ($p=30$), but only a subset of them ($s=2$) are relevant and enter the model both linearly and through nonlinear interaction terms. In practice, the number of variables $p$ doubles with the FD approach described in the panel IV DML Algorithm~\ref{alg:xtdml} because the first-order lags of the included variables are included in the regression.
Panel IV DML estimates the nuisance functions flexibly using Lasso, a single-layer neural network (NNet), and gradient boosting with 100 trees (Boosting). Conventional 2SLS, by contrast, estimates these functions linearly.

Overall, this design allows us to investigate the performance of our panel IV DML estimator over the conventional 2SLS when the true functional form of the covariates is flexible.

\section{Hyperparameter Tuning}\label{sec:tuning}
\renewcommand{\thetable}{F.\arabic{table}}
\setcounter{table}{0}
Hyperparameter tuning is critical for achieving accurate effect estimation in machine learning, regardless of the choice of base learners or estimators. Using default values, whether suggested by software packages or the literature, can substantially limit learner performance and introduce bias into the causal estimand \citep{machlanski2023, machlanski2024, bach2024hyper}. 

In the empirical applications (Section~\ref{sec:empirical}) and Monte Carlo simulations (Section~\ref{sec:mcsimul}), hyperparameters were tuned according to the procedures summarized in Table~\ref{tab:hyperpara}. Specifically,  Lasso penalty parameter is selected to minimize the cross-validated mean error, whereas the hyperparameters for gradient boosting and neural networks are tuned using grid search \citep{bergstra2012}.  We set the hyperparameter optimizer to try five distinct values per hyperparameter, randomly selected from the intervals specified in Table~\ref{tab:hyperpara}, within each evaluation and terminate the optimization at the fifth evaluation. 

\begin{table}[h!]
\centering
 \caption{Hyperparameter tuning}\label{tab:hyperpara}
\scalebox{.75}{
\begin{threeparttable}	%use it for notes
   \begin{tabular}{lllll}  
\hline\hline
Learner&Hyperparamter& Value of parameter in set & Description\\	
\hline
\vspace{-2mm}\\
Lasso&\texttt{lambda.min}& cross-validated&$\lambda$ equivalent to minimum mean cross-validated error\\
\vspace{-2mm}\\

% CART&\texttt{cp}&real value in \{0.001,0.1\}&Prune all nodes with a complexity less than cp from the printout.\\
% &\texttt{minbucket} & default&Minimum number of observations in any terminal <leaf> node.\\
% %&minsplit&minbucket$\times$3&Minimal node size to split at.\\
% %Minimum number of observations that must exist in a node for a split to be attempted. \\
% &\texttt{maxdepth}&integer in \{2,20\}&Maximum depth of any node of the final tree.\\

% \vspace{-2mm}\\
% RF& \texttt{num.trees}&1000&Number of trees in the forest.\\
% &\texttt{min.node.size}&default&Minimal node size to split at. \\
% &\texttt{max.depth}&integer in \{2,10\}&Maximum depth of any node of the final tree.\\
% &\texttt{mtry}& all covariates &The number of covariates, randomly sampled, to split at each node.\\
% % &\texttt{importance}& impurity & The Gini index for classification; the variance of the responses for regression.\\
% \vspace{-2mm}\\

\vspace{-2mm}\\
Gradient Boosting&\texttt{lambda}&real value in \{0,2\}&L2 regularization term on weights.\\
&\texttt{maxdepth}&integer in \{2,10\}&Maximum depth of any node of the final tree.\\
&\texttt{nrounds}& &Number of decision trees in the final model\\
&& 1000&\hspace{3mm} In empirical applications.\\
&& 100&\hspace{3mm} In Monte Carlo simulations.\\

\vspace{-2mm}\\
NNET& \texttt{maxit}&100&Maximum number of iterations.\\
&\texttt{MaxNWts}&2000&The maximum allowable number of weights.\\
&\texttt{trace}&FALSE&Switch for tracing optimization..\\
&\texttt{size}& integer in \{2,10\}&Number of units in the hidden layer.\\
&\texttt{decay}&double in \{0,0.5\}& Parameter for weight decay.\\
\vspace{-2mm}\\
\hline
\end{tabular} 
\begin{tablenotes}[para,flushleft]	
\footnotesize{Note: The hyperparameters of the base learners chosen to model the nuisance functions are tuned in each Monte Carlo replication via grid search, that evaluates each possible combination of hyperparameters' values in the grid  \citep{bergstra2012}. We set the hyperparameter optimizer to try five distinct values per hyperparameter randomly selected from the specified intervals, within each evaluation and terminate the optimization at the fifth evaluation.
}
\end{tablenotes}
% \footnotesize
% \renewcommand{\baselineskip}{11pt}
% \textbf{Note:}  The hyperparameters of the base learners chosen to model the nuisance functions are tuned in each Monte Carlo replication via grid search, that evaluates each possible combination of hyperparameters' values in the grid  \citep{bergstra2012}. We set the hyperparameter optimizer to try 10 distinct values per hyperparameter randomly selected from the specified intervals, within each evaluation and terminate the optimization at the 10th evaluation.
\end{threeparttable}}
\end{table}

\newpage
\section{Data Collection for Survey of AER Articles}\label{sec:meta}
The survey is conducted by manually collecting information on 477 empirical articles published in the American Economic Review (AER) between 2011 and 2018, without the use of automated text-mining techniques.\footnote{This dataset was originally constructed for different projects, several years ago, and now reused for the present study.} Articles classified as purely theoretical (i.e., without any econometric methods) were excluded from the dataset.

For each article, we recorded key information such as the author names, publication date (year and month), type of data analyzed (cross-sectional, panel, time series, or a combination), and the estimation methods employed. We grouped the estimation methods into broad categories: Least Squares (OLS, WLS, and GLS), Maximum Likelihood (probit and logit types regressions), IV methods (GMM, IV, 2SLS, 3SLS), and other non-linear techniques. Any remaining specialized estimators were placed in the other non-linear techniques category.
%%%%%%%%%%%%%
When the analysis in the article used multiple estimation techniques, we recorded both the primary and secondary estimation methods. The primary method is the main technique used (typically for the key regression), while the secondary method is the second-most important. Often the secondary method serves as an alternative estimator for the main relationship or is used in a supplementary analysis. We did not count methods used only for robustness checks as secondary, since those serve solely to support the main analysis. %For instance, 71 articles used OLS as the primary method and an IV approach as the secondary method (46 of which involve panel data)—a common combination employed to address potential endogeneity concerns.

\end{document}